\documentclass[aps,twocolumn,prc,showpacs,preprintnumbers,amssymb,nofootinbib,english]{revtex4-2}
\usepackage{mathrsfs}
\usepackage{hyperref}
\hypersetup{
	colorlinks=true,
	linkcolor=blue,
	filecolor=magenta,      
	urlcolor=cyan,
	pdftitle={Overleaf Example},
	pdfpagemode=FullScreen,
}
\usepackage{graphicx}
\usepackage{amsmath, amssymb}
\usepackage{babel}
\usepackage{color}
\usepackage{slashed}
\usepackage[T1]{fontenc}
\usepackage{titlesec}
\def\beqn{\begin{eqnarray}}
	\def\eeqn{\end{eqnarray}}
\def\barr{\begin{array}}
	\def\earr{\end{array}}
\def\btab{\begin{tabular}}
	\def\etab{\end{tabular}}
\def\bite{\begin{itemize}}
	\def\eite{\end{itemize}}
\def\bcen{\begin{center}}
	\def\ecen{\end{center}}

\newcommand{\p}{\mathbf{p}}
\newcommand{\q}{\mathbf{q}}

\begin{document}
	
	\title{Data-driven reevaluation of $ft$ values in superallowed $\beta$ decays}
	
	\author{Chien-Yeah Seng$^{1,2}$}
	\author{Mikhail Gorchtein$^{3,4}$}

	\affiliation{$^{1}$Facility for Rare Isotope Beams, Michigan State University, East Lansing, MI 48824, USA}
	\affiliation{$^{2}$Department of Physics, University of Washington,
		Seattle, WA 98195-1560, USA}
	\affiliation{$^{3}$Institut f\"ur Kernphysik, Johannes Gutenberg-Universit\"{a}t,\\
		J.J. Becher-Weg 45, 55128 Mainz, Germany}
	\affiliation{$^{4}$PRISMA$^+$ Cluster of Excellence, Johannes Gutenberg-Universit\"{a}t, Mainz, Germany}

	\date{\today}
	
	\begin{abstract}
		
		We present a comprehensive re-evaluation of the $ft$ values in superallowed nuclear $\beta$ decays crucial for the precise determination of $V_{ud}$ and low-energy tests of the electroweak Standard Model. It consists of the first, fully data-driven analysis of the nuclear $\beta$ decay form factor, that utilizes isospin relations to connect the nuclear charged weak distribution to the measurable charge distributions. This prescription supersedes  previous shell-model estimations, and allows for a rigorous quantification of theory uncertainties in $f$ which is absent in the existing literature. Our new evaluation shows an overall downward shift of the central values of $f$ at the level of 0.01\%. 
		
	\end{abstract}
	
	\maketitle
	
	
	\section{Introduction}
	
	The top-row Cabibbo-Kobayashi-Maskawa (CKM) matrix element $V_{ud}$ is a fundamental parameter in the Standard Model (SM) that governs the strength of charged weak interactions involving up and down quarks. Its precise determination constitutes an important component of the low-energy tests of the SM and the search of physics beyond the Standard Model (BSM) at the precision frontier. Currently, beta transitions between isospin $T=1$, and spin-parity $J^P=0^+$ nuclear states (the so-called superallowed nuclear $\beta$ decays) and free neutron are the two competing candidates for the most precise determination of $V_{ud}$. The advantage of the former is the existence of the many nuclear transitions that had been measured over decades and averaged over (see, e.g. Reg.\cite{Hardy:2020qwl} and references therein), but the existence of nuclear structure effects complicates the theory analysis. In contrast, neutron decay is limited by the experimental precision but is free from nuclear uncertainties and is theoretically cleaner. 
	
	The recent improved measurement of the neutron lifetime $\tau_n$ by UCN$\tau$~\cite{UCNt:2021pcg} and the axial coupling constant $g_A$ by PERKEO-III~\cite{Markisch:2018ndu} have made the precision of $V_{ud}$ from neutron $\beta$ decay almost comparable to that from superallowed nuclear $\beta$ decays, but a mild tension starts to develop between the two values\footnote{We have rescaled the result in Ref.\cite{Hardy:2020qwl} by the new averaged nucleus-independent radiative corrections in Ref.\cite{Gorchtein:2023srs}; there are other choices of average, e.g. Ref.\cite{Cirigliano:2022yyo}, and there is an ongoing effort to reach a community consensus on this input~\cite{consensus}.}:
	\begin{equation}
		|V_{ud}|_{0^+}=0.97361(31)~\text{\cite{Hardy:2020qwl}}~,~ |V_{ud}|_n^\text{precise}=0.97404(42)~\text{\cite{Gorchtein:2023srs}} ~.
	\end{equation}
	Notice that, however, there are some consistency issues on experimental inputs to the neutron decay, for example the well-known beam-bottle discrepancy for $\tau_n$ (see, e.g. Ref.\cite{Tan:2023mpj} and references therein) and the discrepancy between the $g_A$ extracted from the $\beta$-asymmetry $A$~\cite{Markisch:2018ndu} and from the electron-neutrino correlation $a$~\cite{Wietfeldt:2023mdb,Beck:2023hnt}. Adopting the latest PDG-averaged values of $\tau_n$ and $g_A$~\cite{ParticleDataGroup:2022pth} results in $|V_{ud}|_n$ with a largely-inflated experimental uncertainty:
	\begin{equation}
		|V_{ud}|_n^\text{PDG-av}=0.97433(87)
	\end{equation}
	which overshadows the tension above.
	
	An inconsistency also occurs in the determination of $V_{us}$: From semileptonic kaon decays one obtains $|V_{us}|_{K_{\ell 3}}=0.22308(55)$~\cite{Seng:2021nar,Seng:2022wcw} (with $N_f=2+1+1$ lattice determination of the $K^0\rightarrow \pi^-$ transition matrix element~\cite{Carrasco:2016kpy,Bazavov:2018kjg,FlavourLatticeAveragingGroupFLAG:2021npn}), whereas from leptonic kaon decays one obtains $|V_{us}|_{K_{\mu 2}}=0.2252(5)$~\cite{ParticleDataGroup:2022pth} which is significantly larger. These different values of $V_{ud}$ and $V_{us}$, combining with $|V_{ub}|_{K_{\ell 3}}=3.82(20)\times10^{-3}$~\cite{ParticleDataGroup:2022pth}, give very different results in the test of the first-row CKM unitarity $|V_{ud}|^2+|V_{us}|^2+|V_{ub}|^2=1$. Just for an illustration, combining $|V_{ud}|_{0^+}$ and $|V_{us}|_{K_\ell 3}$ give a $3.6\sigma$ deficit of the first-row CKM unitarity, but changing $|V_{ud}|_{0+}$ to $|V_{ud}|_{n}^\text{precise}$ the deficit reduces to $1.7\sigma$. A recent global analysis that took into account all these different determinations reported a $2.8\sigma$ unitarity deficit~\cite{Cirigliano:2022yyo}.
	Given its profound impact on the SM precision tests, it is important to understand the origin of discrepancies between different experimental determinations of the first-row CKM matrix elements.

	It is a commonplace for low-energy precision tests that the main  limitation in precision comes from radiative corrections that are sensitive to the effects of strong interaction which is described by Quantum Chromodynamics (QCD). At low energies QCD is nonperturbative, which complicates the uncertainty estimation of theory calculations. 
	In the recent years, effort has been put in developing methods that would allow to compute such corrections to $\beta$ decay with a controlled systematics. 
	The dispersion relation (DR)~\cite{Seng:2018yzq,Seng:2018qru,Shiells:2020fqp}, effective field theory (EFT)~\cite{Cirigliano:2022hob,Cirigliano:2023fnz} and lattice QCD~\cite{Feng:2020zdc,Yoo:2023gln,Ma:2023kfr} analyses have ensured a high-precision determination of the single-nucleon radiative correction. The SM theory uncertainties in the free neutron decay are believed to be firmly under control at the precision level of $10^{-4}$. 
	
	The theory for superallowed nuclear $\beta$ decays is more involved due to the presence of specifically nuclear corrections. 
	This fact is reflected in the master formula for the extraction of $V_{ud}$~\cite{Hardy:1975eq}\footnote{Throughout this paper, we take $\hbar=c=1$, but keep $m_e$ explicit unless mentioned otherwise.},
	\begin{equation}
		|V_{ud}|_{0^+}^2=\frac{\pi^3\ln 2}{G_F^2m_e^5\mathcal{F}t(1+\Delta_R^V)}~.\label{eq:Vudmaster}
	\end{equation}
	Above, $\Delta_R^V$ is the free-nucleon radiative correction which is also present in neutron decay. All nuclear structure effects are absorbed into the so-called $\mathcal{F}t$ value~\cite{Towner:2002rg},
	\begin{equation}
		\mathcal{F}t=ft(1+\delta_\text{R}')(1+\delta_\text{NS}-\delta_\text{C})~,\label{eq:Ft}
	\end{equation}
	where $t$, the partial half-life, can be obtained from the experimental branching ratio after accounting for the small correction from the electron capture fraction~\cite{Hardy:2004id,Bambynek:1977zz}. All remaining quantities in the expression above require nuclear theory inputs at either tree or loop level. First, $\delta_\text{R}'$ is known as the nucleus-dependent {\it outer} radiative correction, which is calculable order-by-order with Quantum Electrodynamics (QED) assuming the nucleus as a point charge~\cite{Sirlin:1967zza,Sirlin:1987sy,Sirlin:1986cc}. The remaining radiative corrections that depend on the  nuclear structure are contained in $\delta_\text{NS}$, which has previously been studied in the nuclear shell model~\cite{Barker:1991tw,Towner:1992xm,Towner:1994mw,Towner:2002rg,Towner:2007np}.
	Furthermore, $\delta_\text{C}$ represents the isospin-symmetry-breaking (ISB) correction to the Fermi matrix element. This correction has been object of study by the nuclear theory and experimental community over the past 6 decades~\cite{MacDonald:1958zz,Towner:2007np,Satula:2016hbs,Ormand:1995df,Liang:2009pf,Auerbach:2008ut,Towner:2010bx,Melconian:2011kk,Park:2014lja,Malbrunot-Ettenauer:2015kda,Xayavong:2017kim,Bencomo:2019qzx,Iacob:2020mhu,Martin:2021bud}. Both $\delta_\text{NS}$ and $\delta_\text{C}$ have recently been under renewed scrutiny~\cite{Miller:2008my,Miller:2009cg,Seng:2018qru,Gorchtein:2018fxl,Condren:2022dji}, and new methods were devised to study them either using nuclear ab-initio methods~\cite{Seng:2022cnq,Seng:2023cvt} or by relating them to  experimental measurements~\cite{Seng:2022epj}, which we will not discuss here. 
	
	The focus of this paper is the statistical rate function $f$ in superallowed $\beta$ decays. It represents the phase space integral over the spectrum of the positron originating from a $\beta$ decay process $\phi_i\rightarrow \phi_f e^+\nu_e$.  At the leading order it is fixed by the atomic mass splitting (i.e. the $Q_\text{EC}$ value). However,  a number of effects that lead to sizable corrections to the spectrum require atomic and nuclear theory inputs, and have to be included in $f$. Among these are the distortion of the outgoing positron wave function in the Coulomb field of the daughter nucleus, the nuclear form factors, screening effects from atomic electrons, recoil corrections etc. In principle, each of these inputs bears its own theory uncertainty which must be accounted for in the total error budget.  
	Unfortunately, in most existing literature, including the series of reviews by Hardy and Towner~\cite{Hardy:2004id,Hardy:2008gy,Hardy:2014qxa,Hardy:2020qwl}, only the experimental uncertainty of $Q_\text{EC}$ is  included in the evaluation of $f$. Here we address the validity of this assumption, given the precision goal of $10^{-4}$ for the extraction of $V_{ud}$.
	
	A quantity of fundamental importance in the determination of the statistical rate function is the charged weak form factor $f_+(q^2)$, defined through the (relativistic) nuclear matrix element of the vector charged current\footnote{Here, the Quantum Field Theory (QFT) plane wave states are normalized as ${}_\text{QFT}\langle \phi(p')|\phi(p)\rangle_\text{QFT}=(2\pi)^32E_p\delta^{(3)}(\vec{p}-\vec{p}')$.}:
	\begin{align}
		&{}_\text{QFT}\langle \phi_f(p_f)|(J_W^{\dagger\mu}(0))_V|\phi_i(p_i)\rangle_\text{QFT} \nonumber\\
		&= f_+(q^2)(p_i+p_f)^\mu+f_-(q^2)(p_i-p_f)^\mu~,\label{eq:betaFF}
	\end{align}
	with $q^2=(p_i-p_f)^2$. In nuclear physics it is common to use the Breit frame where $q$ only has the spatial component. The contribution of $f_-$ to the decay rate is suppressed simultaneously by ISB and kinematics, so only $f_+$ is relevant. 
	After scaling out its $\vec{q\,}^2=0$ value which is just the Fermi matrix element $M_F$ ($=\sqrt{2}$ in the isospin limit), one can perform a Fourier transform\footnote{All distributions in this paper are normalized as $\int_0^\infty 4\pi r^2\rho(r)dr=1$.}
	\begin{equation}
		f_+(q^2)=M_F\int d^3x e^{-i\vec{q}\cdot \vec{x}}\rho_\text{cw}(r)~,\label{eq:rhocw}
	\end{equation}
	which defines the nuclear charged weak
	distribution $\rho_\text{cw}(r)$; it is essentially the distribution of ``active'' protons eligible to transition weakly into a neutron in a nucleus. Obviously, $\rho_\text{cw}(r)$ is a basic property of the nucleus, just like the nuclear charge distribution $\rho_\text{ch}(r)$. Yet, in the literature they are treated with very different levels of rigor: $\rho_\text{ch}(r)$ was deduced from experimental data where uncertainties are (in principle) quantifiable, whereas $\rho_\text{cw}(r)$ is evaluated using simplified nuclear models. This may introduce an uncontrolled systematic uncertainty and neglects the fact that the two distributions are correlated. 
	
	The purpose of this work is a careful re-evaluation of $f$ with a more rigorous, data-driven error analysis. In particular, we adopt the strategy pioneered in Ref.\cite{Seng:2022inj} that connects $\rho_\text{cw}(r)$ to the charge distributions of the members of the superallowed isotriplet using model-independent isospin relations. This prescription transforms the non-quantifiable model uncertainty in the usual approach to $\rho_\text{cw}(r)$ into uncertainty estimates that are derived from experimental ones under the only assumption of an approximate isospin symmetry. Furthermore, the new approach automatically accounts for the correlation between the Fermi function and the decay form factor, and treats their uncertainties on the same footing. We also analyze possible uncertainties from secondary effects, such as the screening corrections by the atomic electrons. A necessary condition to apply the new isospin-based prescription is that at least two out of three nuclear charge radii in a nuclear isotriplet must be experimentally known, which is currently satisfied by 15 measured superallowed transitions. We report the newly-calculated $f$ for these transitions with a more robust uncertainty estimate. Our result lays a foundation for the future, more rigorous extraction of $V_{ud}$ from superallowed $\beta$ decays. 
	
	This work is organized as follows. In Sec.\ref{sec:f} we introduce the statistical rate function specifying various correction terms. A particular emphasis is put on the shape factor that depends on the charged weak distribution. In Sec.\ref{sec:isospin} we describe the isospin relations that connect different electroweak distribution functions. Sec.\ref{sec:distparameter} is the central part of this work, where we describe in full detail our procedure of selecting the nuclear charge distribution data that we use for the data-driven analysis. In Sec.\ref{sec:secondary} we discuss our treatment of the secondary nuclear/atomic structure effects that enter $f$. We present our final results in Sec.\ref{sec:final} and discuss their influence and prospects. Some useful formulas on the solutions of the Dirac equation, the shape factor and the nuclear charge distributions can be found in the Appendix.

	\section{\label{sec:f}Statistical rate function and the shape factor}
	
	We study the superallowed $\beta^+$ decay, $\phi_i\rightarrow\phi_f e^+\nu_e$, where we denote the positron energy and momentum as
	as $E\equiv E_e$ and $\vec{p}\equiv \vec{p}_e$, with $\p=|\vec{p}|$. The positron end-point energy of the decay is given by $E_0^\text{full}\equiv(M_i^2-M_f^2+m_e^2)/(2M_i)$, but upon neglecting recoil corrections it can be approximated as $E_0\equiv M_i-M_f$. Before applying various corrections, the uncorrected differential decay rate is proportional to 
	$\p E(E_0-E)^2$.
	The statistical rate function $f$ is defined as the integrated decay rate in atomic units ($\hbar=c=m_e=1$).
	
	Ref.\cite{Hayen:2017pwg} provided an in-depth survey of some 12 different types of atomic/nuclear corrections that should be applied to formula above for a generic allowed $\beta$ decay. For superallowed decays of $0^+$ nuclei, the number of relevant corrections is reduced. Therefore, following Refs.\cite{Hardy:2004id,Hardy:2008gy},
	we express the statistical rate function as
	\begin{equation}
		f=m_e^{-5}\int\limits_{m_e}^{E_0}\p E(E_0-E)^2F(E)C(E)Q(E)R(E)r(E)dE\,,\label{eq:fformula}
	\end{equation}
	where we have arranged the correction factors in decreasing degrees of importance: (1) The Fermi function $F(E)$, (2) The shape factor $C(E)$, (3) The atomic shadowing correction $Q(E)$, (4) The kinematic recoil correction $R(E)$, and (5) The atomic overlap correction $r(E)$. 
	All five corrections depend on the nucleus, which is usually denoted by carrying the daughter nucleus charge $Z$ as a second argument, but we suppress this dependence for compactness.
	In this work we classify the former two corrections as {\it primary}, as their sizes are the largest and, more importantly, they are sensitive to the details of nuclear charge distributions. These corrections will be evaluated consistently using the most recent nuclear distribution data. 
	The latter three corrections, on the other hand, are classified as {\it secondary} as their sizes are smaller and are insensitive to the shape of the nuclear charge distribution. We will not treat these corrections differently than in the literature (except for a more careful account for theory uncertainties).

	\begin{figure}
		\centering
		\includegraphics[width=0.8\columnwidth]{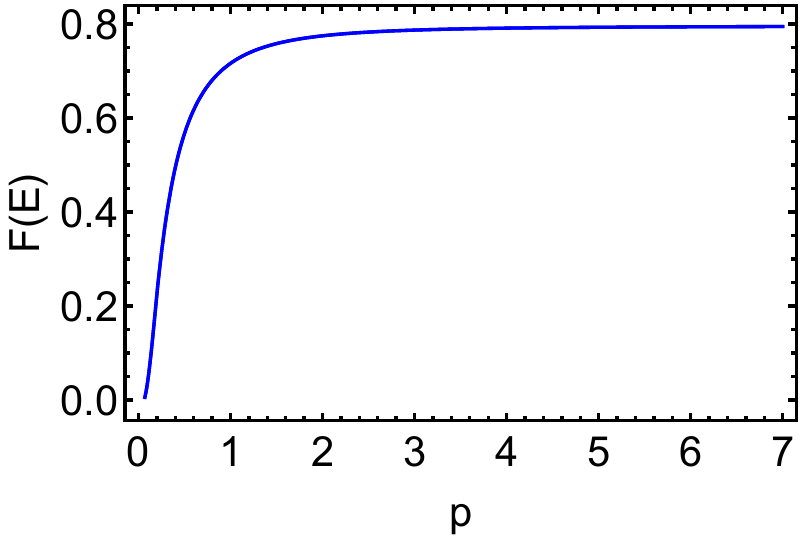}
		\caption{A plot of the Fermi function $F(E)$ with respect to the magnitude of the positron momentum $\p$ (in units of $m_e$) for $^{22}$Mg. The error band due to uncertainties from the nuclear charged distribution parameters is too small to be visible. }
		\label{fig:Fermi}
	\end{figure}
	
	We start from the largest correction, the Fermi function $F(E)$ that accounts for the Coulomb interaction between the outgoing positron and the \textit{daughter} nucleus~\cite{Fermi:1934hr}. Historically, it was first derived by solving the Dirac equation of the charged lepton under the Coulomb potential of a point-like nucleus, for which an analytic solution exists. This solution diverges at $r=0$ where the charge density is infinite, so it was instead evaluated at an arbitrarily-chosen nuclear radius $R$~\cite{konopinski1941fermi}. Corrections due to the finite nuclear charge density could then be added on top of it~\cite{behrens1969landolt,Calaprice:1976jbi}. Here we do not adopt this two-step approach, but rather solving the full Dirac equation numerically with a given nuclear charge distribution. The numerical solution is finite at $r=0$, from which we can define the Fermi function as: 
	\begin{equation}
		F(E)=\frac{f_{+1}^2(0)+g_{-1}^2(0)}{2\p^2}=\frac{\alpha_{+1}^2+\alpha_{-1}^2}{2\p ^2}~,\label{eq:Fermi}
	\end{equation}
	where the coefficients $\alpha_{\pm k}$ ($k=1$ in this case) come from the solution of the radial Dirac equation, detailed in Appendix~\ref{sec:radial} and \ref{sec:innersol}. Fig.\ref{fig:Fermi} shows the typical shape of the Fermi function for $\beta^+$ decay: since the Coulomb force is repulsive for a positron, the probability of its existence at $r=0$ with low energy is suppressed. 
	
	The second largest correction is the shape factor $C(E)$, which incorporates the influence of the $\beta$ decay form factor in Eq.\eqref{eq:betaFF} (or equivalently, the charged weak distribution in Eq.\eqref{eq:rhocw}). A closed expression was  obtained by Behrens and B\"{u}hring~\cite{behrens1982electron}:
	\begin{equation}
		C(E)=\sum_k\lambda_k\left\{M_0^2(k)+m_0^2(k)-\frac{2\mu_k\gamma_k}{kE}M_0(k)m_0(k)\right\}~,\label{eq:shape}
	\end{equation}
	where $k=+1,+2,...$.
	The involved Coulomb functions are
	\begin{equation}
		\lambda_k=\frac{\alpha_{-k}^2+\alpha_{+k}^2}{\alpha_{-1}^2+\alpha_{+1}^2}~,~\mu_k=\frac{\alpha_{-k}^2-\alpha_{+k}^2}{\alpha_{-k}^2+\alpha_{+k}^2}\frac{kE}{\gamma_k}~,
	\end{equation}
	where $\gamma_k=\sqrt{k^2-\alpha^2Z_f^2}$, with $Z_f$ the atomic number of the \textit{daughter} nucleus.
	The functions that depend on $\rho_\text{cw}(r)$ are:
	\begin{eqnarray}
		M_0(k)&=&\frac{\sqrt{k}}{(2k-1)!!}\int_0^\infty 4\pi r^2dr\rho_\text{cw}(r)(\p r)^{k-1}\nonumber\\
		&&\times\left[H_k(r)j_{k-1}(E_\nu r)-\frac{r}{R}D_k(r)j_k(E_\nu r)\right]\nonumber\\
		m_0(k)&=&\frac{\sqrt{k}}{(2k-1)!!}\int_0^\infty 4\pi r^2dr\rho_\text{cw}(r)(\p r)^{k-1}\nonumber\\
		&&\times\left[h_k(r)j_{k-1}(E_\nu r)-\frac{r}{R}d_k(r)j_k(E_\nu r)\right]~\label{eq:M0km0k}
	\end{eqnarray}
	where $E_\nu\approx E_0-E$ is the neutrino energy, and the functions $H_k$, $h_k$, $D_k$ and $d_k$ are defined in Eq.\eqref{eq:HhDd}. Notice that the overall Fermi matrix element has been factored out from the definitions above. The derivation of this master formula can be found in Appendix~\ref{sec:master}.
	One can also check that it reduces to the simple expression in Ref.\cite{Seng:2022inj} upon switching off the electromagnetic interaction. 
	
	The series in Eq.\ref{eq:shape} converges very fast. In fact, explicit calculation shows that the $k=2$ correction to $f$ is smaller than 0.0003\% for all measured transitions (i.e. up to $A=74$), therefore it is sufficient retain only the $k=1$ term which greatly simplifies the analysis. 
	
	\section{\label{sec:isospin}Isospin formalism}
	
	\begin{figure}
		\centering
		\includegraphics[width=0.8\columnwidth]{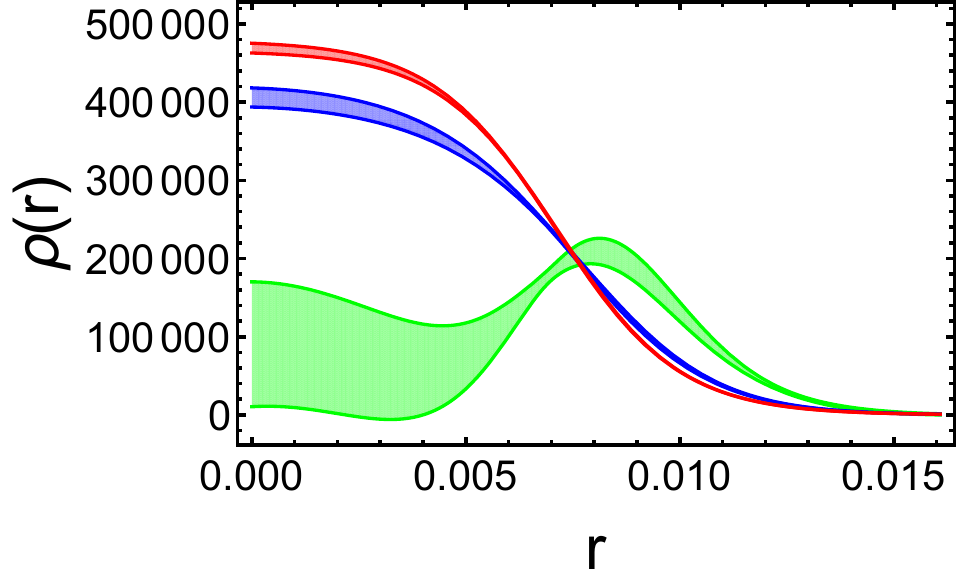}
		\caption{Plot of the nuclear charge distribution $\rho_\text{ch}(r)$ in atomic units ($\hbar=c=m_e=1$) for $^{22}$Mg (blue), $^{22}$Ne (red), and the corresponding charged weak distribution $\rho_\text{cw}(r)$ (green).  The selection of nuclear charge distribution data is explained in Sec.\ref{sec:distns}.}
		\label{fig:distributions}
	\end{figure}
	
	In the survey by Hardy and Towner~\cite{Hardy:2004id}, the weak form factor was evaluated in the impulse approximation where nucleus is treated as a collection of non-interacting nucleons.
	In this formalism, the nuclear matrix element of the weak transition operator $\hat{O}$ reads
	\begin{equation}
		\langle \phi_f|\hat{O}|\phi_i\rangle=\sum_{\alpha\beta}\langle\alpha|\hat{O}|\beta\rangle \langle \phi_f|a_\alpha^\dagger a_\beta|\phi_i\rangle~,
	\end{equation}
	where $\{\alpha,\beta\}$ are single-nucleon states, $\{a_\alpha^\dagger,a_\beta\}$ are their corresponding creation and annihilation operator, $\langle \alpha|\hat{O}|\beta\rangle$ is the single-nucleon matrix element, and $\langle \phi_f|a_\alpha^\dagger a_\beta|\phi_i\rangle$ the one-body density matrix element evaluated with shell model. In this formalism the Fermi function and the shape factor are completely decoupled, and the theory error from the shell model calculation is not quantifiable. 
	
	An alternative approach was adopted by Wilkinson in Ref.\cite{Wilkinson:1993hx}. It consists of first identifying $\rho_\text{cw}$ with $\rho_\text{ch}$ at zeroth order and adding a correction that is assumed to be small,
	\begin{equation}
		\rho_\text{cw}(r)=\rho_\text{ch}(r)+\delta\rho(r)~,
	\end{equation}
	with the latter estimated in the nuclear shell model. Ref.\cite{Hayen:2017pwg} interpreted $\delta\rho (r)$ as a consequence of ISB (Section F) and assumed to be small. However, we will show that the size of $\delta \rho(r)$ is enhanced and is comparable to  $\rho_\text{ch}(r)$, hence it cannot be taken as a small correction.

	In this work we perform a consistent treatment of $F(E)$ and $C(E)$ using the isospin formalism, coined in the earlier days the conserved vector current (CVC) hypothesis~\cite{Gershtein:1955fb,Feynman:1958ty,Holstein:1974zf}. It arises from the expressions of the vector charged weak and electromagnetic current:
	\begin{align}
		(J_W^{\dagger \mu})_V&=\bar{d}\gamma^\mu u\,,\\
		J_\text{em}^\mu &=\frac{1}{6}(\bar{u}\gamma^\mu u+\bar{d}\gamma^\mu d)+\frac{1}{2}(\bar{u}\gamma^\mu u-\bar{d}\gamma^\mu d)\,,\nonumber
	\end{align}
	where the former is purely isovector, while the latter has both isoscalar and isovector components. It is therefore the presence of the isoscalar electromagnetic current that gives rise to a non-zero $\delta\rho(r)$, even in absence of ISB. To connect the nuclear matrix elements of the two currents, we construct linear combinations which subtract out the matrix element of the isoscalar current. 
	%
		%
		%
	To that end, we apply the Wigner-Eckart theorem in the isospin space to the members of the $0^+$ isotriplet $T_f=T_i=1$, 
	\begin{align}
		\langle T_f,T_{z,f}|O_{T_z}^T|T_i,T_{z,i}\rangle&=(-1)^{T_{f}-T_{z,f}}\label{eq:WET}\\
		&\times\left(\begin{array}{ccc}
			T_{f} & T & T_{i}\\
			-T_{z,f} & T_{z} & T_{z,i}
		\end{array}\right)\langle T_{f}||O^{T}||T_{i}\rangle\,,\nonumber
	\end{align}
	where $\langle T_f||O^T||T_i\rangle$ is a reduced matrix element. Expressing now the 
	time component of the electroweak currents as tensors in the isospin space, 
	\begin{equation}
		J_\text{em}^0=O^0_0-\frac{1}{2}O^1_0~,~(J_W^{\dagger 0})_V=-\frac{1}{\sqrt{2}}O^{1}_{1}~,
	\end{equation}
	we obtain the electromagnetic and charged weak form factors as:
	\begin{align}
		Z_{T_z}F_{\text{ch},T_z}^0&=\langle 1,T_z|J_\text{em}^0|1,T_z\rangle\\
		&=-\frac{T_z}{2\sqrt{6}}\langle 1||O^1||1\rangle + \frac{1}{\sqrt{3}}\langle 1||O^0||1\rangle \nonumber\\
		M_F^0 F_\text{cw}^0&=\langle 1,T_{zf}|(J_W^{\dagger 0})_V|1,T_{zi}\rangle=\frac{1}{2\sqrt{3}}\langle 1||O^1||1\rangle\,,\nonumber
	\end{align}
	with $M_F^0=\sqrt{2}$, and $Z_{T_z}$ is the atomic number of the nucleus within isospin quantum numbers $(1,T_z)$\footnote{In this paper we adopt the nuclear physics's convention of isospin, i.e. $T_z(p)=-1/2$.}. Fourier-transforming the form factors into the coordinate space gives: 
	\begin{align}
		\rho_\text{cw}(r)&=\rho_\text{ch,1}(r)+Z_0\left(\rho_\text{ch,0}(r)-\rho_\text{ch,1}(r)\right)\nonumber\\
		&=\rho_\text{ch,1}(r)+\frac{Z_{-1}}{2}\left(\rho_\text{ch,-1}(r)-\rho_\text{ch,1}(r)\right)~.\label{eq:isospin}
	\end{align} 
	This means, taking $\rho_\text{ch,1}$ as a reference distribution (since the neutron-rich nucleus is always the most stable), one has $\delta \rho=Z_0(\rho_\text{ch,0}-\rho_\text{ch,1})=Z_{-1}(\rho_\text{ch,-1}-\rho_\text{ch,1})/2$. 
	
	In Fig.\ref{fig:distributions} we show the plot of two $\rho_\text{ch}$ along with $\rho_\text{cw}$ in the $A=22$ isotriplet, the later obtained from the isospin relation~\eqref{eq:isospin}. A few features are observed:
	\begin{enumerate}
		\item $\rho_\text{cw}$ differs significantly from all $\rho_\text{ch}$, and $\delta\rho$ cannot be taken as a small perturbation.
		\item Unlike the ordinary charge distributions that falls off monotonically with increasing $r$, $\rho_\text{cw}$ is peaked at a larger value of $r$. This can be qualitatively understood in a shell-model picture: While a photon couples equally to all protons inside the nucleus, a $W$-boson can only couple to a proton in the outermost shell because the corresponding neutron state in an inner shell is filled. 
		\item The error band of $\rho_\text{cw}$ deduced from the isospin relation is much larger than that of the individual $\rho_\text{ch}$ due to the enhancement of the $Z$-factor in $\delta \rho$. 
	\end{enumerate}
	In short, the isospin relation \eqref{eq:isospin} allows us to evaluate $F(E)$ and $C(E)$ simultaneously with reduced model dependence and a fully-correlated error analysis.

	Finally, isospin symmetry also relates the three charge distributions within an isotriplet:
	\begin{equation}
		2Z_0\rho_\text{ch,0}(r)= Z_1\rho_\text{ch,1}(r)+Z_{-1}\rho_\text{ch,-1}(r)~.\label{eq:CHisospin}
	\end{equation}
	Therefore, if the charge distribution of a particular daughter nucleus is unknown, one can still obtain it if the other two charge distributions within the isotriplet are. For example, the unknown charge distribution of  ${}^{18}\text{Fe(ex)}$ can be deduced from the data of ${}^{18}\text{Ne}$ and ${}^{18}\text{O}$ using the formula above.
	
	\section{\label{sec:distns}Selection of nuclear charge distribution data}

	\begin{table*}[t]
		\begin{tabular}{|c|c|c|c|}
			\hline 
			$A$ & $\langle r_\text{ch,-1}^2\rangle^{1/2}$ (fm) & $\langle r_\text{ch,0}^2\rangle^{1/2}$ (fm) & $\langle r_\text{ch,1}^2\rangle^{1/2}$ (fm) \tabularnewline
			\hline 
			\hline 
			10 & $_{6}^{10}$C & $_{5}^{10}$B(ex) & $_{4}^{10}$Be: 2.3550(170)$^{a}$\tabularnewline
			\hline 
			14 & $_{8}^{14}$O & $_{7}^{14}$N(ex) & $_{6}^{14}$C: 2.5025(87)$^{a}$\tabularnewline
			\hline 
			18 & $_{10}^{18}$Ne: 2.9714(76)$^{a}$ & $_{9}^{18}$F(ex) & $_{8}^{18}$O: 2.7726(56)$^{a}$\tabularnewline
			\hline 
			22 & $_{12}^{22}$Mg: 3.0691(89)$^{b}$ & $_{11}^{22}$Na(ex) & $_{10}^{22}$Ne: 2.9525(40)$^{a}$\tabularnewline
			\hline 
			26 & $_{14}^{26}$Si & $_{13}^{26m}$Al: 3.130(15)$^f$ & $_{12}^{26}$Mg: 3.0337(18)$^{a}$\tabularnewline
			\hline 
			30 & $_{16}^{30}$S & $_{15}^{30}$P(ex) & $_{14}^{30}$Si: 3.1336(40)$^{a}$\tabularnewline
			\hline 
			34 & $_{18}^{34}$Ar: 3.3654(40)$^{a}$ & $_{17}^{34}$Cl & $_{16}^{34}$S: 3.2847(21)$^{a}$\tabularnewline
			\hline 
			38 & $_{20}^{38}$Ca: 3.467(1)$^{c}$ & $_{19}^{38m}$K: 3.437(4)$^{d}$ & $_{18}^{38}$Ar: 3.4028(19)$^{a}$\tabularnewline
			\hline 
			42 & $_{22}^{42}$Ti & $_{21}^{42}$Sc: 3.5702(238)$^{a}$ & $_{20}^{42}$Ca: 3.5081(21)$^{a}$\tabularnewline
			\hline 
			46 & $_{24}^{46}$Cr & $_{23}^{46}$V & $_{22}^{46}$Ti: 3.6070(22)$^{a}$ \tabularnewline
			\hline 
			50 & $_{26}^{50}$Fe & $_{25}^{50}$Mn: 3.7120(196)$^{a}$ & $_{24}^{50}$Cr: 3.6588(65)$^{a}$\tabularnewline
			\hline 
			54 & $_{28}^{54}$Ni: 3.738(4)$^{e}$ & $_{27}^{54}$Co & $_{26}^{54}$Fe: 3.6933(19)$^{a}$\tabularnewline
			\hline 
			62 & $_{32}^{62}$Ge & $_{31}^{62}$Ga & $_{30}^{62}$Zn: 3.9031(69)$^{b}$\tabularnewline
			\hline 
			66 & $_{34}^{66}$Se & $_{33}^{66}$As & $_{32}^{66}$Ge\tabularnewline
			\hline 
			70 & $_{36}^{70}$Kr & $_{35}^{70}$Br & $_{34}^{70}$Se\tabularnewline
			\hline 
			74 & $_{38}^{74}$Sr & $_{37}^{74}$Rb: 4.1935(172)$^{b}$ & $_{36}^{74}$Kr: 4.1870(41)$^{a}$\tabularnewline
			\hline 
		\end{tabular}
		\par
		\caption{\label{tab:RCh}Available data on nuclear RMS charge radii for isotriplets in measured superallowed decays. Superscripts denote the source of data: Ref.\cite{Angeli:2013epw}$^a$, Ref.\cite{Li:2021fmk}$^b$, Ref.\cite{miller2019proton}$^c$, Ref.\cite{Bissell:2014vva}$^d$, Ref.\cite{Pineda:2021shy}$^e$ and Ref.\cite{Plattner:2023fmu}$^f$.}
	\end{table*}
	
	A comprehensive data-driven analysis of $f$ using the isospin formalism requires a careful selection of nuclear charge distribution data. The most important distribution parameter is the root-mean-square (RMS) charge radius,
	\begin{equation}
		r_\text{RMS}\equiv \langle r^2_\text{ch}\rangle^{1/2}=\left[\int_0^\infty 4\pi r^2\:r^2\rho_\text{ch}(r)dr\right]^{1/2}~.
	\end{equation}
	For stable nuclei it can be extracted from  elastic electron scattering or from spectra of muonic atoms. For unstable ones it can be deduced from the field shift relative to a stable reference nucleus.
	Many compilations of nuclear charge radii are available, including Fricke, Heilig and Schopper~\cite{fricke2004nuclear}, Angeli and Marinova~\cite{Angeli:2013epw}, and Li \textit{et al.}~\cite{Li:2021fmk}. While the data analysis in Ref.\cite{fricke2004nuclear} is more transparent, Refs.\cite{Angeli:2013epw,Li:2021fmk} cover more nuclei and will be adopted in this paper, alongside with several new measurements~\cite{miller2019proton,Bissell:2014vva,Pineda:2021shy,Plattner:2023fmu}. We summarize the available data of RMS nuclear radii relevant to superallowed transitions in Table~\ref{tab:RCh}.

	The full functional form of the nuclear charge distribution beyond the RMS charge radius can only be extracted from electron scattering off stable nuclei, where the available data is quite limited. The most recent compilation by de Vries \textit{et al.}, which will be our main source of reference, dates back to 1987~\cite{de1987nuclear}.
	In Appendix~\ref{sec:distparameter}, we summarize the most commonly used parameterizations in that compilation: The two-parameter Fermi (2pF), three-parameter Fermi (3pF), three-parameter Gaussian (3pG) and harmonic oscillator (HO).
	For each distribution, we define the ``primary'' distribution parameter which is just $\langle r_\text{ch}^2\rangle^{1/2}$, and one or two independent, ``secondary'' distribution parameters ($a$ for 2pF, $a$ and $w$ for 3pF and 3pG, $\alpha_\text{HO}$ for HO). The primary parameter is always taken from Table~\ref{tab:RCh}, whereas the secondary parameters are taken from the compilation by de Vries \textit{et al.}. The analytic expressions of $\langle r_\text{ch}^2\rangle$ given in Appendix~\ref{sec:distparameter} then allow us to fix the remaining, non-independent parameters ($c$ for 2pF, 3pF and 3pG, $b$ for HO). 
	
	Given the limited information, we must develop a selection criteria in order to make full use of the data Ref.\cite{de1987nuclear} to determine all the (independent) secondary distribution parameters. Inspired by Ref.\cite{Hardy:2004id}, we adopt the following prescription:
	\begin{enumerate}
		\item If the data for a desired nucleus is  available in Ref.\cite{de1987nuclear}, we use the secondary parameter(s) listed there.
		\item If the data of a particular nucleus is not available, we take the secondary parameter(s) from the nearest isotope.
		\item If no data of any isotope exists, we take the secondary parameter(s) from an  available nucleus with the closest mass number $A$.
	\end{enumerate}
	
	For some nuclei there are more than one set of distribution parameters given in Ref.\cite{de1987nuclear}. In that case we need to choose the ``best'' set of data which we evaluate according to the following criterion. First, we compare the quoted central value of $r_\text{RMS}$ for an available nucleus in de Vries's compilation (not necessarily one that participates in a superallowed decay) with those in Angeli's review~\cite{Angeli:2013epw}. The latter typically has a smaller uncertainty. We then use $|r_\text{deVries}-r_\text{Angeli}|$ as a measure of the accuracy of de Vries's fitting. At the same time, we use the quoted uncertainty of $r_\text{RMS}$ in de Vries's compilation, $\delta r_\text{deVries}$, 
	as a measure of its precision. Then, we may select a set of data which has the best overall accuracy and precision by requiring:
	\begin{equation}
		\Delta\equiv ((r_\text{deVries}-r_\text{Angeli})^2+(\delta r_\text{deVries})^2)=\text{min}~.
	\end{equation}
	
	Finally, we are only interested in those nuclear isotriplets where at least two charge radii are measured, such that the isospin formalism can be applied. This includes 9 nuclear isotriplets and covers 15 superallowed transitions. In what follows we summarize, for all such nuclei, the charge distribution parameters that we chose for the evaluation of the Fermi function and the shape function, and explain the reasoning of our choice. 
	
	\subsection{$A=18$}
	\begin{itemize}
		\item For $^{18}$Ne, we take $\langle r_\text{ch}^2\rangle^{1/2}=2.9714(76)$~fm~\cite{Angeli:2013epw}. The nearest isotope of which charge distribution data exists in Ref.\cite{de1987nuclear} is $^{20}$Ne, with three parameterizations: 2pF (1971)~\cite{moreira1971charge}, 2pF (1981)~\cite{knight1981elastic} and 3pF (1985)~\cite{Be85}. We adopt the secondary parameters from 3pF (1985): $a=0.698(5)$~fm, $w=-0.168(8)$ which returns the smallest $\Delta$. 
		\item For $^{18}$O, we take $\langle r_\text{ch}^2\rangle^{1/2}=2.7726(56)$~fm~\cite{Angeli:2013epw}. The charge distribution data exists in Ref.\cite{de1987nuclear}, from which we take the secondary parameter: HO (1970), $\alpha_\text{HO}=1.513$~\cite{singhal1970rms}.  
	\end{itemize}
	
	\subsection{$A=22$}
	\begin{itemize}
		\item For $^{22}$Mg, we take $\langle r_\text{ch}^2\rangle^{1/2}=3.0691(89)$~fm~\cite{Li:2021fmk}. The nearest isotope of which charge distribution data exists in Ref.\cite{de1987nuclear} is $^{24}$Mg, with three parameterizations: 3pF (1974)~\cite{Av74}, 3pF (1974v2)~\cite{li1974high} and 2pF (1976)~\cite{lees1976elastic}. We adopt the secondary parameters from 3pF (1974): $a=0.607(9)$~fm, $w=-0.163(30)$ which returns the smallest $\Delta$. 
		\item For $^{22}$Ne, we take $\langle r_\text{ch}^2\rangle^{1/2}=2.9525(40)$~fm~\cite{Angeli:2013epw}. The charge distribution data exists Ref.\cite{de1987nuclear}, from which we take the secondary parameter: 2pF (1971), $a=0.549(4)$~fm~\cite{moreira1971charge}.
	\end{itemize}
	
	\subsection{$A=26$}
	\begin{itemize}
		\item For $^{26m}$Al, we take $\langle r_\text{ch}^2\rangle^{1/2}=3.130(15)$~fm~\cite{Plattner:2023fmu}. The nearest isotope of which charge distribution data exists in Ref.\cite{de1987nuclear} is $^{27}$Al, with two parameterizations: 2pF (1967)~\cite{Lombard:1967skb} and 2pF (1973)~\cite{fey1973nuclear}. We adopt the secondary parameters from 2pF (1973): $a=0.569$~fm which returns the smallest $\Delta$. 
		\item For $^{26}$Mg, we take $\langle r_\text{ch}^2\rangle^{1/2}=3.0337(18)$~fm~\cite{Angeli:2013epw}. The charge distribution data exists Ref.\cite{de1987nuclear}, from which we take the secondary parameter: 2pF (1976), $a=0.523(32)$~fm~\cite{lees1976elastic}.
	\end{itemize}
	
	\subsection{$A=34$}
	\begin{itemize}
		\item For $^{34}$Ar, we take $\langle r_\text{ch}^2\rangle^{1/2}=3.3654(40)$~fm~\cite{Angeli:2013epw}. The nearest isotope of which charge distribution data exists in Ref.\cite{de1987nuclear} is $^{36}$Ar, from which we take the secondary parameter: 2pF (1976), $a=0.507(15)$~fm~\cite{finn1976elastic}. 
		\item For $^{34}$S, we take $\langle r_\text{ch}^2\rangle^{1/2}=3.2847(21)$~fm~\cite{Angeli:2013epw}. The nearest isotope of which charge distribution data exists in Ref.\cite{de1987nuclear} is $^{32}$S, from which we take the secondary parameters: 3pG (1974), $a=2.191(10)$~fm, $w=0.160(12)$~\cite{fivozinsky1974electron}.
	\end{itemize}
	
	\subsection{$A=38$}
	
	\begin{itemize}
		\item For $^{38}$Ca, we take $\langle r_\text{ch}^2\rangle^{1/2}=3.467(1)$~fm~\cite{miller2019proton}. The nearest isotope of which charge distribution data exists in Ref.\cite{de1987nuclear} is $^{40}$Ca, from which we take the secondary parameters: 3pF (1973), $a=0.586(5)$~fm, $w=-0.161(23)$~\cite{sinha1973nuclear}.
		\item For $^{38m}$K, its RMS charge radius is experimentally known: $\langle r_\text{ch}^2\rangle^{1/2}=3.437(4)$~fm~\cite{Bissell:2014vva} but is the least precise among all three in the isotriplet. So we obtain instead the radius and charge distributions of this nucleus using the isospin relation~\eqref{eq:CHisospin}: $\langle r_\text{ch}^2\rangle^{1/2}=3.4367(10)$~fm. 
		\item For $^{38}$Ar, we take $\langle r_\text{ch}^2\rangle^{1/2}=3.4028(19)$~fm~\cite{Angeli:2013epw}. The nearest isotope of which charge distribution data exists in Ref.\cite{de1987nuclear} is $^{36}$Ar, which we mentioned above.
	\end{itemize}
	
	We emphasize that this nuclear isotriplet plays a special role as it is the only isotriplet where all three nuclear charge radii are measured. This allows us to test the validity of the CVC assumption we used in deducing $\rho_\text{cw}$. As pointed out in  Ref.\cite{Seng:2022epj}, a non-zero value of the quantity
	\begin{equation}
		\Delta M_B^{(1)}\equiv \frac{1}{2}\left(Z_1\langle r_\text{ch,1}\rangle^2+Z_{-1}\langle r_\text{ch,-1}\rangle^2\right)-Z_0\langle r_\text{ch,0}\rangle^2
	\end{equation}
	measures nuclear isospin mixing effect not probed by the nuclear mass splitting. Using Table~\ref{tab:RCh}, we obtain $\Delta M_B^{(1)}=-0.03(54)$fm$^2$ which is consistent with zero. This shows that the current experimental precision of radii observables is not yet enough to resolve ISB effect; this also validates our strategy to use CVC with experimental data.

	\subsection{$A=42$}
	\begin{itemize}
		\item For $^{42}$Sc, we take $\langle r_\text{ch}^2\rangle^{1/2}=3.5702(238)$~fm~\cite{Angeli:2013epw}. No data of charge distributions on Sc isotopes exists in Ref.\cite{de1987nuclear}, so we pick the available nucleus of nearest mass number, $^{40}$Ca, which we mentioned above.
		\item For $^{42}$Ca, we take $\langle r_\text{ch}^2\rangle^{1/2}=3.5081(21)$~fm~\cite{Angeli:2013epw}. The nearest isotope of which charge distribution data exists in Ref.\cite{de1987nuclear} is $^{40}$Ca, which was already mentioned before.
	\end{itemize}
	
	\subsection{$A=50$}
	\begin{itemize}
		\item For $^{50}$Mn, we take $\langle r_\text{ch}^2\rangle^{1/2}=3.7120(196)$~fm~\cite{Angeli:2013epw}. The nearest isotope of which charge distribution data exists in Ref.\cite{de1987nuclear} is $^{55}$Mn, from which we take the secondary parameter: 2pF (1969), $a=0.567$~fm~\cite{theissen1969elastic}.  
		\item For $^{50}$Cr, we take $\langle r_\text{ch}^2\rangle^{1/2}=3.6588(65)$~fm~\cite{Angeli:2013epw}. The charge distribution data exists in Ref.\cite{de1987nuclear}, with two parameterizations: 2pF (1976)~\cite{La76} and 2pF (1978)~\cite{shevchenko1978charge}. We adopt the secondary parameter from 2pF (1976): $a=0.520(13)$~fm which returns the smallest $\Delta$. 
	\end{itemize}
	
	\subsection{$A=54$}
	\begin{itemize}
		\item For $^{54}$Ni, we take $\langle r_\text{ch}^2\rangle^{1/2}=3.738(4)$~fm~\cite{Pineda:2021shy}. The nearest isotope of which charge distribution data exists in Ref.\cite{de1987nuclear} is $^{58}$Ni, from which we take the secondary parameters: 3pF (1970), $a=0.5169$~fm, $w=-0.1308$~\cite{ficenec1970elastic}.
		\item For $^{54}$Fe, we take $\langle r_\text{ch}^2\rangle^{1/2}=3.6933(19)$~fm~\cite{Angeli:2013epw}. The charge distribution data exists in Ref.\cite{de1987nuclear}, with three parameterizations: 3pG (1976)~\cite{Wo76}, 2pF (1976)~\cite{La76} and 2pF (1978)~\cite{shevchenko1978charge}. We adopt the secondary parameters from 3pG (1976): $a=2.270(12)$~fm, $w=0.403(15)$ which return the smallest $\Delta$.  
	\end{itemize}
	
	\subsection{$A=74$}
	\begin{itemize}
		\item For $^{74}$Rb, we take $\langle r_\text{ch}^2\rangle^{1/2}=4.1935(172)$~fm~\cite{Li:2021fmk}. No data of charge distributions on Rb isotopes exists in Ref.\cite{de1987nuclear}, so we pick the available nucleus of nearest mass number, $^{72}$Ge, from which we take the secondary parameter: 2pF (1975), $a=0.573(7)$~fm~\cite{kline1975electron}.  
		\item For $^{74}$Kr, we take $\langle r_\text{ch}^2\rangle^{1/2}=4.1870(41)$~fm~\cite{Angeli:2013epw}. No data of charge distributions on Kr isotopes exists in Ref.\cite{de1987nuclear}, so we pick the available nucleus of nearest mass number, $^{72}$Ge, which was already mentioned before.
	\end{itemize}
	With the information above, one can now evaluate $F(E)$ and $C(E)$ simultaneously using Eqs.\eqref{eq:Fermi}, \eqref{eq:shape} with a fully-correlated error analysis. 
	
	\section{\label{sec:secondary}Secondary corrections}
	
	This section outlines the procedure we adopt to compute the remaining, ``secondary'' corrections to $f$ in Eq.\eqref{eq:fformula}.
	
	\subsection{Screening correction}

	\begin{table*}
		\begin{centering}
			\begin{tabular}{|ccccccccccccccccc|}
				\hline 
				$Z_{i}$ & $\mathfrak{N}(Z_{i})$ &  & $Z_{i}$ & $\mathfrak{N}(Z_{i})$ &  & $Z_{i}$ & $\mathfrak{N}(Z_{i})$ &  & $Z_{i}$ & $\mathfrak{N}(Z_{i})$ &  & $Z_{i}$ & $\mathfrak{N}(Z_{i})$ &  & $Z_{i}$ & $\mathfrak{N}(Z_{i})$\tabularnewline
				\hline 
				\hline 
				1 & 1.000 &  & 14 & 1.481 &  & 25 & 1.513 &  & 39 & 1.553 &  & 60 & 1.572 &  & 80 & 1.599\tabularnewline
				7 & 1.399 &  & 15 & 1.484 &  & 27 & 1.518 &  & 45 & 1.561 &  & 64 & 1.577 &  & 86 & 1.600\tabularnewline
				8 & 1.420 &  & 16 & 1.488 &  & 30 & 1.540 &  & 49 & 1.566 &  & 66 & 1.579 &  & 92 & 1.601\tabularnewline
				9 & 1.444 &  & 17 & 1.494 &  & 32 & 1.556 &  & 52 & 1.567 &  & 68 & 1.586 &  & 94 & 1.603\tabularnewline
				10 & 1.471 &  & 18 & 1.496 &  & 35 & 1.550 &  & 53 & 1.568 &  & 70 & 1.590 &  &  & \tabularnewline
				11 & 1.476 &  & 20 & 1.495 &  & 36 & 1.551 &  & 54 & 1.568 &  & 74 & 1.593 &  &  & \tabularnewline
				12 & 1.474 &  & 23 & 1.504 &  & 38 & 1.552 &  & 55 & 1.567 &  & 76 & 1.595 &  &  & \tabularnewline
				\hline 
			\end{tabular}
			\par\end{centering}
		\caption{\label{tab:nZi} Hartree-Fock calculation of $\mathfrak{N}(Z_i)$ from Ref.\cite{GarrettBhalla}.}
		
	\end{table*}
	
	The presence of atomic electrons that reside around the atomic radius $r_A\sim 1$\AA{} alters the nuclear potential felt by the outgoing positron; namely, at very large $r$ the positron feels not the point-like Coulomb potential $V(r)=|Z|\alpha/r$ (where $\alpha$ is the fine-structure constant) but a screened version. 
	To estimate this correction, we use the simple formula by Rose~\cite{Rose:1936zz} derived from the Wentzel–Kramers–Brillouin (WKB) approximation:
	\begin{equation}
		Q(E)=\frac{\tilde{\p}\tilde{E}}{\p E}\frac{F(\tilde{E})}{F(E)}~,
	\end{equation}
	with $\tilde{E}=E-V_0$, $\tilde{\p}=\sqrt{\tilde{E}^2-m_e^2}$, $V_0=\alpha^2Z_i^{4/3}\mathfrak{N}(Z_i)$, where $Z_i$ is the atomic number of the \textit{parent} nucleus, and the function $\mathfrak{N}(Z_i)$ can be computed approximately using Hartree-Fock wavefunctions; here we obtain its functional form by interpolating the discrete points in Ref.\cite{GarrettBhalla}, which we reproduce in Table~\ref{tab:nZi} for the convenience of the readers.
	
	The size of the screening correction is of the order $10^{-3}$, but the simplified formula above does not permit a rigorous quantification of its uncertainty. Nevertheless, one could gain some insights by comparing the outcomes of different models. Ref.\cite{Hayen:2017pwg} compared the simple Rose formula to the solution of a more sophisticated potential by Salvat \textit{et al.}~\cite{Salvat:1987zz} (which they adopted); they found that the two are practically indistinguishable except at very small $E$, see Fig.5 of their paper. For $\beta^+$ decay, the small-$E$ contribution to $f$ is suppressed not only by the kinematic factor $\p E$ but also by the Fermi function, see Fig.\ref{fig:Fermi}. Therefore, it is reasonable to believe that the simple Rose formula is sufficient to meet our precision goal. Nevertheless, we will assign a 10\% uncertainty to the total screening correction to $f$ to stay on the safe side.

	\subsection{Kinematic recoil correction}
	
	The kinematic recoil correction factor $R(E)$ in Eq.\eqref{eq:fformula}
	takes into account two effects: (1) the difference between $E_0^\text{full}$ and $E_0$ in the upper limit of the $E$-integration, and (2) the $1/M$-suppressed terms in the tree-level squared amplitude, with $M$ the average nuclear mass. One may derive its expression starting from the exact, relativistic phase space formula for the decay of spinless particles, see, e.g. Appendix A in Ref.\cite{Seng:2022cnq}. Retaining terms up to $\mathcal{O}(1/M)$ gives:
	\begin{equation}
		R(E)\approx 1+\frac{2E^3-2E_0E^2+E_0^2E-m_e^2E}{E(E-E_0)M}~.
	\end{equation} 
	Ref.\cite{Hardy:2004id} adopted a simpler, $E$-independent form, which is equivalent to the expression above to $\mathcal{O}(1/M)$ after integrating over $E$:
	\begin{equation}
		R_\text{HT}(E_0)\approx 1-\frac{3E_0}{2M}~.
	\end{equation} 
	The size of this correction is $\sim10^{-4}$, so there is no need to assign an uncertainty of it. 
	
	It is worth noticing that, depending on whether $E_0$ or $E_0^\text{full}$ is used in the ``zeroth order'' expression of $f$, the expression of $R(E)$ will appear differently, e.g. between Ref.\cite{Hardy:2004id} and Ref.\cite{Hayen:2017pwg}, which is a minor detail often not clearly explained in literature. 
	
	\subsection{Atomic overlap correction}
	
	The last structure-dependent correction in Eq.\eqref{eq:fformula} is the atomic overlap correction $r(E)$ which accounts for the mismatch between the initial and the final atomic states in the $\beta$ decay; it is of the order $\lesssim 10^{-4}$. We evaluate this correction using the empirical formula in Ref.\cite{Hardy:2008gy}:
	\begin{equation}
		r(E)=1-\frac{1}{E_0-E}\frac{\partial^2}{\partial Z_i^2}B(G)~,
	\end{equation}
	with\begin{eqnarray}
		B(G) & = & \left\{ \begin{array}{ccc}
			13.080Z_{i}^{2.42}\text{eV} & , & 6\leq Z_{i}\leq10\\
			14.945Z_{i}^{2.37}\text{eV} & , & 11\leq Z_{i}\leq30\\
			11.435Z_{i}^{2.45}\text{eV} & , & 31\leq Z_{i}\leq39
		\end{array}\right.~,
	\end{eqnarray}
	where $Z_i$ is again the atomic number of the \textit{parent} nucleus. 
	Similarly, it is unnecessary to assign an uncertainty due to its smallness.  
	
	\section{\label{sec:final}Final results and discussions}
	
	\begin{table*}
		\begin{centering}		\begin{tabular}{|c|c|c|c|}
				\hline 
				Transition & $f_{\text{\text{new}}}$ & $f_{\text{HT}}$ & $\frac{f_{\text{new}}-f_{\text{HT}}}{f_{\text{new}}}$ (\%)\tabularnewline
				\hline 
				\hline 
				$^{18}$Ne$\rightarrow$$^{18}$F & $134.62(0)_{\text{rad}}(0)_{\text{shape}}(2)_\text{scr}(17)_{Q_{\text{EC}}}$ & $134.64(17)_{Q_{\text{EC}}}$ & $-0.01(0)_{\text{rad}}(0)_{\text{shape}}(2)_\text{scr}$\tabularnewline
				\hline 
				$^{22}$Mg$\rightarrow$$^{22}$Na & $418.27(1)_{\text{rad}}(1)_{\text{shape}}(7)_\text{scr}(13)_{Q_{\text{EC}}}$ & $418.35(13)_{Q_{\text{EC}}}$ & $-0.02(0)_{\text{rad}}(0)_{\text{shape}}(2)_\text{scr}$\tabularnewline
				\hline 
				$^{26}$Si$\rightarrow$$^{26m}$Al & $1027.52(15)_{\text{rad}}(12)_{\text{shape}}(17)_\text{scr}(12)_{Q_{\text{EC}}}$ & $1028.03(12)_{Q_{\text{EC}}}$ & $-0.05(1)_{\text{rad}}(1)_{\text{shape}}(2)_\text{scr}$\tabularnewline
				\hline 
				$^{34}$Ar$\rightarrow$$^{34}$Cl & $3409.89(16)_{\text{rad}}(18)_{\text{shape}}(60)_\text{scr}(25)_{Q_{\text{EC}}}$ & $3410.85(25)_{Q_{\text{EC}}}$ & $-0.03(0)_{\text{rad}}(1)_{\text{shape}}(2)_\text{scr}$\tabularnewline
				\hline 
				$^{38}$Ca$\rightarrow$$^{38m}$K & $5327.49(14)_{\text{rad}}(36)_{\text{shape}}(98)_\text{scr}(31)_{Q_{\text{EC}}}$ & $5328.88(31)_{Q_{\text{EC}}}$ & $-0.03(0)_{\text{rad}}(1)_{\text{shape}}(2)_\text{scr}$\tabularnewline
				\hline 
				$^{42}$Ti$\rightarrow$$^{42}$Sc & $7124.3(57)_{\text{rad}}(8)_{\text{shape}}(14)_\text{scr}(14)_{Q_{\text{EC}}}$ & $7130.1(14)_{Q_{\text{EC}}}$ & $-0.08(8)_{\text{rad}}(1)_{\text{shape}}(2)_\text{scr}$\tabularnewline
				\hline 
				$^{50}$Fe$\rightarrow$$^{50}$Mn & $15053(18)_{\text{rad}}(3)_{\text{shape}}(3)_\text{scr}(60)_{Q_{\text{EC}}}$ & $15060(60)_{Q_{\text{EC}}}$ & $-0.04(12)_{\text{rad}}(2)_{\text{shape}}(2)_\text{scr}$\tabularnewline
				\hline 
				$^{54}$Ni$\rightarrow$$^{54}$Co & $21137(3)_{\text{rad}}(1)_{\text{shape}}(5)_\text{scr}(52)_{Q_{\text{EC}}}$ & $21137(57)_{Q_{\text{EC}}}$ & $+0.00(2)_{\text{rad}}(0)_{\text{shape}}(2)_\text{scr}$\tabularnewline
				\hline 
				$^{26m}$Al$\rightarrow$$^{26}$Mg & $478.097(60)_{\text{rad}}(54)_{\text{shape}}(82)_\text{scr}(100)_{Q_{\text{EC}}}$ & $478.270(98)_{Q_{\text{EC}}}$ & $-0.04(1)_{\text{rad}}(1)_{\text{shape}}(2)_\text{scr}$\tabularnewline
				\hline 
				$^{34}$Cl$\rightarrow$$^{34}$S & $1995.076(81)_{\text{rad}}(103)_{\text{shape}}(364)_\text{scr}(94)_{Q_{\text{EC}}}$ & $1996.003(96)_{Q_{\text{EC}}}$ & $-0.05(0)_{\text{rad}}(1)_{\text{shape}}(2)_\text{scr}$\tabularnewline
				\hline 
				$^{38m}$K$\rightarrow$$^{38}$Ar & $3296.32(8)_{\text{rad}}(21)_{\text{shape}}(63)_\text{scr}(15)_{Q_{\text{EC}}}$ & $3297.39(15)_{Q_{\text{EC}}}$ & $-0.03(0)_{\text{rad}}(1)_{\text{shape}}(2)_\text{scr}$\tabularnewline
				\hline 
				$^{42}$Sc$\rightarrow$$^{42}$Ca & $4468.53(336)_{\text{rad}}(52)_{\text{shape}}(91)_\text{scr}(46)_{Q_{\text{EC}}}$ & $4472.46(46)_{Q_{\text{EC}}}$ & $-0.09(8)_{\text{rad}}(1)_{\text{shape}}(2)_\text{scr}$\tabularnewline
				\hline 
				$^{50}$Mn$\rightarrow$$^{50}$Cr & $10737.93(1150)_{\text{rad}}(202)_{\text{shape}}(229)_\text{scr}(50)_{Q_{\text{EC}}}$ & $10745.99(49)_{Q_{\text{EC}}}$ & $-0.08(11)_{\text{rad}}(2)_{\text{shape}}(2)_\text{scr}$\tabularnewline
				\hline 
				$^{54}$Co$\rightarrow$$^{54}$Fe & $15769.4(23)_{\text{rad}}(7)_{\text{shape}}(34)_\text{scr}(27)_{Q_{\text{EC}}}$ & $15766.8(27)_{Q_{\text{EC}}}$ & $+0.02(1)_{\text{rad}}(0)_{\text{shape}}(2)_\text{scr}$\tabularnewline
				\hline 
				$^{74}$Rb$\rightarrow$$^{74}$Kr & $47326(127)_{\text{rad}}(18)_{\text{shape}}(12)_\text{scr}(94)_{Q_{\text{EC}}}$ & $47281(93)_{Q_{\text{EC}}}$ & $+0.10(27)_{\text{rad}}(4)_{\text{shape}}(3)_\text{scr}$\tabularnewline
				\hline 
			\end{tabular}
			\par	   
		\end{centering}
		
		\caption{\label{tab:final}Comparison between new and old results of $f$. Notation: 123.12(234) means $123.12\pm 2.34$.}
		
	\end{table*}
	
	\begin{table}
		\begin{centering}
			\begin{tabular}{|c|c|c|c|}
				\hline 
				Transition & $t$ (ms) & $(ft)_{\text{HT}}$ (s) & $(ft)_{\text{new}} (s)$\tabularnewline
				\hline 
				\hline 
				$^{18}$Ne$\rightarrow$$^{18}$F & $21630\pm590$ & $2912\pm79$ & $2912\pm80$\tabularnewline
				\hline 
				$^{22}$Mg$\rightarrow$$^{22}$Na & $7293\pm16$ & $3051.1\pm6.9$ & $3050.4\pm6.8$\tabularnewline
				\hline 
				$^{26}$Si$\rightarrow$$^{26m}$Al & $2969.0\pm5.4$ & $3052.2\pm5.6$ & $3050.7\pm5.6$\tabularnewline
				\hline 
				$^{34}$Ar$\rightarrow$$^{34}$Cl & $896.55\pm0.81$ & $3058.0\pm2.8$ & $3057.1\pm2.8$\tabularnewline
				\hline 
				$^{38}$Ca$\rightarrow$$^{38m}$K & $574.8\pm1.1$ & $3062.8\pm6.0$ & $3062.2\pm5.9$\tabularnewline
				\hline 
				$^{42}$Ti$\rightarrow$$^{42}$Sc & $433\pm12$ & $3090\pm88$ & $3085\pm86$\tabularnewline
				\hline 
				$^{50}$Fe$\rightarrow$$^{50}$Mn & $205.8\pm4.7$ & $3099\pm71$ & $3098\pm72$\tabularnewline
				\hline 
				$^{54}$Ni$\rightarrow$$^{54}$Co & $144.9\pm2.3$ & $3062\pm50$ & $3063\pm49$\tabularnewline
				\hline 
				$^{26m}$Al$\rightarrow$$^{26}$Mg & $6351.24_{-0.54}^{+0.55}$ & $3037.61\pm0.67$ & $3036.5\pm1.0$\tabularnewline
				\hline 
				$^{34}$Cl$\rightarrow$$^{34}$S & $1527.77_{-0.44}^{+0.47}$ & $3049.43_{-0.88}^{+0.95}$ & $3048.0\pm1.1$\tabularnewline
				\hline 
				$^{38m}$K$\rightarrow$$^{38}$Ar & $925.42\pm0.28$ & $3051.45\pm0.92$ & $3050.5\pm1.1$\tabularnewline
				\hline 
				$^{42}$Sc$\rightarrow$$^{42}$Ca & $681.44\pm0.26$ & $3047.7\pm1.2$ & $3045.0\pm2.7$\tabularnewline
				\hline 
				$^{50}$Mn$\rightarrow$$^{50}$Cr & $283.68\pm0.11$ & $3048.4\pm1.2$ & $3046.1\pm3.6$\tabularnewline
				\hline 
				$^{54}$Co$\rightarrow$$^{54}$Fe & $193.495_{-0.063}^{+0.086}$ & $3050.8_{-1.1}^{+1.4}$ & $3051.3_{-1.4}^{+1.7}$\tabularnewline
				\hline 
				$^{74}$Rb$\rightarrow$$^{74}$Kr & $65.201\pm0.047$ & $3082.8\pm6.5$ & $3086\pm11$\tabularnewline
				\hline 
			\end{tabular}
			\par\end{centering}
		\caption{\label{tab:ft}Summary of the experimental results of the partial half-life $t$ and the previous $ft$ determination, both from Ref.\cite{Hardy:2020qwl}, and our updated $ft$ values for 15 superallowed transitions.}
		
	\end{table}
	
	Our final results of the statistical rate function (denoted as $f_\text{new}$) are summarized in Table~\ref{tab:final}, alongside the latest compilation by Hardy and Towner~\cite{Hardy:2020qwl} (denoted as $f_\text{HT}$). In contrast to the latter that quoted only the experimental uncertainty from the $Q_\text{EC}$ values, our results fully account for the theory uncertainties from the Fermi function, the shape factor and the screening correction (scr). The errors from the former two are fully correlated and stem from the radial (rad) and higher-order shape parameters (shape) in the nuclear charge distribution functions. It is apparent from our analysis that in many cases the total theory uncertainty (rad + shape + scr) is larger than the experimental ones ($Q_\text{EC}$).  Based on this we deem that Ref.\cite{Hardy:2020qwl} has  underestimated the errors in $f$. To be complete, we also compare the old and new determination of the full $ft$ value in Table~\ref{tab:ft}.

	It is interesting to study the shift of the central value of $f$ from the previous determination. It was shown in Ref.\cite{Seng:2022inj}, by inspecting the analytic formula of the ``pure-QCD'' shape factor $C_\text{QCD}(E)$ in the absence of electromagnetic interaction, that an increase of $\langle r_\text{cw}^2\rangle^{1/2}$, the MS radius characterizing $\rho_\text{cw}$, in general leads smaller values of $f$. Indeed, from the last column in Table~\ref{tab:final} we see that in most cases our new evaluation reduces the central value of $f$ at the level of 0.01\%, although some of such shifts are within the quoted (theory) uncertainties. The magnitude of the shift obtained in this work is in general smaller than those estimated in Ref.\cite{Seng:2022inj} upon accounting for the correlated effects with the Fermi function. Nevertheless, according to Eq.\eqref{eq:Vudmaster}, a coherent downward shift of $f$ may lead to an upward shift of $V_{ud}$, which could partially alleviate the current CKM unitarity deficit. 
	
	We refrain from quoting immediately an updated value of $V_{ud}$ based on the new values of $f$ for several reasons:
	\begin{enumerate}
		\item In this work we only improved the control over the nuclear structure effects that reside in the statistical rate function, but not in other pieces of Eq.\eqref{eq:Ft}, especially $\delta_\text{NS}$ and $\delta_\text{C}$.  Before similar theory progress on these two quantities (which can be expected in the next few years), any update on $\mathcal{F}t$ values would be preliminary. 
		\item With the existing data on nuclear charge radii, we are only able to re-evaluate $f$ for 15 out of the 25 measured superallowed transitions. Furthermore, most of the information of the secondary charge distribution parameters in these 15 transitions are not directly measured but inferred from the nearest isotopes. The effects of isotope shifts to the secondary parameters are not systematically accounted for.
		\item Moreover, the experimental determination of the nuclear charge radii is not unambiguous. In some cases  electron scattering and atomic spectroscopy disagree with each other. In addition, the extraction of nuclear radii from data relies on the removal of higher-order corrections, most notably the nuclear polarization correction. 
		In the nuclear radii compilation by Fricke and Heilig~\cite{fricke2004nuclear} this correction is taken from older calculations~\cite{Rinker:1978kh} from the 1970's. Meanwhile, the compilation by Angeli and Marinova~\cite{Angeli:2013epw} does not quote neither the value nor the source of the nuclear polarization correction used. Thus, one may not be able to claim to have gained a full control over all theory systematics until these ambiguities are fully resolved. 
	\end{enumerate}
	
	With the above caveats in mind, our work represents an important first step towards a fully data-driven analysis of $ft$ values based on available data of nuclear charge distributions. Our approach offers a well-defined prescription to rigorously quantify the theory uncertainties, both in the Fermi function and in the shape factor. It also helps to identify some of the most urgently needed experimental measurements for future improvements. For instance, one extra measurement of nuclear charge radius in each of the $A$=10, 14, 30, 46, 62 nuclear isotriplets will activate the data-driven analysis on these systems based on the isospin formalism, and for $A$=66 and 70, two measurements on each isotriplet are needed. Also, the continuation of this work would greatly benefit from a more coherent, comprehensive and transparent compilation of nuclear charge radii and their uncertainties, so it provides extra motivations for such effort in the future. 

	\begin{acknowledgments}
		
		We acknowledge the participation of Giovanni Carotenuto, Michela Sestu, Matteo Cadeddu and Nicola Cargioli at  earlier stages of this project.
		The work of C.-Y.S. is supported in
		part by the U.S. Department of Energy (DOE), Office of Science, Office of Nuclear Physics, under the FRIB Theory Alliance award DE-SC0013617, by the DOE grant DE-FG02-97ER41014, and by the DOE Topical Collaboration ``Nuclear Theory for New Physics'', award No.
		DE-SC0023663. M.G. acknowledges support by EU Horizon 2020 research and innovation programme, STRONG-2020 project
		under grant agreement No 824093, and by the Deutsche Forschungsgemeinschaft (DFG) under the grant agreement GO 2604/3-1. 
		
	\end{acknowledgments}
	
	\begin{appendix}
		
		\section{Radial solutions of the Dirac equation\label{sec:radial}}
		
		For a nucleus of charge $Z$ and charge distribution $\rho_\text{ch}(r)$, the potential experienced by an electron reads:
		\begin{align}
			V(r)&=-4\pi Z\alpha\left[\frac{1}{r}\int_0^r dr'\rho_\text{ch}(r')r^{\prime 2}+\int_r^\infty dr'\rho_\text{ch}(r')r'\right]\,.\nonumber\\
			&
		\end{align}
		The radial Dirac equations are:
		\begin{align}
			f_\kappa'(r)&=\frac{\kappa-1}{r}f_\kappa(r)-(E-m_e-V(r))g_\kappa(r)\nonumber\\
			g_\kappa'(r)&=(E+m_e-V(r))f_\kappa(r)-\frac{\kappa+1}{r}g_\kappa(r)~.
		\end{align}
		We choose the normalization such that, when $V(r)=0$ the unbounded radial functions read:
		\begin{equation}
			\left(\begin{array}{c}
				g_{\kappa}^\text{free}(r)\\
				f_{\kappa}^\text{free}(r)
			\end{array}\right)=\p\left(\begin{array}{c}
				\sqrt{\frac{E+m_{e}}{E}}j_\ell(\p r)\\
				\text{sgn}(\kappa)\sqrt{\frac{E-m_{e}}{E}}j_{\bar{\ell}}(\p r)
			\end{array}\right)~,
		\end{equation}
		where $j_\ell$ is the spherical Bessel function, with:
		\begin{equation}
			\ell=\left\{ \begin{array}{ccc}
				\kappa & , & \kappa>0\\
				-\kappa-1 & , & \kappa<0
			\end{array}\right.\,,~~\bar{\ell}=\left\{ \begin{array}{ccc}
				\kappa-1 & , & \kappa>0\\
				-\kappa & , & \kappa<0
			\end{array}\right.~.
		\end{equation}

		It is beneficial to define $k\equiv|\kappa|$. With that, one defines four new types of radial functions $H_k$, $h_k$, $D_k$ and $d_k$ as:
		\begin{align}
			f_{+k}(r)&\equiv\frac{\alpha_{+k}}{(2k-1)!!}(\p r)^{k-1}\left\{H_k(r)+h_k(r)\right\}\nonumber\\
			g_{-k}(r)&\equiv\frac{\alpha_{-k}}{(2k-1)!!}(\p r)^{k-1}\left\{H_k(r)-h_k(r)\right\}\nonumber\\
			f_{-k}(r)&\equiv-\frac{\alpha_{-k}}{(2k-1)!!}(\p r)^{k-1}\frac{r}{R}\left\{D_k(r)-d_k(r)\right\}\nonumber\\
			g_{+k}(r)&\equiv\frac{\alpha_{+k}}{(2k-1)!!}(\p r)^{k-1}\frac{r}{R}\left\{D_k(r)+d_k(r)\right\}\,,\label{eq:HhDd}
		\end{align}
		with the normalization $H_k(0)\equiv 1$, $h_k(0)\equiv 0$, and $R$ is an arbitrarily-chosen radius parameter such that the nuclear charge is practically zero at $r>R$. These definitions, together with the normalization of $f_\kappa(r)$, $g_\kappa(r)$, fully define the parameters $\alpha_{\pm k}$.
		
		A particularly important case is the point-like Coulomb potential:
		\begin{equation}
			V(r)=-\frac{Z\alpha}{r}~,\label{eq:pointCoulomb}
		\end{equation}
		where there are two sets of solutions, the ``regular'' and ``irregular'' ones. The ``regular'' solution reads:
		\begin{equation}
			\left(\begin{array}{c}
				g_{\kappa}^\text{reg}(r)\\
				f_{\kappa}^\text{reg}(r)
			\end{array}\right)=\p\left(\begin{array}{c}
				\sqrt{\frac{E+m_{e}}{E}}\mathfrak{Re}\\
				-\sqrt{\frac{E-m_{e}}{E}}\mathfrak{Im}
			\end{array}\right)Q_{\kappa}(r)~,
		\end{equation}
		where 
		\begin{align}
			&Q_\kappa(r)\equiv 2e^{\frac{\pi y}{2}}\frac{\left|\Gamma(\gamma_\kappa+iy)\right|}{\Gamma(2\gamma_\kappa+1)}(\gamma_\kappa+iy)(2\p r)^{\gamma_\kappa-1}\nonumber\\
			&\times e^{-i\p r+i\eta_\kappa}{}_1F_1(\gamma_\kappa+1+iy;2\gamma_\kappa+1;2i\p r)\,,
		\end{align}
		with 
		\begin{align}
			\gamma_\kappa&=\sqrt{\kappa^2-\alpha^2Z^2}\,,~y=\frac{Z\alpha E}{\p}\,,\nonumber\\
			\eta_\kappa&=\text{sgn}(\kappa Z)\sin^{-1}\sqrt{\frac{1}{2}\left(1+\frac{\kappa \gamma_\kappa-y^2m_e/E}{\gamma_\kappa^2+y^2}\right)}~.
		\end{align}
		Meanwhile, the ``irregular'' solution reads:
		\begin{equation}
			\left(\begin{array}{c}
				g_{\kappa}^\text{irreg}(r)\\
				f_{\kappa}^\text{irreg}(r)
			\end{array}\right)=\p\left(\begin{array}{c}
				\sqrt{\frac{E+m_{e}}{E}}\mathfrak{Re}\\
				-\sqrt{\frac{E-m_{e}}{E}}\mathfrak{Im}
			\end{array}\right)\bar{Q}_{\kappa}(r)~,
		\end{equation}
		where $\bar{Q}_\kappa(r)$ is obtained from $Q_\kappa(r)$ by simply switching $\gamma_\kappa\rightarrow-\gamma_\kappa$. 
		
		When $r\rightarrow \infty$, the regular solution takes the following asymptotic form:
		\begin{equation}
			\left(\begin{array}{c}
				g_{\kappa}^\text{reg}(r)\\
				f_{\kappa}^\text{reg}(r)
			\end{array}\right)\rightarrow\frac{1}{r}\left(\begin{array}{c}
				\sqrt{\frac{E+m_{e}}{E}}\cos(\p r+\delta_\kappa)\\
				-\sqrt{\frac{E-m_{e}}{E}}\sin(\p r+\delta_\kappa)
			\end{array}\right)~,
		\end{equation}
		where
		\begin{equation}
			\delta_\kappa=y\ln(2\p r)-\arg\Gamma(\gamma_\kappa+iy)+\eta_\kappa-\frac{\pi\gamma_\kappa}{2}~
		\end{equation}
		is the phase shift for the Coulomb potential. The corresponding phase shift for the irregular solution is $\bar{\delta}_\kappa$, which is again obtained by taking $\gamma_\kappa\rightarrow-\gamma_\kappa$.
		
		If the point-like Coulomb potential holds for all distances (i.e. from $r=0$ to $r\rightarrow\infty$), then only the regular solutions survive because the irregular solutions blow up at $r\rightarrow 0$. However, in reality the nuclear charge is distributed over a finite space, so Eq.\eqref{eq:pointCoulomb} only holds at $r>R$. Therefore, since the analytic solutions never apply to $r=0$, we must retain both the regular and irregular solutions. To be more specific, the radial function at $r>R$ (which we call the ``outer solution'') is a linear combination of the two: 
		\begin{equation}
			\left(\begin{array}{c}
				g_{\kappa}(r)\\
				f_{\kappa}(r)
			\end{array}\right)=A_\kappa\left(\begin{array}{c}
				g_{\kappa}^\text{reg}(r)\\
				f_{\kappa}^\text{reg}(r)
			\end{array}\right)+B_\kappa\left(\begin{array}{c}
				g_{\kappa}^\text{irreg}(r)\\
				f_{\kappa}^\text{irreg}(r)
			\end{array}\right)~,~r>R\label{eq:fullouter}
		\end{equation} 
		where the coefficients satisfy the following normalization condition, which we express in terms of matrix product for future benefit~\cite{BhallaRose}:
		\begin{equation}
			\left(\begin{array}{c}
				A_{\kappa}\\
				B_{\kappa}
			\end{array}\right)^T
			\left(\begin{array}{cc}
				1 & \cos(\delta_{\kappa}-\bar{\delta}_{\kappa})\\
				\cos(\delta_{\kappa}-\bar{\delta}_{\kappa}) & 1
			\end{array}\right)\left(\begin{array}{c}
				A_{\kappa}\\
				B_{\kappa}
			\end{array}\right)=1~.\label{eq:AkBk}
		\end{equation}
		The other condition comes from the matching with the inner solution (i.e. the $r<R$ solution) at $r=R$, which we will describe later.
		
		Finally, to obtain radial functions for the positron, one simply switches $Z\rightarrow -Z$.
		
		\section{\label{sec:innersol}Obtaining the inner solution}
		
		Here we outline the procedure to obtain the inner solution as well as the matching to the outer solution. We start with the $\kappa=+k$ functions, and define:
		\begin{align}
			f_{+k}&\equiv\frac{\alpha_{+k}}{(2k-1)!!}(\p r)^{k-1}\bar{f}_{+k}\,,\nonumber\\
			g_{+k}&\equiv\frac{\alpha_{+k}}{(2k-1)!!}(\p r)^{k-1}\frac{r}{R}\bar{g}_{+k}~,\label{eq:fg+k}
		\end{align}
		where $\bar{f}_{+k}=H_k+h_k$, $\bar{g}_{+k}=D_k+d_k$. They satisfy the following radial equations:
		\begin{align}
			\bar{f}_{+k}'(r)&=-(E-m_e-V(r))\frac{r}{R}\bar{g}_{+k}(r)\\
			\frac{r}{R}\bar{g}_{+k}'(r)&=(E+m_e-V(r))\bar{f}_{+k}(r)-\frac{2k+1}{R}\bar{g}_{+k}(r)\,,\nonumber
		\end{align}
		with the normalization condition $\bar{f}_{+k}(0)=H_k(0)+h_k(0)=1$. It is easy to see that this one normalization condition completely fixes both functions; for instance, taking $r=0$ at both sides of the second differential equation gives $\bar{g}_{+k}(0)=R(E+m_e-V(0))/(2k+1)$, so we now know the values of both functions at $r=0$. The values of their first derivative at $r=0$ are then given immediately by the differential equations, so on and so forth. Similarly, for the $\kappa=-k$ radial functions, we define:
		\begin{align}
			g_{-k}&\equiv\frac{\alpha_{-k}}{(2k-1)!!}(\p r)^{k-1}\bar{g}_{-k}\,,\nonumber\\
			f_{-k}&\equiv-\frac{\alpha_{-k}}{(2k-1)!!}(\p r)^{k-1}\frac{r}{R}\bar{f}_{-k}~,\label{eq:fg-k}
		\end{align}
		where $\bar{g}_{-k}=H_k-h_k$, $\bar{f}_{-k}=D_k-d_k$.
		They satisfy the following radial equations:
		\begin{align}
			\bar{g}_{-k}'(r)&=-(E+m_e-V(r))\frac{r}{R}\bar{f}_{-k}(r)\\
			\frac{r}{R}\bar{f}_{-k}'(r)&=(E-m_e-V(r))\bar{g}_{-k}(r)-\frac{2k+1}{R}\bar{f}_{-k}(r)\,,\nonumber
		\end{align}
		with the normalization condition $\bar{g}_{-k}(0)=H_k(0)-h_k(0)=1$.
		
		Given a choice of nuclear charge distribution (which fixes the potential $V(r)$), we can solve for the functions $\bar{g}_{\pm k}(r)$, $\bar{f}_{\pm k}(r)$ numerically from $r=0$ to $r=R$. Then, at $r=R$, we match them to the analytic expressions of the outer solutions. Combining Eqs.\eqref{eq:fullouter}, \eqref{eq:fg+k} and \eqref{eq:fg-k}, the matching gives:
		\begin{align}
			\left(\begin{array}{c}
				A_{\kappa}\\
				B_{\kappa}
			\end{array}\right)&=\frac{\alpha_{\kappa}(\p R)^{k-1}}{(2k-1)!!\,\p}
			\left(\begin{array}{cc}
				\mathfrak{Re}Q_{\kappa}(R) & \mathfrak{Re}\bar{Q}_{\kappa}(R)\\
				\mathfrak{Im}Q_{\kappa}(R) & \mathfrak{Im}\bar{Q}_{\kappa}(R)
			\end{array}\right)^{-1}\nonumber\\
			&\times\left(\begin{array}{c}
				\sqrt{\frac{E}{E+m_{e}}}\bar{g}_{\kappa}(R)\\
				-\text{sgn}(\kappa)\sqrt{\frac{E}{E-m_{e}}}\bar{f}_{\kappa}(R)
			\end{array}\right)~,\label{eq:solveAB}
		\end{align}
		where $\kappa=\pm k$. Substituting this to Eq.\eqref{eq:AkBk} gives:
		\begin{align}
			\alpha_{\kappa}^{-2} & = \left(\frac{(\p R)^{k-1}}{(2k-1)!!\p}\right)^{2}\left[\left(\begin{array}{cc}
				\mathfrak{Re}Q_{\kappa}(R) & \mathfrak{Re}\bar{Q}_{\kappa}(R)\\
				\mathfrak{Im}Q_{\kappa}(R) & \mathfrak{Im}\bar{Q}_{\kappa}(R)
			\end{array}\right)^{-1}\right.\nonumber\\
			&\times
			\left.\left(\begin{array}{c}
				\sqrt{\frac{E}{E+m_{e}}}\bar{g}_{\kappa}(R)\\
				-\text{sgn}(\kappa)\sqrt{\frac{E}{E-m_{e}}}\bar{f}_{\kappa}(R)
			\end{array}\right)\right]^{T}\nonumber\\
			&   \times\left(\begin{array}{cc}
				1 & \cos(\delta_{\kappa}-\bar{\delta}_{\kappa})\\
				\cos(\delta_{\kappa}-\bar{\delta}_{\kappa}) & 1
			\end{array}\right)\nonumber\\
			&\times\left(\begin{array}{cc}
				\mathfrak{Re}Q_{\kappa}(R) & \mathfrak{Re}\bar{Q}_{\kappa}(R)\\
				\mathfrak{Im}Q_{\kappa}(R) & \mathfrak{Im}\bar{Q}_{\kappa}(R)
			\end{array}\right)^{-1}\nonumber\\
			&\times\left(\begin{array}{c}
				\sqrt{\frac{E}{E+m_{e}}}\bar{g}_{\kappa}(R)\\
				-\text{sgn}(\kappa)\sqrt{\frac{E}{E-m_{e}}}\bar{f}_{\kappa}(R)
			\end{array}\right)~.\nonumber\\
			\label{eq:solvealpha}
		\end{align}
		Thus, with the numerical solutions of $\bar{f}_{\pm k}(R)$ and $\bar{g}_{\pm k}(R)$, Eqs.\eqref{eq:solveAB}, \eqref{eq:solvealpha} give spontaneously the coefficients $\alpha_{\pm k}$ and $\{A_{\pm k},B_{\pm k}\}$; the former give all the Coulomb functions while the latter determine the full radial functions at $r>R$. 
		
		\section{\label{sec:master}Derivation of the master formula of shape factor}
		
		In this appendix we briefly outline the derivation of the master formula of the shape factor, Eq.\eqref{eq:shape}, based on the formalism by Behrens and B\"{u}hring~\cite{behrens1982electron}. To match their notations, we adopt the following normalization of states:
		\begin{equation}
			\langle \vec{k}'|\vec{k}\rangle=(2\pi)^3\delta^{(3)}(\vec{k}-\vec{k}')~,
		\end{equation}
		i.e. the states are rescaled with respect to the QFT states in the introduction as $|\vec{k}\rangle=(1/2E_k)|\vec{k}\rangle_\text{QFT}\approx (1/2M)|\vec{k}\rangle_\text{QFT}$.

		We start by introducing the Behrens-B\"{u}hring form factors in terms of the nuclear matrix element of the charged weak current:
		\begin{align}
			&\langle J_W^{\dagger0}(0)\rangle_{fi}  =  \sum_{Lm_{L}}\sqrt{4\pi(2J_{i}+1)}(-1)^{J_{f}-m_{J_{f}}}
			\\ &\times\left(\begin{array}{ccc}
				J_{f} & L & J_{i}\\
				-m_{J_{f}} & m_{L} & m_{J_{i}}
			\end{array}\right)Y_{Lm_{L}}^{*}(\hat{q})\frac{(\q R)^{L}}{(2L+1)!!}F_{L}(\q^{2})\nonumber\\
			\nonumber\\
			&\langle\vec{J}_W^\dagger(0)\rangle_{fi}  =  \sum_{KLm_{K}}\sqrt{4\pi(2J_{i}+1)}(-1)^{J_{f}-m_{J_{f}}}\nonumber\\
			&\times \left(\begin{array}{ccc}
				J_{f} & K & J_{i}\\
				-m_{J_{f}} & m_{K} & m_{J_{i}}
			\end{array}\right)\vec{Y}_{KL}^{m_{K}*}(\hat{q})\frac{(\q R)^{L}}{(2L+1)!!}F_{KL}(\q^{2})\,.\nonumber
		\end{align}
		where $q=p_f-p_i$, $\q=|\vec{q}|$, with $Y_{Lm_L}$ and $\vec{Y}_{KL}^{m_K}$ the spherical harmonics and the vector spherical tensor respectively. 
		When $J_i=J_f=0$, only the $F_0$ and $F_{01}$ form factors survive, but the latter is proportional to $f_-(q^2)$ (in the Breit frame) which vanishes in the isospin limit. The former gives:
		\begin{equation}
			\langle J_W^{\dagger 0}(0)\rangle_{fi}=F_0(\q^2)~,
		\end{equation}
		where $\q\rightarrow 0$ limit gives the Fermi matrix element: $F_0(0)=M_F$. 
		
		The differential rate of the tree-level decay $\phi_i(p_i)\rightarrow \phi_f(p_f)e^+(p_e)\nu_e(p_\nu)$ is given by:
		\begin{equation}
			d\Gamma=\frac{d^3p_f}{(2\pi)^3}\frac{d^3p_e}{(2\pi)^3}\frac{d^3p_\nu}{(2\pi)^3}(2\pi)^4\delta^{(4)}(p_i-p_f-p_e-p_\nu)\sum_{\lambda_e\lambda_\nu}|\mathcal{T}|^2~.
		\end{equation}
		The amplitude, using the lepton current in configuration space, reads:
		\begin{align}
			\mathcal{T}&=-\frac{G_FV_{ud}}{\sqrt{2}}\int\frac{d^3q'}{(2\pi)^3}\langle \phi_f(\vec{q}')|J_W^{\dagger\mu}(0)|\phi_i(\vec{0})\rangle\nonumber\\
			&\quad\times
			\int d^3x e^{-i\vec{q}'\cdot\vec{x}}\bar{\psi}_{\nu,\vec{p}_\nu}(\vec{x})\gamma_\mu(1-\gamma_5)\psi_{e^+,\vec{p}}(\vec{x})\nonumber\\
			&\rightarrow-\frac{G_FV_{ud}}{\sqrt{2}}\frac{1}{2\pi^2}\int_0^\infty d\q' \q^{\prime 2}F_0(\q^{\prime 2})\nonumber\\
			&\quad\times\int d^3xj_0(\q'r)\psi_{\nu,\vec{p}_\nu}^{\lambda_\nu\dagger}(\vec{x})(1-\gamma_5)\psi_{e^+,\vec{p}}^{\lambda_e}(\vec{x})~,\label{eq:Tamp}
		\end{align}
		the second expression applies to $J_i=J_f=0$ decays, where $\lambda_e,\lambda_\nu$ denote the lepton spin orientations. This representation is particularly convenient for the implementation of Coulomb effects, as we just need to take the lepton wavefunctions as the solution of the Dirac equation. To that end, we shall expand these wavefunctions in terms of spherical waves:
		\begin{align}
			\psi_{\nu,\vec{p}_\nu}^{\lambda_\nu}(\vec{x})
			&=
			\sum_{\kappa_\nu\mu_\nu}i^{l_\nu}b_{\kappa_\nu \mu_\nu}^{\lambda_\nu}\psi_{\nu,\kappa_\nu}^{\mu_\nu}(\vec{x})\,,\nonumber\\
			\psi_{e^+,\vec{p}}^{\lambda_e}(\vec{x})&=\sum_{\kappa_e\mu_e}(-1)^{j_e+\mu_e}i^{l_e}a_{\kappa_e \mu_e}^{\lambda_e *}\psi_{e^+,\kappa_e}^{-\mu_e}(\vec{x})~.\label{eq:spexpansion}
		\end{align} 
		The spherical waves read,
		\begin{align}
			\psi_{\nu,\kappa_{\nu}}^{\mu_{\nu}}(\vec{x})
			&=
			\left(\begin{array}{c}
				j_{l_{\nu}}(E_{\nu}r)\chi_{\kappa_{\nu}}^{\mu_{\nu}}(\hat{r})\\
				i\,\text{sgn}(\kappa_{\nu})j_{\bar{l}_{\nu}}(E_{\nu}r)\chi_{-\kappa_{\nu}}^{\mu_{\nu}}(\hat{r})
			\end{array}\right)
			\,,\nonumber\\
			\psi_{e^{+},\kappa_{e}}^{-\mu_{e}}(\vec{x})&=\left(\begin{array}{c}
				if_{\kappa_{e}}(r)\chi_{-\kappa_{e}}^{-\mu_{e}}(\hat{r})\\
				-g_{\kappa_{e}}(r)\chi_{\kappa_{e}}^{-\mu_{e}}(\hat{r})
			\end{array}\right)~,
		\end{align}
		where
		\begin{equation}
			\chi_\kappa^\mu\equiv \sum_m C_{\ell\:\mu-m; \frac{1}{2}\:m}^{j\:\mu} Y_{\ell\:\mu-m}(\hat{r})\chi_m
		\end{equation}
		is a two-component spinor, with $C_{\ell\:\mu-m;\frac{1}{2}\:m}^{j\:\mu}$ the Clebsch-Gordan coefficients. The expansion coefficients read,
		\begin{align}
			b_{\kappa_\nu\mu_\nu}^{\lambda_\nu}
			&=
			\frac{4\pi}{\sqrt{2}}C_{l_\nu\:\mu_\nu-\lambda_\nu;\frac{1}{2}\:\lambda_\nu}^{j_\nu\:\mu_\nu}Y_{l_\nu\:\mu_\nu-\lambda_\nu}^*(\hat{p}_\nu)\:,\nonumber\\
			a_{\kappa_e\mu_e}^{\lambda_e}
			&=
			\frac{4\pi}{\sqrt{2}\p }C_{l_e\:\mu_e-\lambda_e;\frac{1}{2}\:\lambda_e}^{j_e\:\mu_e}Y_{l_e\:\mu_e-\lambda_e}^*(\hat{p}_e)e^{i\Delta_{\kappa_e}}~,
		\end{align}
		with $\Delta_{\kappa_e}$ an extra phase due to the distortion by the nuclear charge. 
		
		Substituting Eq.\eqref{eq:spexpansion} into Eq.\eqref{eq:Tamp}, one may perform the angular integration to obtain:
		\begin{align}
			\mathcal{T}&=-\frac{G_FV_{ud}}{\sqrt{2}}\frac{1}{2\pi^2}\int_0^\infty d\q'\q^{\prime 2}F_0(\q^{\prime 2})\int_0^\infty dr r^2 j_0(\q'r)
			\nonumber\\
			&\times
			\sum_{\kappa_e\mu_e \kappa_\nu\mu_\nu}(-1)^{j_e-\mu_e+1}a_{\kappa_e\mu_e}^{\lambda_e *}b_{\kappa_\nu\mu_\nu}^{\lambda_\nu *}\delta_{\mu_e,-\mu_\nu}\nonumber\\
			&\Bigl\{g_{\kappa_e}(r)[j_{l_\nu}(E_\nu r)\delta_{\kappa_e,\kappa_\nu}+j_{\bar{l}_\nu}(E_\nu r)\delta_{\kappa_e,-\kappa_\nu}]\nonumber\\
			&\quad-\text{sgn}(\kappa_e)f_{\kappa_e}(r)[j_{l_\nu}(E_\nu r)\delta_{-\kappa_e,\kappa_\nu}\nonumber\\ &\quad +j_{\bar{l}_\nu}(E_\nu r)\delta_{-\kappa_e,-\kappa_\nu}]\Bigr\}~.\label{eq:Tampstep2}
		\end{align}
		Now, we may express $g_{\kappa_e}$ and $f_{\kappa_e}$ in terms of $H,h,D,d$ as we defined in Appendix~\ref{sec:radial}, which allows us to introduce the Behrens-B\"{u}hring's shape factor functions $M_K(k_e,k_\nu)$ and $m_K(k_e,k_\nu)$. In superallowed decays, we only need the $K=L=S=0$ functions:
		\begin{align}
			M_0(k_e,k_\nu)&=\frac{2}{\pi M_F}\int_0^\infty d\q'\q^{\prime 2}\int_0^\infty dr r^2 j_0(\q'r)F_0(\q^{\prime 2})\nonumber\\
			&\times
			\frac{(\p r)^{k_e-1}}{(2k_e-1)!!}\sqrt{\frac{2j_e+1}{2}}\delta_{k_ek_\nu}\nonumber\\
			&\times\left\{H_{k_e}(r)j_{k_\nu-1}(E_\nu r)-\frac{r}{R}D_{k_e}(r)j_{k_\nu}(E_\nu r)\right\}\nonumber\\
			m_0(k_e,k_\nu)&=\frac{2}{\pi M_F}\int_0^\infty d\q'\q^{\prime 2}\int_0^\infty dr r^2 j_0(\q'r)F_0(\q^{\prime 2})\nonumber\\
			&\times
			\frac{(\p r)^{k_e-1}}{(2k_e-1)!!}\sqrt{\frac{2j_e+1}{2}}\delta_{k_ek_\nu}\nonumber\\
			&\times\left\{h_{k_e}(r)j_{k_\nu-1}(E_\nu r)-\frac{r}{R}d_{k_e}(r)j_{k_\nu}(E_\nu r)\right\}~,\nonumber\\
			\label{eq:M0m0}
		\end{align}
		where we have rescaled the functions by  $1/F_0(0)=1/M_F$. With them we can rewrite Eq.\eqref{eq:Tampstep2}, after some algebra, as:
		\begin{align}
			\mathcal{T}&=\frac{G_FV_{ud}}{4\pi}M_F
			\sum_{\kappa_e\mu_e\kappa_\nu\mu_\nu}\frac{(-1)^{j_e-\mu_e+1}}{\sqrt{2j_e+1}}\nonumber\\
			&\times
			a_{\kappa_e\mu_e}^{\lambda_e *}
			b_{\kappa_\nu \mu_\nu}^{\lambda_\nu *}
			\delta_{\mu_e,-\mu_\nu}\alpha_{\kappa_e}\nonumber\\
			&\times\left\{\text{sgn}(\kappa_e)M_0(k_e,k_\nu)+m_0(k_e,k_\nu)\right\}~.
		\end{align}

		Next we evaluate the squared amplitude and perform the phase-space integration. Neglecting kinematic recoil corrections, one can easily show that,
		\begin{equation}
			\frac{d\Gamma}{dE}\approx \frac{1}{(2\pi)^5}E\p (E_0-E)^2\int d\Omega_e\int d\Omega_\nu\sum_{\lambda_e\lambda_\nu}|\mathcal{T}|^2~.\label{eq:dGammadE}
		\end{equation}
		The angular integration and summation over lepton spin act only on the expansion coefficients $a_{\kappa_e\mu_e}^{\lambda_e},b_{\kappa_\mu\mu_\nu}^{\lambda_\nu}$:
		\begin{align}
			\sum_{\lambda_e}&\int d\Omega_e a_{\kappa_e\mu_e}^{\lambda_e *}a_{\kappa_e'\mu_e'}^{\lambda_e}=\frac{8\pi^2}{\p ^2}\delta_{\kappa_e\kappa_e'}\delta_{\mu_e\mu_e'}\,,\nonumber\\
			\sum_{\lambda_\nu}&\int d\Omega_\nu b_{\kappa_\nu \mu_\nu}^{\lambda_\nu *}b_{\kappa_\nu'\mu_\nu'}^{\lambda_\nu}=8\pi^2\delta_{\kappa_\nu\kappa_\nu'}\delta_{\mu_\nu\mu_\nu'}~.
		\end{align}
		We can further simplify Eq.\eqref{eq:M0m0}: Since both $M_0(k_e,k_\nu)$ and $m_0(k_e,k_\mu)$ are proportional to $\delta_{k_e k_\nu}$, we can define:
		\begin{align}
			M_0(k_e,k_\nu)&\equiv\delta_{k_ek_\nu}M_0(k),\nonumber\\
			m_0(k_e,k_\nu)&\equiv\delta_{k_ek_\nu}m_0(k)~,
		\end{align}
		where $k_e=k_\nu\equiv k$. Furthermore, we know that the Fourier transform of $F_0(\q^{\prime 2})$
		gives the charged weak distribution function:
		\begin{equation}
			\int_0^\infty d\q' \q^{\prime 2} F_0(\q^{\prime 2})j_0(\q' r)=2\pi^2 M_F\rho_\text{cw}(r)~,
		\end{equation}
		this leads us to the expressions of $M_0(k)$ and $m_0(k)$ in Eq.\eqref{eq:M0km0k}. Finally, plugging everything into Eq.\eqref{eq:dGammadE} 
		gives:
		\begin{equation}
			\frac{d\Gamma}{dE}\approx \frac{G_F^2V_{ud}^2}{2\pi^3}M_F^2\p E(E_0-E)^2F(E)C(E)~,
		\end{equation}
		where the Fermi function $F(E)$ and the shape factor $C(E)$ are exactly those given by Eqs.\eqref{eq:Fermi}, \eqref{eq:shape} respectively. 
		
		\section{\label{sec:distparameter}Parameterizations of nuclear charge distributions}
		
		Here we summarize the few parameterizations of nuclear charge distributions used in Ref.\cite{de1987nuclear}. 
		
		\begin{itemize}
			\item Two-parameter Fermi (2pF):
			\begin{equation}
				\rho_\text{ch}(r)=\frac{\rho_0}{1+\exp\{(r-c)/a\}}
			\end{equation}
			where 
			\begin{equation}
				\rho_0=-\frac{1}{8\pi a^3\text{Li}_3(-\exp\{c/a\})}
			\end{equation}
			and the MS charge radius:
			\begin{equation}
				\langle r_\text{ch}^2\rangle =\frac{12a^2\text{Li}_5(-\exp\{c/a\})}{\text{Li}_3(-\exp\{c/a\})}~.
			\end{equation}
			\item Three-parameter Fermi (3pF):
			\begin{equation}
				\rho_\text{ch}(r)=\frac{\rho_0(1+wr^2/c^2)}{1+\exp\{(r-c)/a\}}
			\end{equation}
			where
			\begin{equation}
				\rho_0=-\frac{1}{8\pi a^3\left[\text{Li}_3(-e^{c/a})+(12a^2w/c^2)\text{Li}_5(-e^{c/a})\right]}
			\end{equation}
			and
			\begin{equation}
				\langle r_\text{ch}^2\rangle=\frac{12\left[30a^4w\text{Li}_7(-\exp\{c/a\})+a^2c^2\text{Li}_5(-e^{c/a})\right]}{12a^2w\text{Li}_5(-e^{c/a})+c^2\text{Li}_3(-e^{c/a})}~,
			\end{equation}
			where $\text{Li}_s(z)$ is the polylogarithm function.
			\item Three-parameter Gaussian (3pG):
			\begin{equation}
				\rho_\text{ch}(r)=\frac{\rho_0(1+wr^2/c^2)}{1+\exp\{(r^2-c^2)/a^2\}}
			\end{equation}
			where
			\begin{equation}
				\rho_0=-\frac{2c^2}{\pi^{3/2} a^3\left[3a^2w\text{Li}_{5/2}(-e^{c^2/a^2})+2c^2\text{Li}_{3/2}(-e^{c^2/a^2})\right]}
			\end{equation}
			and
			\begin{equation}
				\langle r_\text{ch}^2\rangle=\frac{6a^2c^2\text{Li}_{5/2}(-e^{c^2/a^2})+15a^4w\text{Li}_{7/2}(-e^{c^2/a^2})}{6a^2w\text{Li}_{5/2}(-e^{c^2/a^2})+4c^2\text{Li}_{3/2}(-e^{c^2/a^2})}~.
			\end{equation}
			\item Harmonic oscillator (HO):
			\begin{equation}
				\rho_\text{ch}(r)=\rho_0\left(1+\alpha_\text{HO} r^2/b^2\right)\exp\{-r^2/b^2\}
			\end{equation}
			where
			\begin{equation}
				\rho_0=\frac{2}{\pi^{3/2}(3\alpha_\text{HO}+2)b^3}
			\end{equation}
			and
			\begin{equation}
				\langle r_\text{ch}^2\rangle =\frac{3(5\alpha_\text{HO}+2)b^2}{6\alpha_\text{HO}+4}~.
			\end{equation}
		\end{itemize}

	\end{appendix}

	\bibliography{ft_ref}

\begin{thebibliography}{98}%
\makeatletter
\providecommand \@ifxundefined [1]{%
 \@ifx{#1\undefined}
}%
\providecommand \@ifnum [1]{%
 \ifnum #1\expandafter \@firstoftwo
 \else \expandafter \@secondoftwo
 \fi
}%
\providecommand \@ifx [1]{%
 \ifx #1\expandafter \@firstoftwo
 \else \expandafter \@secondoftwo
 \fi
}%
\providecommand \natexlab [1]{#1}%
\providecommand \enquote  [1]{``#1''}%
\providecommand \bibnamefont  [1]{#1}%
\providecommand \bibfnamefont [1]{#1}%
\providecommand \citenamefont [1]{#1}%
\providecommand \href@noop [0]{\@secondoftwo}%
\providecommand \href [0]{\begingroup \@sanitize@url \@href}%
\providecommand \@href[1]{\@@startlink{#1}\@@href}%
\providecommand \@@href[1]{\endgroup#1\@@endlink}%
\providecommand \@sanitize@url [0]{\catcode `\\12\catcode `\$12\catcode
  `\&12\catcode `\#12\catcode `\^12\catcode `\_12\catcode `\%12\relax}%
\providecommand \@@startlink[1]{}%
\providecommand \@@endlink[0]{}%
\providecommand \url  [0]{\begingroup\@sanitize@url \@url }%
\providecommand \@url [1]{\endgroup\@href {#1}{\urlprefix }}%
\providecommand \urlprefix  [0]{URL }%
\providecommand \Eprint [0]{\href }%
\providecommand \doibase [0]{https://doi.org/}%
\providecommand \selectlanguage [0]{\@gobble}%
\providecommand \bibinfo  [0]{\@secondoftwo}%
\providecommand \bibfield  [0]{\@secondoftwo}%
\providecommand \translation [1]{[#1]}%
\providecommand \BibitemOpen [0]{}%
\providecommand \bibitemStop [0]{}%
\providecommand \bibitemNoStop [0]{.\EOS\space}%
\providecommand \EOS [0]{\spacefactor3000\relax}%
\providecommand \BibitemShut  [1]{\csname bibitem#1\endcsname}%
\let\auto@bib@innerbib\@empty
\bibitem [{\citenamefont {Hardy}\ and\ \citenamefont
  {Towner}(2020)}]{Hardy:2020qwl}%
  \BibitemOpen
  \bibfield  {author} {\bibinfo {author} {\bibfnamefont {J.~C.}\ \bibnamefont
  {Hardy}}\ and\ \bibinfo {author} {\bibfnamefont {I.~S.}\ \bibnamefont
  {Towner}},\ }\bibfield  {title} {\bibinfo {title} {{Superallowed $0^+ \to
  0^+$ nuclear $\beta$ decays: 2020 critical survey, with implications for
  V$_{ud}$ and CKM unitarity}},\ }\href
  {https://doi.org/10.1103/PhysRevC.102.045501} {\bibfield  {journal} {\bibinfo
   {journal} {Phys. Rev. C}\ }\textbf {\bibinfo {volume} {102}},\ \bibinfo
  {pages} {045501} (\bibinfo {year} {2020})}\BibitemShut {NoStop}%
\bibitem [{\citenamefont {Gonzalez}\ \emph {et~al.}(2021)\citenamefont
  {Gonzalez} \emph {et~al.}}]{UCNt:2021pcg}%
  \BibitemOpen
  \bibfield  {author} {\bibinfo {author} {\bibfnamefont {F.~M.}\ \bibnamefont
  {Gonzalez}} \emph {et~al.} (\bibinfo {collaboration}
  {UCN\ensuremath{\tau}}),\ }\bibfield  {title} {\bibinfo {title} {{Improved
  Neutron Lifetime Measurement with UCN\ensuremath{\tau}}},\ }\href
  {https://doi.org/10.1103/PhysRevLett.127.162501} {\bibfield  {journal}
  {\bibinfo  {journal} {Phys. Rev. Lett.}\ }\textbf {\bibinfo {volume} {127}},\
  \bibinfo {pages} {162501} (\bibinfo {year} {2021})},\ \Eprint
  {https://arxiv.org/abs/2106.10375} {arXiv:2106.10375 [nucl-ex]} \BibitemShut
  {NoStop}%
\bibitem [{\citenamefont {M\"arkisch}\ \emph {et~al.}(2019)\citenamefont
  {M\"arkisch} \emph {et~al.}}]{Markisch:2018ndu}%
  \BibitemOpen
  \bibfield  {author} {\bibinfo {author} {\bibfnamefont {B.}~\bibnamefont
  {M\"arkisch}} \emph {et~al.},\ }\bibfield  {title} {\bibinfo {title}
  {{Measurement of the Weak Axial-Vector Coupling Constant in the Decay of Free
  Neutrons Using a Pulsed Cold Neutron Beam}},\ }\href
  {https://doi.org/10.1103/PhysRevLett.122.242501} {\bibfield  {journal}
  {\bibinfo  {journal} {Phys. Rev. Lett.}\ }\textbf {\bibinfo {volume} {122}},\
  \bibinfo {pages} {242501} (\bibinfo {year} {2019})},\ \Eprint
  {https://arxiv.org/abs/1812.04666} {arXiv:1812.04666 [nucl-ex]} \BibitemShut
  {NoStop}%
\bibitem [{\citenamefont {Gorchtein}\ and\ \citenamefont
  {Seng}(2023)}]{Gorchtein:2023srs}%
  \BibitemOpen
  \bibfield  {author} {\bibinfo {author} {\bibfnamefont {M.}~\bibnamefont
  {Gorchtein}}\ and\ \bibinfo {author} {\bibfnamefont {C.-Y.}\ \bibnamefont
  {Seng}},\ }\bibfield  {title} {\bibinfo {title} {{The Standard Model Theory
  of Neutron Beta Decay}},\ }\href {https://doi.org/10.3390/universe9090422}
  {\bibfield  {journal} {\bibinfo  {journal} {Universe}\ }\textbf {\bibinfo
  {volume} {9}},\ \bibinfo {pages} {422} (\bibinfo {year} {2023})},\ \Eprint
  {https://arxiv.org/abs/2307.01145} {arXiv:2307.01145 [hep-ph]} \BibitemShut
  {NoStop}%
\bibitem [{\citenamefont {Cirigliano}\ \emph
  {et~al.}(2023{\natexlab{a}})\citenamefont {Cirigliano}, \citenamefont
  {Crivellin}, \citenamefont {Hoferichter},\ and\ \citenamefont
  {Moulson}}]{Cirigliano:2022yyo}%
  \BibitemOpen
  \bibfield  {author} {\bibinfo {author} {\bibfnamefont {V.}~\bibnamefont
  {Cirigliano}}, \bibinfo {author} {\bibfnamefont {A.}~\bibnamefont
  {Crivellin}}, \bibinfo {author} {\bibfnamefont {M.}~\bibnamefont
  {Hoferichter}},\ and\ \bibinfo {author} {\bibfnamefont {M.}~\bibnamefont
  {Moulson}},\ }\bibfield  {title} {\bibinfo {title} {{Scrutinizing CKM
  unitarity with a new measurement of the K\ensuremath{\mu}3/K\ensuremath{\mu}2
  branching fraction}},\ }\href
  {https://doi.org/10.1016/j.physletb.2023.137748} {\bibfield  {journal}
  {\bibinfo  {journal} {Phys. Lett. B}\ }\textbf {\bibinfo {volume} {838}},\
  \bibinfo {pages} {137748} (\bibinfo {year} {2023}{\natexlab{a}})},\ \Eprint
  {https://arxiv.org/abs/2208.11707} {arXiv:2208.11707 [hep-ph]} \BibitemShut
  {NoStop}%
\bibitem [{con()}]{consensus}%
  \BibitemOpen
  \href@noop {} {}\bibinfo {note} {Martin Hoferichter, private
  communication.}\BibitemShut {Stop}%
\bibitem [{\citenamefont {Tan}(2023)}]{Tan:2023mpj}%
  \BibitemOpen
  \bibfield  {author} {\bibinfo {author} {\bibfnamefont {W.}~\bibnamefont
  {Tan}},\ }\bibfield  {title} {\bibinfo {title} {{Neutron Lifetime Anomaly and
  Mirror Matter Theory}},\ }\href {https://doi.org/10.3390/universe9040180}
  {\bibfield  {journal} {\bibinfo  {journal} {Universe}\ }\textbf {\bibinfo
  {volume} {9}},\ \bibinfo {pages} {180} (\bibinfo {year} {2023})},\ \Eprint
  {https://arxiv.org/abs/2302.07805} {arXiv:2302.07805 [nucl-ex]} \BibitemShut
  {NoStop}%
\bibitem [{\citenamefont {Wietfeldt}\ \emph {et~al.}(2023)\citenamefont
  {Wietfeldt}, \citenamefont {Byron}, \citenamefont {Collett}, \citenamefont
  {Dewey}, \citenamefont {Gentile}, \citenamefont {Hassan}, \citenamefont
  {Jones}, \citenamefont {Komives}, \citenamefont {Nico},\ and\ \citenamefont
  {Stephenson}}]{Wietfeldt:2023mdb}%
  \BibitemOpen
  \bibfield  {author} {\bibinfo {author} {\bibfnamefont {F.~E.}\ \bibnamefont
  {Wietfeldt}}, \bibinfo {author} {\bibfnamefont {W.~A.}\ \bibnamefont
  {Byron}}, \bibinfo {author} {\bibfnamefont {B.}~\bibnamefont {Collett}},
  \bibinfo {author} {\bibfnamefont {M.~S.}\ \bibnamefont {Dewey}}, \bibinfo
  {author} {\bibfnamefont {T.~R.}\ \bibnamefont {Gentile}}, \bibinfo {author}
  {\bibfnamefont {M.~T.}\ \bibnamefont {Hassan}}, \bibinfo {author}
  {\bibfnamefont {G.~L.}\ \bibnamefont {Jones}}, \bibinfo {author}
  {\bibfnamefont {A.}~\bibnamefont {Komives}}, \bibinfo {author} {\bibfnamefont
  {J.~S.}\ \bibnamefont {Nico}},\ and\ \bibinfo {author} {\bibfnamefont
  {E.~J.}\ \bibnamefont {Stephenson}},\ }\bibfield  {title} {\bibinfo {title}
  {{Recoil-Order and Radiative Corrections to the aCORN Experiment}},\
  }\href@noop {} {\  (\bibinfo {year} {2023})},\ \Eprint
  {https://arxiv.org/abs/2306.15042} {arXiv:2306.15042 [nucl-ex]} \BibitemShut
  {NoStop}%
\bibitem [{\citenamefont {Beck}\ \emph {et~al.}(2024)\citenamefont {Beck},
  \citenamefont {Heil}, \citenamefont {Schmidt}, \citenamefont {Bae\ss{}ler},
  \citenamefont {Gl\"uck}, \citenamefont {Konrad},\ and\ \citenamefont
  {Schmidt}}]{Beck:2023hnt}%
  \BibitemOpen
  \bibfield  {author} {\bibinfo {author} {\bibfnamefont {M.}~\bibnamefont
  {Beck}}, \bibinfo {author} {\bibfnamefont {W.}~\bibnamefont {Heil}}, \bibinfo
  {author} {\bibfnamefont {C.}~\bibnamefont {Schmidt}}, \bibinfo {author}
  {\bibfnamefont {S.}~\bibnamefont {Bae\ss{}ler}}, \bibinfo {author}
  {\bibfnamefont {F.}~\bibnamefont {Gl\"uck}}, \bibinfo {author} {\bibfnamefont
  {G.}~\bibnamefont {Konrad}},\ and\ \bibinfo {author} {\bibfnamefont
  {U.}~\bibnamefont {Schmidt}},\ }\bibfield  {title} {\bibinfo {title}
  {{Reanalysis of the
  \ensuremath{\beta}\ensuremath{-}\ensuremath{\nu}\textasciimacron{}e Angular
  Correlation Measurement from the aSPECT Experiment with New Constraints on
  Fierz Interference}},\ }\href
  {https://doi.org/10.1103/PhysRevLett.132.102501} {\bibfield  {journal}
  {\bibinfo  {journal} {Phys. Rev. Lett.}\ }\textbf {\bibinfo {volume} {132}},\
  \bibinfo {pages} {102501} (\bibinfo {year} {2024})},\ \Eprint
  {https://arxiv.org/abs/2308.16170} {arXiv:2308.16170 [nucl-ex]} \BibitemShut
  {NoStop}%
\bibitem [{\citenamefont {Workman}\ \emph {et~al.}(2022)\citenamefont {Workman}
  \emph {et~al.}}]{ParticleDataGroup:2022pth}%
  \BibitemOpen
  \bibfield  {author} {\bibinfo {author} {\bibfnamefont {R.~L.}\ \bibnamefont
  {Workman}} \emph {et~al.} (\bibinfo {collaboration} {Particle Data Group}),\
  }\bibfield  {title} {\bibinfo {title} {{Review of Particle Physics}},\ }\href
  {https://doi.org/10.1093/ptep/ptac097} {\bibfield  {journal} {\bibinfo
  {journal} {PTEP}\ }\textbf {\bibinfo {volume} {2022}},\ \bibinfo {pages}
  {083C01} (\bibinfo {year} {2022})}\BibitemShut {NoStop}%
\bibitem [{\citenamefont {Seng}\ \emph
  {et~al.}(2022{\natexlab{a}})\citenamefont {Seng}, \citenamefont {Galviz},
  \citenamefont {Marciano},\ and\ \citenamefont {Mei\ss{}ner}}]{Seng:2021nar}%
  \BibitemOpen
  \bibfield  {author} {\bibinfo {author} {\bibfnamefont {C.-Y.}\ \bibnamefont
  {Seng}}, \bibinfo {author} {\bibfnamefont {D.}~\bibnamefont {Galviz}},
  \bibinfo {author} {\bibfnamefont {W.~J.}\ \bibnamefont {Marciano}},\ and\
  \bibinfo {author} {\bibfnamefont {U.-G.}\ \bibnamefont {Mei\ss{}ner}},\
  }\bibfield  {title} {\bibinfo {title} {{Update on $|V_{us}|$ and
  $|V_{us}/V_{ud}|$ from semileptonic kaon and pion decays}},\ }\href
  {https://doi.org/10.1103/PhysRevD.105.013005} {\bibfield  {journal} {\bibinfo
   {journal} {Phys. Rev. D}\ }\textbf {\bibinfo {volume} {105}},\ \bibinfo
  {pages} {013005} (\bibinfo {year} {2022}{\natexlab{a}})},\ \Eprint
  {https://arxiv.org/abs/2107.14708} {arXiv:2107.14708 [hep-ph]} \BibitemShut
  {NoStop}%
\bibitem [{\citenamefont {Seng}\ \emph
  {et~al.}(2022{\natexlab{b}})\citenamefont {Seng}, \citenamefont {Galviz},
  \citenamefont {Gorchtein},\ and\ \citenamefont {Mei\ss{}ner}}]{Seng:2022wcw}%
  \BibitemOpen
  \bibfield  {author} {\bibinfo {author} {\bibfnamefont {C.-Y.}\ \bibnamefont
  {Seng}}, \bibinfo {author} {\bibfnamefont {D.}~\bibnamefont {Galviz}},
  \bibinfo {author} {\bibfnamefont {M.}~\bibnamefont {Gorchtein}},\ and\
  \bibinfo {author} {\bibfnamefont {U.-G.}\ \bibnamefont {Mei\ss{}ner}},\
  }\bibfield  {title} {\bibinfo {title} {{Complete theory of radiative
  corrections to K$_{\ell 3}$ decays and the V$_{us}$ update}},\ }\href
  {https://doi.org/10.1007/JHEP07(2022)071} {\bibfield  {journal} {\bibinfo
  {journal} {JHEP}\ }\textbf {\bibinfo {volume} {07}},\ \bibinfo {pages}
  {071}},\ \Eprint {https://arxiv.org/abs/2203.05217} {arXiv:2203.05217
  [hep-ph]} \BibitemShut {NoStop}%
\bibitem [{\citenamefont {Carrasco}\ \emph {et~al.}(2016)\citenamefont
  {Carrasco}, \citenamefont {Lami}, \citenamefont {Lubicz}, \citenamefont
  {Riggio}, \citenamefont {Simula},\ and\ \citenamefont
  {Tarantino}}]{Carrasco:2016kpy}%
  \BibitemOpen
  \bibfield  {author} {\bibinfo {author} {\bibfnamefont {N.}~\bibnamefont
  {Carrasco}}, \bibinfo {author} {\bibfnamefont {P.}~\bibnamefont {Lami}},
  \bibinfo {author} {\bibfnamefont {V.}~\bibnamefont {Lubicz}}, \bibinfo
  {author} {\bibfnamefont {L.}~\bibnamefont {Riggio}}, \bibinfo {author}
  {\bibfnamefont {S.}~\bibnamefont {Simula}},\ and\ \bibinfo {author}
  {\bibfnamefont {C.}~\bibnamefont {Tarantino}},\ }\bibfield  {title} {\bibinfo
  {title} {{$K \to \pi$ semileptonic form factors with $N_f=2+1+1$ twisted mass
  fermions}},\ }\href {https://doi.org/10.1103/PhysRevD.93.114512} {\bibfield
  {journal} {\bibinfo  {journal} {Phys. Rev. D}\ }\textbf {\bibinfo {volume}
  {93}},\ \bibinfo {pages} {114512} (\bibinfo {year} {2016})},\ \Eprint
  {https://arxiv.org/abs/1602.04113} {arXiv:1602.04113 [hep-lat]} \BibitemShut
  {NoStop}%
\bibitem [{\citenamefont {Bazavov}\ \emph {et~al.}(2019)\citenamefont {Bazavov}
  \emph {et~al.}}]{Bazavov:2018kjg}%
  \BibitemOpen
  \bibfield  {author} {\bibinfo {author} {\bibfnamefont {A.}~\bibnamefont
  {Bazavov}} \emph {et~al.} (\bibinfo {collaboration} {Fermilab Lattice,
  MILC}),\ }\bibfield  {title} {\bibinfo {title} {{$|V_{us}|$ from $K_{\ell 3}$
  decay and four-flavor lattice QCD}},\ }\href
  {https://doi.org/10.1103/PhysRevD.99.114509} {\bibfield  {journal} {\bibinfo
  {journal} {Phys. Rev.}\ }\textbf {\bibinfo {volume} {D99}},\ \bibinfo {pages}
  {114509} (\bibinfo {year} {2019})},\ \Eprint
  {https://arxiv.org/abs/1809.02827} {arXiv:1809.02827 [hep-lat]} \BibitemShut
  {NoStop}%
\bibitem [{\citenamefont {Aoki}\ \emph {et~al.}(2022)\citenamefont {Aoki} \emph
  {et~al.}}]{FlavourLatticeAveragingGroupFLAG:2021npn}%
  \BibitemOpen
  \bibfield  {author} {\bibinfo {author} {\bibfnamefont {Y.}~\bibnamefont
  {Aoki}} \emph {et~al.} (\bibinfo {collaboration} {Flavour Lattice Averaging
  Group (FLAG)}),\ }\bibfield  {title} {\bibinfo {title} {{FLAG Review 2021}},\
  }\href {https://doi.org/10.1140/epjc/s10052-022-10536-1} {\bibfield
  {journal} {\bibinfo  {journal} {Eur. Phys. J. C}\ }\textbf {\bibinfo {volume}
  {82}},\ \bibinfo {pages} {869} (\bibinfo {year} {2022})},\ \Eprint
  {https://arxiv.org/abs/2111.09849} {arXiv:2111.09849 [hep-lat]} \BibitemShut
  {NoStop}%
\bibitem [{\citenamefont {Seng}\ \emph {et~al.}(2018)\citenamefont {Seng},
  \citenamefont {Gorchtein}, \citenamefont {Patel},\ and\ \citenamefont
  {Ramsey-Musolf}}]{Seng:2018yzq}%
  \BibitemOpen
  \bibfield  {author} {\bibinfo {author} {\bibfnamefont {C.-Y.}\ \bibnamefont
  {Seng}}, \bibinfo {author} {\bibfnamefont {M.}~\bibnamefont {Gorchtein}},
  \bibinfo {author} {\bibfnamefont {H.~H.}\ \bibnamefont {Patel}},\ and\
  \bibinfo {author} {\bibfnamefont {M.~J.}\ \bibnamefont {Ramsey-Musolf}},\
  }\bibfield  {title} {\bibinfo {title} {{Reduced Hadronic Uncertainty in the
  Determination of $V_{ud}$}},\ }\href
  {https://doi.org/10.1103/PhysRevLett.121.241804} {\bibfield  {journal}
  {\bibinfo  {journal} {Phys. Rev. Lett.}\ }\textbf {\bibinfo {volume} {121}},\
  \bibinfo {pages} {241804} (\bibinfo {year} {2018})},\ \Eprint
  {https://arxiv.org/abs/1807.10197} {arXiv:1807.10197 [hep-ph]} \BibitemShut
  {NoStop}%
\bibitem [{\citenamefont {Seng}\ \emph {et~al.}(2019)\citenamefont {Seng},
  \citenamefont {Gorchtein},\ and\ \citenamefont
  {Ramsey-Musolf}}]{Seng:2018qru}%
  \BibitemOpen
  \bibfield  {author} {\bibinfo {author} {\bibfnamefont {C.~Y.}\ \bibnamefont
  {Seng}}, \bibinfo {author} {\bibfnamefont {M.}~\bibnamefont {Gorchtein}},\
  and\ \bibinfo {author} {\bibfnamefont {M.~J.}\ \bibnamefont
  {Ramsey-Musolf}},\ }\bibfield  {title} {\bibinfo {title} {{Dispersive
  evaluation of the inner radiative correction in neutron and nuclear $\beta$
  decay}},\ }\href {https://doi.org/10.1103/PhysRevD.100.013001} {\bibfield
  {journal} {\bibinfo  {journal} {Phys. Rev.}\ }\textbf {\bibinfo {volume}
  {D100}},\ \bibinfo {pages} {013001} (\bibinfo {year} {2019})},\ \Eprint
  {https://arxiv.org/abs/1812.03352} {arXiv:1812.03352 [nucl-th]} \BibitemShut
  {NoStop}%
\bibitem [{\citenamefont {Shiells}\ \emph {et~al.}(2021)\citenamefont
  {Shiells}, \citenamefont {Blunden},\ and\ \citenamefont
  {Melnitchouk}}]{Shiells:2020fqp}%
  \BibitemOpen
  \bibfield  {author} {\bibinfo {author} {\bibfnamefont {K.}~\bibnamefont
  {Shiells}}, \bibinfo {author} {\bibfnamefont {P.~G.}\ \bibnamefont
  {Blunden}},\ and\ \bibinfo {author} {\bibfnamefont {W.}~\bibnamefont
  {Melnitchouk}},\ }\bibfield  {title} {\bibinfo {title} {{Electroweak axial
  structure functions and improved extraction of the Vud CKM matrix element}},\
  }\href {https://doi.org/10.1103/PhysRevD.104.033003} {\bibfield  {journal}
  {\bibinfo  {journal} {Phys. Rev. D}\ }\textbf {\bibinfo {volume} {104}},\
  \bibinfo {pages} {033003} (\bibinfo {year} {2021})},\ \Eprint
  {https://arxiv.org/abs/2012.01580} {arXiv:2012.01580 [hep-ph]} \BibitemShut
  {NoStop}%
\bibitem [{\citenamefont {Cirigliano}\ \emph {et~al.}(2022)\citenamefont
  {Cirigliano}, \citenamefont {de~Vries}, \citenamefont {Hayen}, \citenamefont
  {Mereghetti},\ and\ \citenamefont {Walker-Loud}}]{Cirigliano:2022hob}%
  \BibitemOpen
  \bibfield  {author} {\bibinfo {author} {\bibfnamefont {V.}~\bibnamefont
  {Cirigliano}}, \bibinfo {author} {\bibfnamefont {J.}~\bibnamefont
  {de~Vries}}, \bibinfo {author} {\bibfnamefont {L.}~\bibnamefont {Hayen}},
  \bibinfo {author} {\bibfnamefont {E.}~\bibnamefont {Mereghetti}},\ and\
  \bibinfo {author} {\bibfnamefont {A.}~\bibnamefont {Walker-Loud}},\
  }\bibfield  {title} {\bibinfo {title} {{Pion-Induced Radiative Corrections to
  Neutron \ensuremath{\beta} Decay}},\ }\href
  {https://doi.org/10.1103/PhysRevLett.129.121801} {\bibfield  {journal}
  {\bibinfo  {journal} {Phys. Rev. Lett.}\ }\textbf {\bibinfo {volume} {129}},\
  \bibinfo {pages} {121801} (\bibinfo {year} {2022})},\ \Eprint
  {https://arxiv.org/abs/2202.10439} {arXiv:2202.10439 [nucl-th]} \BibitemShut
  {NoStop}%
\bibitem [{\citenamefont {Cirigliano}\ \emph
  {et~al.}(2023{\natexlab{b}})\citenamefont {Cirigliano}, \citenamefont
  {Dekens}, \citenamefont {Mereghetti},\ and\ \citenamefont
  {Tomalak}}]{Cirigliano:2023fnz}%
  \BibitemOpen
  \bibfield  {author} {\bibinfo {author} {\bibfnamefont {V.}~\bibnamefont
  {Cirigliano}}, \bibinfo {author} {\bibfnamefont {W.}~\bibnamefont {Dekens}},
  \bibinfo {author} {\bibfnamefont {E.}~\bibnamefont {Mereghetti}},\ and\
  \bibinfo {author} {\bibfnamefont {O.}~\bibnamefont {Tomalak}},\ }\bibfield
  {title} {\bibinfo {title} {{Effective field theory for radiative corrections
  to charged-current processes I: Vector coupling}},\ }\href@noop {} {\
  (\bibinfo {year} {2023}{\natexlab{b}})},\ \Eprint
  {https://arxiv.org/abs/2306.03138} {arXiv:2306.03138 [hep-ph]} \BibitemShut
  {NoStop}%
\bibitem [{\citenamefont {Feng}\ \emph {et~al.}(2020)\citenamefont {Feng},
  \citenamefont {Gorchtein}, \citenamefont {Jin}, \citenamefont {Ma},\ and\
  \citenamefont {Seng}}]{Feng:2020zdc}%
  \BibitemOpen
  \bibfield  {author} {\bibinfo {author} {\bibfnamefont {X.}~\bibnamefont
  {Feng}}, \bibinfo {author} {\bibfnamefont {M.}~\bibnamefont {Gorchtein}},
  \bibinfo {author} {\bibfnamefont {L.-C.}\ \bibnamefont {Jin}}, \bibinfo
  {author} {\bibfnamefont {P.-X.}\ \bibnamefont {Ma}},\ and\ \bibinfo {author}
  {\bibfnamefont {C.-Y.}\ \bibnamefont {Seng}},\ }\bibfield  {title} {\bibinfo
  {title} {{First-principles calculation of electroweak box diagrams from
  lattice QCD}},\ }\href {https://doi.org/10.1103/PhysRevLett.124.192002}
  {\bibfield  {journal} {\bibinfo  {journal} {Phys. Rev. Lett.}\ }\textbf
  {\bibinfo {volume} {124}},\ \bibinfo {pages} {192002} (\bibinfo {year}
  {2020})},\ \Eprint {https://arxiv.org/abs/2003.09798} {arXiv:2003.09798
  [hep-lat]} \BibitemShut {NoStop}%
\bibitem [{\citenamefont {Yoo}\ \emph {et~al.}(2023)\citenamefont {Yoo},
  \citenamefont {Bhattacharya}, \citenamefont {Gupta}, \citenamefont {Mondal},\
  and\ \citenamefont {Yoon}}]{Yoo:2023gln}%
  \BibitemOpen
  \bibfield  {author} {\bibinfo {author} {\bibfnamefont {J.-S.}\ \bibnamefont
  {Yoo}}, \bibinfo {author} {\bibfnamefont {T.}~\bibnamefont {Bhattacharya}},
  \bibinfo {author} {\bibfnamefont {R.}~\bibnamefont {Gupta}}, \bibinfo
  {author} {\bibfnamefont {S.}~\bibnamefont {Mondal}},\ and\ \bibinfo {author}
  {\bibfnamefont {B.}~\bibnamefont {Yoon}},\ }\bibfield  {title} {\bibinfo
  {title} {{Electroweak box diagram contribution for pion and kaon decay from
  lattice QCD}},\ }\href {https://doi.org/10.1103/PhysRevD.108.034508}
  {\bibfield  {journal} {\bibinfo  {journal} {Phys. Rev. D}\ }\textbf {\bibinfo
  {volume} {108}},\ \bibinfo {pages} {034508} (\bibinfo {year} {2023})},\
  \Eprint {https://arxiv.org/abs/2305.03198} {arXiv:2305.03198 [hep-lat]}
  \BibitemShut {NoStop}%
\bibitem [{\citenamefont {Ma}\ \emph {et~al.}(2023)\citenamefont {Ma},
  \citenamefont {Feng}, \citenamefont {Gorchtein}, \citenamefont {Jin},
  \citenamefont {Liu}, \citenamefont {Seng}, \citenamefont {Wang},\ and\
  \citenamefont {Zhang}}]{Ma:2023kfr}%
  \BibitemOpen
  \bibfield  {author} {\bibinfo {author} {\bibfnamefont {P.-X.}\ \bibnamefont
  {Ma}}, \bibinfo {author} {\bibfnamefont {X.}~\bibnamefont {Feng}}, \bibinfo
  {author} {\bibfnamefont {M.}~\bibnamefont {Gorchtein}}, \bibinfo {author}
  {\bibfnamefont {L.-C.}\ \bibnamefont {Jin}}, \bibinfo {author} {\bibfnamefont
  {K.-F.}\ \bibnamefont {Liu}}, \bibinfo {author} {\bibfnamefont {C.-Y.}\
  \bibnamefont {Seng}}, \bibinfo {author} {\bibfnamefont {B.-G.}\ \bibnamefont
  {Wang}},\ and\ \bibinfo {author} {\bibfnamefont {Z.-L.}\ \bibnamefont
  {Zhang}},\ }\bibfield  {title} {\bibinfo {title} {{Lattice QCD Calculation of
  Electroweak Box Contributions to Superallowed Nuclear and Neutron Beta
  Decays}},\ }\href@noop {} {\  (\bibinfo {year} {2023})},\ \Eprint
  {https://arxiv.org/abs/2308.16755} {arXiv:2308.16755 [hep-lat]} \BibitemShut
  {NoStop}%
\bibitem [{\citenamefont {Hardy}\ and\ \citenamefont
  {Towner}(1975)}]{Hardy:1975eq}%
  \BibitemOpen
  \bibfield  {author} {\bibinfo {author} {\bibfnamefont {J.~C.}\ \bibnamefont
  {Hardy}}\ and\ \bibinfo {author} {\bibfnamefont {I.~S.}\ \bibnamefont
  {Towner}},\ }\bibfield  {title} {\bibinfo {title} {{Superallowed 0+
  --\ensuremath{>} 0+ Nuclear beta Decays and Cabibbo Universality}},\ }\href
  {https://doi.org/10.1016/0375-9474(75)90214-6} {\bibfield  {journal}
  {\bibinfo  {journal} {Nucl. Phys. A}\ }\textbf {\bibinfo {volume} {254}},\
  \bibinfo {pages} {221} (\bibinfo {year} {1975})}\BibitemShut {NoStop}%
\bibitem [{\citenamefont {Towner}\ and\ \citenamefont
  {Hardy}(2002)}]{Towner:2002rg}%
  \BibitemOpen
  \bibfield  {author} {\bibinfo {author} {\bibfnamefont {I.~S.}\ \bibnamefont
  {Towner}}\ and\ \bibinfo {author} {\bibfnamefont {J.~C.}\ \bibnamefont
  {Hardy}},\ }\bibfield  {title} {\bibinfo {title} {{Calculated corrections to
  superallowed Fermi beta decay: New evaluation of the nuclear structure
  dependent terms}},\ }\href {https://doi.org/10.1103/PhysRevC.66.035501}
  {\bibfield  {journal} {\bibinfo  {journal} {Phys. Rev. C}\ }\textbf {\bibinfo
  {volume} {66}},\ \bibinfo {pages} {035501} (\bibinfo {year} {2002})},\
  \Eprint {https://arxiv.org/abs/nucl-th/0209014} {arXiv:nucl-th/0209014}
  \BibitemShut {NoStop}%
\bibitem [{\citenamefont {Hardy}\ and\ \citenamefont
  {Towner}(2005)}]{Hardy:2004id}%
  \BibitemOpen
  \bibfield  {author} {\bibinfo {author} {\bibfnamefont {J.~C.}\ \bibnamefont
  {Hardy}}\ and\ \bibinfo {author} {\bibfnamefont {I.~S.}\ \bibnamefont
  {Towner}},\ }\bibfield  {title} {\bibinfo {title} {{Superallowed 0+
  ---\ensuremath{>} 0+ nuclear beta decays: A Critical survey with tests of CVC
  and the standard model}},\ }\href
  {https://doi.org/10.1103/PhysRevC.71.055501} {\bibfield  {journal} {\bibinfo
  {journal} {Phys. Rev. C}\ }\textbf {\bibinfo {volume} {71}},\ \bibinfo
  {pages} {055501} (\bibinfo {year} {2005})},\ \Eprint
  {https://arxiv.org/abs/nucl-th/0412056} {arXiv:nucl-th/0412056} \BibitemShut
  {NoStop}%
\bibitem [{\citenamefont {Bambynek}\ \emph {et~al.}(1977)\citenamefont
  {Bambynek}, \citenamefont {Behrens}, \citenamefont {Chen}, \citenamefont
  {Crasemann}, \citenamefont {Fitzpatrick}, \citenamefont {Ledingham},
  \citenamefont {Genz}, \citenamefont {Mutterer},\ and\ \citenamefont
  {Intemann}}]{Bambynek:1977zz}%
  \BibitemOpen
  \bibfield  {author} {\bibinfo {author} {\bibfnamefont {W.}~\bibnamefont
  {Bambynek}}, \bibinfo {author} {\bibfnamefont {H.}~\bibnamefont {Behrens}},
  \bibinfo {author} {\bibfnamefont {M.~H.}\ \bibnamefont {Chen}}, \bibinfo
  {author} {\bibfnamefont {B.}~\bibnamefont {Crasemann}}, \bibinfo {author}
  {\bibfnamefont {M.~L.}\ \bibnamefont {Fitzpatrick}}, \bibinfo {author}
  {\bibfnamefont {K.~W.~D.}\ \bibnamefont {Ledingham}}, \bibinfo {author}
  {\bibfnamefont {H.}~\bibnamefont {Genz}}, \bibinfo {author} {\bibfnamefont
  {M.}~\bibnamefont {Mutterer}},\ and\ \bibinfo {author} {\bibfnamefont
  {R.~L.}\ \bibnamefont {Intemann}},\ }\bibfield  {title} {\bibinfo {title}
  {{Orbital electron capture by the nucleus}},\ }\href
  {https://doi.org/10.1103/RevModPhys.49.77} {\bibfield  {journal} {\bibinfo
  {journal} {Rev. Mod. Phys.}\ }\textbf {\bibinfo {volume} {49}},\ \bibinfo
  {pages} {77} (\bibinfo {year} {1977})},\ \bibinfo {note} {[Erratum:
  Rev.Mod.Phys. 49, 961--962 (1977)]}\BibitemShut {NoStop}%
\bibitem [{\citenamefont {Sirlin}(1967)}]{Sirlin:1967zza}%
  \BibitemOpen
  \bibfield  {author} {\bibinfo {author} {\bibfnamefont {A.}~\bibnamefont
  {Sirlin}},\ }\bibfield  {title} {\bibinfo {title} {{General Properties of the
  Electromagnetic Corrections to the Beta Decay of a Physical Nucleon}},\
  }\href {https://doi.org/10.1103/PhysRev.164.1767} {\bibfield  {journal}
  {\bibinfo  {journal} {Phys. Rev.}\ }\textbf {\bibinfo {volume} {164}},\
  \bibinfo {pages} {1767} (\bibinfo {year} {1967})}\BibitemShut {NoStop}%
\bibitem [{\citenamefont {Sirlin}(1987)}]{Sirlin:1987sy}%
  \BibitemOpen
  \bibfield  {author} {\bibinfo {author} {\bibfnamefont {A.}~\bibnamefont
  {Sirlin}},\ }\bibfield  {title} {\bibinfo {title} {{Remarks Concerning the
  O(z alpha**2) Corrections to Fermi Decays, Conserved Vector Current
  Predictions and Universality}},\ }\href
  {https://doi.org/10.1103/PhysRevD.35.3423} {\bibfield  {journal} {\bibinfo
  {journal} {Phys. Rev. D}\ }\textbf {\bibinfo {volume} {35}},\ \bibinfo
  {pages} {3423} (\bibinfo {year} {1987})}\BibitemShut {NoStop}%
\bibitem [{\citenamefont {Sirlin}\ and\ \citenamefont
  {Zucchini}(1986)}]{Sirlin:1986cc}%
  \BibitemOpen
  \bibfield  {author} {\bibinfo {author} {\bibfnamefont {A.}~\bibnamefont
  {Sirlin}}\ and\ \bibinfo {author} {\bibfnamefont {R.}~\bibnamefont
  {Zucchini}},\ }\bibfield  {title} {\bibinfo {title} {{Accurate Verification
  of the Conserved Vector Current and Standard Model Predictions}},\ }\href
  {https://doi.org/10.1103/PhysRevLett.57.1994} {\bibfield  {journal} {\bibinfo
   {journal} {Phys. Rev. Lett.}\ }\textbf {\bibinfo {volume} {57}},\ \bibinfo
  {pages} {1994} (\bibinfo {year} {1986})}\BibitemShut {NoStop}%
\bibitem [{\citenamefont {Barker}\ \emph {et~al.}(1992)\citenamefont {Barker},
  \citenamefont {Brown}, \citenamefont {Jaus},\ and\ \citenamefont
  {Rasche}}]{Barker:1991tw}%
  \BibitemOpen
  \bibfield  {author} {\bibinfo {author} {\bibfnamefont {F.~C.}\ \bibnamefont
  {Barker}}, \bibinfo {author} {\bibfnamefont {B.~A.}\ \bibnamefont {Brown}},
  \bibinfo {author} {\bibfnamefont {W.}~\bibnamefont {Jaus}},\ and\ \bibinfo
  {author} {\bibfnamefont {G.}~\bibnamefont {Rasche}},\ }\bibfield  {title}
  {\bibinfo {title} {{Determination of V (ud) from Fermi decays and the
  unitarity of the KM mixing matrix}},\ }\href
  {https://doi.org/10.1016/0375-9474(92)90171-F} {\bibfield  {journal}
  {\bibinfo  {journal} {Nucl. Phys. A}\ }\textbf {\bibinfo {volume} {540}},\
  \bibinfo {pages} {501} (\bibinfo {year} {1992})}\BibitemShut {NoStop}%
\bibitem [{\citenamefont {Towner}(1992)}]{Towner:1992xm}%
  \BibitemOpen
  \bibfield  {author} {\bibinfo {author} {\bibfnamefont {I.~S.}\ \bibnamefont
  {Towner}},\ }\bibfield  {title} {\bibinfo {title} {{The Nuclear structure
  dependence of radiative corrections in superallowed Fermi beta decay}},\
  }\href {https://doi.org/10.1016/0375-9474(92)90170-O} {\bibfield  {journal}
  {\bibinfo  {journal} {Nucl. Phys. A}\ }\textbf {\bibinfo {volume} {540}},\
  \bibinfo {pages} {478} (\bibinfo {year} {1992})}\BibitemShut {NoStop}%
\bibitem [{\citenamefont {Towner}(1994)}]{Towner:1994mw}%
  \BibitemOpen
  \bibfield  {author} {\bibinfo {author} {\bibfnamefont {I.~S.}\ \bibnamefont
  {Towner}},\ }\bibfield  {title} {\bibinfo {title} {{Quenching of spin
  operators in the calculation of radiative corrections for nuclear beta
  decay}},\ }\href {https://doi.org/10.1016/0370-2693(94)91000-6} {\bibfield
  {journal} {\bibinfo  {journal} {Phys. Lett. B}\ }\textbf {\bibinfo {volume}
  {333}},\ \bibinfo {pages} {13} (\bibinfo {year} {1994})},\ \Eprint
  {https://arxiv.org/abs/nucl-th/9405031} {arXiv:nucl-th/9405031} \BibitemShut
  {NoStop}%
\bibitem [{\citenamefont {Towner}\ and\ \citenamefont
  {Hardy}(2008)}]{Towner:2007np}%
  \BibitemOpen
  \bibfield  {author} {\bibinfo {author} {\bibfnamefont {I.~S.}\ \bibnamefont
  {Towner}}\ and\ \bibinfo {author} {\bibfnamefont {J.~C.}\ \bibnamefont
  {Hardy}},\ }\bibfield  {title} {\bibinfo {title} {{An Improved calculation of
  the isospin-symmetry-breaking corrections to superallowed Fermi beta
  decay}},\ }\href {https://doi.org/10.1103/PhysRevC.77.025501} {\bibfield
  {journal} {\bibinfo  {journal} {Phys. Rev. C}\ }\textbf {\bibinfo {volume}
  {77}},\ \bibinfo {pages} {025501} (\bibinfo {year} {2008})},\ \Eprint
  {https://arxiv.org/abs/0710.3181} {arXiv:0710.3181 [nucl-th]} \BibitemShut
  {NoStop}%
\bibitem [{\citenamefont {MacDonald}(1958)}]{MacDonald:1958zz}%
  \BibitemOpen
  \bibfield  {author} {\bibinfo {author} {\bibfnamefont {W.~M.}\ \bibnamefont
  {MacDonald}},\ }\bibfield  {title} {\bibinfo {title} {{Coulomb Corrections to
  the Fermi Nuclear Matrix Element}},\ }\href
  {https://doi.org/10.1103/PhysRev.110.1420} {\bibfield  {journal} {\bibinfo
  {journal} {Phys. Rev.}\ }\textbf {\bibinfo {volume} {110}},\ \bibinfo {pages}
  {1420} (\bibinfo {year} {1958})}\BibitemShut {NoStop}%
\bibitem [{\citenamefont {Satu\l{}a}\ \emph {et~al.}(2016)\citenamefont
  {Satu\l{}a}, \citenamefont {B\k{a}czyk}, \citenamefont {Dobaczewski},\ and\
  \citenamefont {Konieczka}}]{Satula:2016hbs}%
  \BibitemOpen
  \bibfield  {author} {\bibinfo {author} {\bibfnamefont {W.}~\bibnamefont
  {Satu\l{}a}}, \bibinfo {author} {\bibfnamefont {P.}~\bibnamefont
  {B\k{a}czyk}}, \bibinfo {author} {\bibfnamefont {J.}~\bibnamefont
  {Dobaczewski}},\ and\ \bibinfo {author} {\bibfnamefont {M.}~\bibnamefont
  {Konieczka}},\ }\bibfield  {title} {\bibinfo {title} {{No-core
  configuration-interaction model for the isospin- and
  angular-momentum-projected states}},\ }\href
  {https://doi.org/10.1103/PhysRevC.94.024306} {\bibfield  {journal} {\bibinfo
  {journal} {Phys. Rev. C}\ }\textbf {\bibinfo {volume} {94}},\ \bibinfo
  {pages} {024306} (\bibinfo {year} {2016})},\ \Eprint
  {https://arxiv.org/abs/1601.03593} {arXiv:1601.03593 [nucl-th]} \BibitemShut
  {NoStop}%
\bibitem [{\citenamefont {Ormand}\ and\ \citenamefont
  {Brown}(1995)}]{Ormand:1995df}%
  \BibitemOpen
  \bibfield  {author} {\bibinfo {author} {\bibfnamefont {W.~E.}\ \bibnamefont
  {Ormand}}\ and\ \bibinfo {author} {\bibfnamefont {B.~A.}\ \bibnamefont
  {Brown}},\ }\bibfield  {title} {\bibinfo {title} {{Isospin-mixing corrections
  for fp-shell Fermi transitions}},\ }\href
  {https://doi.org/10.1103/PhysRevC.52.2455} {\bibfield  {journal} {\bibinfo
  {journal} {Phys. Rev. C}\ }\textbf {\bibinfo {volume} {52}},\ \bibinfo
  {pages} {2455} (\bibinfo {year} {1995})},\ \Eprint
  {https://arxiv.org/abs/nucl-th/9504017} {arXiv:nucl-th/9504017} \BibitemShut
  {NoStop}%
\bibitem [{\citenamefont {Liang}\ \emph {et~al.}(2009)\citenamefont {Liang},
  \citenamefont {Van~Giai},\ and\ \citenamefont {Meng}}]{Liang:2009pf}%
  \BibitemOpen
  \bibfield  {author} {\bibinfo {author} {\bibfnamefont {H.}~\bibnamefont
  {Liang}}, \bibinfo {author} {\bibfnamefont {N.}~\bibnamefont {Van~Giai}},\
  and\ \bibinfo {author} {\bibfnamefont {J.}~\bibnamefont {Meng}},\ }\bibfield
  {title} {\bibinfo {title} {{Isospin corrections for superallowed Fermi beta
  decay in self-consistent relativistic random-phase approximation
  approaches}},\ }\href {https://doi.org/10.1103/PhysRevC.79.064316} {\bibfield
   {journal} {\bibinfo  {journal} {Phys. Rev. C}\ }\textbf {\bibinfo {volume}
  {79}},\ \bibinfo {pages} {064316} (\bibinfo {year} {2009})},\ \Eprint
  {https://arxiv.org/abs/0904.3673} {arXiv:0904.3673 [nucl-th]} \BibitemShut
  {NoStop}%
\bibitem [{\citenamefont {Auerbach}(2009)}]{Auerbach:2008ut}%
  \BibitemOpen
  \bibfield  {author} {\bibinfo {author} {\bibfnamefont {N.}~\bibnamefont
  {Auerbach}},\ }\bibfield  {title} {\bibinfo {title} {{Coulomb corrections to
  superallowed beta decay in nuclei}},\ }\href
  {https://doi.org/10.1103/PhysRevC.79.035502} {\bibfield  {journal} {\bibinfo
  {journal} {Phys. Rev. C}\ }\textbf {\bibinfo {volume} {79}},\ \bibinfo
  {pages} {035502} (\bibinfo {year} {2009})},\ \Eprint
  {https://arxiv.org/abs/0811.4742} {arXiv:0811.4742 [nucl-th]} \BibitemShut
  {NoStop}%
\bibitem [{\citenamefont {Towner}\ and\ \citenamefont
  {Hardy}(2010)}]{Towner:2010bx}%
  \BibitemOpen
  \bibfield  {author} {\bibinfo {author} {\bibfnamefont {I.~S.}\ \bibnamefont
  {Towner}}\ and\ \bibinfo {author} {\bibfnamefont {J.~C.}\ \bibnamefont
  {Hardy}},\ }\bibfield  {title} {\bibinfo {title} {{Comparative tests of
  isospin-symmetry-breaking corrections to superallowed $0^+ -> 0^+$ nuclear
  $\beta$ decay}},\ }\href {https://doi.org/10.1103/PhysRevC.82.065501}
  {\bibfield  {journal} {\bibinfo  {journal} {Phys. Rev. C}\ }\textbf {\bibinfo
  {volume} {82}},\ \bibinfo {pages} {065501} (\bibinfo {year} {2010})},\
  \Eprint {https://arxiv.org/abs/1007.5343} {arXiv:1007.5343 [nucl-th]}
  \BibitemShut {NoStop}%
\bibitem [{\citenamefont {Melconian}\ \emph {et~al.}(2011)\citenamefont
  {Melconian} \emph {et~al.}}]{Melconian:2011kk}%
  \BibitemOpen
  \bibfield  {author} {\bibinfo {author} {\bibfnamefont {D.}~\bibnamefont
  {Melconian}} \emph {et~al.},\ }\bibfield  {title} {\bibinfo {title}
  {{Experimental Validation of the Largest Calculated Isospin-Symmetry-Breaking
  Effect in a Superallowed Fermi Decay}},\ }\href
  {https://doi.org/10.1103/PhysRevLett.107.182301} {\bibfield  {journal}
  {\bibinfo  {journal} {Phys. Rev. Lett.}\ }\textbf {\bibinfo {volume} {107}},\
  \bibinfo {pages} {182301} (\bibinfo {year} {2011})},\ \Eprint
  {https://arxiv.org/abs/1108.5312} {arXiv:1108.5312 [nucl-ex]} \BibitemShut
  {NoStop}%
\bibitem [{\citenamefont {Park}\ \emph {et~al.}(2014)\citenamefont {Park} \emph
  {et~al.}}]{Park:2014lja}%
  \BibitemOpen
  \bibfield  {author} {\bibinfo {author} {\bibfnamefont {H.~I.}\ \bibnamefont
  {Park}} \emph {et~al.},\ }\bibfield  {title} {\bibinfo {title}
  {{\ensuremath{\beta} Decay of $^{38}Ca$ : Sensitive test of Isospin
  Symmetry-Breaking Corrections from Mirror Superallowed $0^+ → 0^+$
  Transitions}},\ }\href {https://doi.org/10.1103/PhysRevLett.112.102502}
  {\bibfield  {journal} {\bibinfo  {journal} {Phys. Rev. Lett.}\ }\textbf
  {\bibinfo {volume} {112}},\ \bibinfo {pages} {102502} (\bibinfo {year}
  {2014})},\ \Eprint {https://arxiv.org/abs/1401.2114} {arXiv:1401.2114
  [nucl-ex]} \BibitemShut {NoStop}%
\bibitem [{\citenamefont {Malbrunot-Ettenauer}\ \emph
  {et~al.}(2015)\citenamefont {Malbrunot-Ettenauer} \emph
  {et~al.}}]{Malbrunot-Ettenauer:2015kda}%
  \BibitemOpen
  \bibfield  {author} {\bibinfo {author} {\bibfnamefont {S.}~\bibnamefont
  {Malbrunot-Ettenauer}} \emph {et~al.},\ }\bibfield  {title} {\bibinfo {title}
  {{Penning trap mass measurements utilizing highly charged ions as a path to
  benchmark isospin-symmetry breaking corrections in Rb74}},\ }\href
  {https://doi.org/10.1103/PhysRevC.91.045504} {\bibfield  {journal} {\bibinfo
  {journal} {Phys. Rev. C}\ }\textbf {\bibinfo {volume} {91}},\ \bibinfo
  {pages} {045504} (\bibinfo {year} {2015})}\BibitemShut {NoStop}%
\bibitem [{\citenamefont {Xayavong}\ and\ \citenamefont
  {Smirnova}(2018)}]{Xayavong:2017kim}%
  \BibitemOpen
  \bibfield  {author} {\bibinfo {author} {\bibfnamefont {L.}~\bibnamefont
  {Xayavong}}\ and\ \bibinfo {author} {\bibfnamefont {N.~A.}\ \bibnamefont
  {Smirnova}},\ }\bibfield  {title} {\bibinfo {title} {{Radial overlap
  correction to superallowed $0^+ \to 0^+$ $\beta $ decay reexamined}},\ }\href
  {https://doi.org/10.1103/PhysRevC.97.024324} {\bibfield  {journal} {\bibinfo
  {journal} {Phys. Rev. C}\ }\textbf {\bibinfo {volume} {97}},\ \bibinfo
  {pages} {024324} (\bibinfo {year} {2018})},\ \Eprint
  {https://arxiv.org/abs/1708.00616} {arXiv:1708.00616 [nucl-th]} \BibitemShut
  {NoStop}%
\bibitem [{\citenamefont {Bencomo}\ \emph {et~al.}(2019)\citenamefont
  {Bencomo}, \citenamefont {Hardy}, \citenamefont {Iacob}, \citenamefont
  {Park}, \citenamefont {Chen}, \citenamefont {Horvat}, \citenamefont {Nica},
  \citenamefont {Roeder}, \citenamefont {Saastamoinen},\ and\ \citenamefont
  {Towner}}]{Bencomo:2019qzx}%
  \BibitemOpen
  \bibfield  {author} {\bibinfo {author} {\bibfnamefont {M.}~\bibnamefont
  {Bencomo}}, \bibinfo {author} {\bibfnamefont {J.~C.}\ \bibnamefont {Hardy}},
  \bibinfo {author} {\bibfnamefont {V.~E.}\ \bibnamefont {Iacob}}, \bibinfo
  {author} {\bibfnamefont {H.~I.}\ \bibnamefont {Park}}, \bibinfo {author}
  {\bibfnamefont {L.}~\bibnamefont {Chen}}, \bibinfo {author} {\bibfnamefont
  {V.}~\bibnamefont {Horvat}}, \bibinfo {author} {\bibfnamefont
  {N.}~\bibnamefont {Nica}}, \bibinfo {author} {\bibfnamefont {B.~T.}\
  \bibnamefont {Roeder}}, \bibinfo {author} {\bibfnamefont {A.}~\bibnamefont
  {Saastamoinen}},\ and\ \bibinfo {author} {\bibfnamefont {I.~S.}\ \bibnamefont
  {Towner}},\ }\bibfield  {title} {\bibinfo {title} {{Precise branching ratio
  measurement for the superallowed \ensuremath{\beta}+ decay of Si26 :
  Completion of a second mirror pair}},\ }\href
  {https://doi.org/10.1103/PhysRevC.100.015503} {\bibfield  {journal} {\bibinfo
   {journal} {Phys. Rev. C}\ }\textbf {\bibinfo {volume} {100}},\ \bibinfo
  {pages} {015503} (\bibinfo {year} {2019})}\BibitemShut {NoStop}%
\bibitem [{\citenamefont {Iacob}\ \emph {et~al.}(2020)\citenamefont {Iacob},
  \citenamefont {Hardy}, \citenamefont {Park}, \citenamefont {Bencomo},
  \citenamefont {Chen}, \citenamefont {Horvat}, \citenamefont {Nica},
  \citenamefont {Roeder}, \citenamefont {Saastamoinen},\ and\ \citenamefont
  {Towner}}]{Iacob:2020mhu}%
  \BibitemOpen
  \bibfield  {author} {\bibinfo {author} {\bibfnamefont {V.~E.}\ \bibnamefont
  {Iacob}}, \bibinfo {author} {\bibfnamefont {J.~C.}\ \bibnamefont {Hardy}},
  \bibinfo {author} {\bibfnamefont {H.~I.}\ \bibnamefont {Park}}, \bibinfo
  {author} {\bibfnamefont {M.}~\bibnamefont {Bencomo}}, \bibinfo {author}
  {\bibfnamefont {L.}~\bibnamefont {Chen}}, \bibinfo {author} {\bibfnamefont
  {V.}~\bibnamefont {Horvat}}, \bibinfo {author} {\bibfnamefont
  {N.}~\bibnamefont {Nica}}, \bibinfo {author} {\bibfnamefont {B.~T.}\
  \bibnamefont {Roeder}}, \bibinfo {author} {\bibfnamefont {A.}~\bibnamefont
  {Saastamoinen}},\ and\ \bibinfo {author} {\bibfnamefont {I.~S.}\ \bibnamefont
  {Towner}},\ }\bibfield  {title} {\bibinfo {title} {{Precise
  \ensuremath{\beta} branching-ratio measurement for the 0+\textrightarrow{}0+
  superallowed decay of Ar34}},\ }\href
  {https://doi.org/10.1103/PhysRevC.101.045501} {\bibfield  {journal} {\bibinfo
   {journal} {Phys. Rev. C}\ }\textbf {\bibinfo {volume} {101}},\ \bibinfo
  {pages} {045501} (\bibinfo {year} {2020})}\BibitemShut {NoStop}%
\bibitem [{\citenamefont {Martin}\ \emph {et~al.}(2021)\citenamefont {Martin},
  \citenamefont {Stroberg}, \citenamefont {Holt},\ and\ \citenamefont
  {Leach}}]{Martin:2021bud}%
  \BibitemOpen
  \bibfield  {author} {\bibinfo {author} {\bibfnamefont {M.~S.}\ \bibnamefont
  {Martin}}, \bibinfo {author} {\bibfnamefont {S.~R.}\ \bibnamefont
  {Stroberg}}, \bibinfo {author} {\bibfnamefont {J.~D.}\ \bibnamefont {Holt}},\
  and\ \bibinfo {author} {\bibfnamefont {K.~G.}\ \bibnamefont {Leach}},\
  }\bibfield  {title} {\bibinfo {title} {{Testing isospin symmetry breaking in
  ab initio nuclear theory}},\ }\href
  {https://doi.org/10.1103/PhysRevC.104.014324} {\bibfield  {journal} {\bibinfo
   {journal} {Phys. Rev. C}\ }\textbf {\bibinfo {volume} {104}},\ \bibinfo
  {pages} {014324} (\bibinfo {year} {2021})},\ \Eprint
  {https://arxiv.org/abs/2101.11826} {arXiv:2101.11826 [nucl-th]} \BibitemShut
  {NoStop}%
\bibitem [{\citenamefont {Miller}\ and\ \citenamefont
  {Schwenk}(2008)}]{Miller:2008my}%
  \BibitemOpen
  \bibfield  {author} {\bibinfo {author} {\bibfnamefont {G.~A.}\ \bibnamefont
  {Miller}}\ and\ \bibinfo {author} {\bibfnamefont {A.}~\bibnamefont
  {Schwenk}},\ }\bibfield  {title} {\bibinfo {title}
  {{Isospin-symmetry-breaking corrections to superallowed Fermi beta decay:
  Formalism and schematic models}},\ }\href
  {https://doi.org/10.1103/PhysRevC.78.035501} {\bibfield  {journal} {\bibinfo
  {journal} {Phys. Rev. C}\ }\textbf {\bibinfo {volume} {78}},\ \bibinfo
  {pages} {035501} (\bibinfo {year} {2008})},\ \Eprint
  {https://arxiv.org/abs/0805.0603} {arXiv:0805.0603 [nucl-th]} \BibitemShut
  {NoStop}%
\bibitem [{\citenamefont {Miller}\ and\ \citenamefont
  {Schwenk}(2009)}]{Miller:2009cg}%
  \BibitemOpen
  \bibfield  {author} {\bibinfo {author} {\bibfnamefont {G.~A.}\ \bibnamefont
  {Miller}}\ and\ \bibinfo {author} {\bibfnamefont {A.}~\bibnamefont
  {Schwenk}},\ }\bibfield  {title} {\bibinfo {title}
  {{Isospin-symmetry-breaking corrections to superallowed Fermi beta decay:
  Radial excitations}},\ }\href {https://doi.org/10.1103/PhysRevC.80.064319}
  {\bibfield  {journal} {\bibinfo  {journal} {Phys. Rev. C}\ }\textbf {\bibinfo
  {volume} {80}},\ \bibinfo {pages} {064319} (\bibinfo {year} {2009})},\
  \Eprint {https://arxiv.org/abs/0910.2790} {arXiv:0910.2790 [nucl-th]}
  \BibitemShut {NoStop}%
\bibitem [{\citenamefont {Gorchtein}(2019)}]{Gorchtein:2018fxl}%
  \BibitemOpen
  \bibfield  {author} {\bibinfo {author} {\bibfnamefont {M.}~\bibnamefont
  {Gorchtein}},\ }\bibfield  {title} {\bibinfo {title} {{$\gamma W$ Box Inside
  Out: Nuclear Polarizabilities Distort the Beta Decay Spectrum}},\ }\href
  {https://doi.org/10.1103/PhysRevLett.123.042503} {\bibfield  {journal}
  {\bibinfo  {journal} {Phys. Rev. Lett.}\ }\textbf {\bibinfo {volume} {123}},\
  \bibinfo {pages} {042503} (\bibinfo {year} {2019})},\ \Eprint
  {https://arxiv.org/abs/1812.04229} {arXiv:1812.04229 [nucl-th]} \BibitemShut
  {NoStop}%
\bibitem [{\citenamefont {Condren}\ and\ \citenamefont
  {Miller}(2022)}]{Condren:2022dji}%
  \BibitemOpen
  \bibfield  {author} {\bibinfo {author} {\bibfnamefont {L.}~\bibnamefont
  {Condren}}\ and\ \bibinfo {author} {\bibfnamefont {G.~A.}\ \bibnamefont
  {Miller}},\ }\bibfield  {title} {\bibinfo {title} {{Nucleon-nucleon
  short-ranged correlations, \ensuremath{\beta} decay, and the unitarity of the
  CKM matrix}},\ }\href {https://doi.org/10.1103/PhysRevC.106.L062501}
  {\bibfield  {journal} {\bibinfo  {journal} {Phys. Rev. C}\ }\textbf {\bibinfo
  {volume} {106}},\ \bibinfo {pages} {L062501} (\bibinfo {year} {2022})},\
  \Eprint {https://arxiv.org/abs/2201.10651} {arXiv:2201.10651 [nucl-th]}
  \BibitemShut {NoStop}%
\bibitem [{\citenamefont {Seng}\ and\ \citenamefont
  {Gorchtein}(2023{\natexlab{a}})}]{Seng:2022cnq}%
  \BibitemOpen
  \bibfield  {author} {\bibinfo {author} {\bibfnamefont {C.-Y.}\ \bibnamefont
  {Seng}}\ and\ \bibinfo {author} {\bibfnamefont {M.}~\bibnamefont
  {Gorchtein}},\ }\bibfield  {title} {\bibinfo {title} {{Dispersive formalism
  for the nuclear structure correction \ensuremath{\delta}NS to the
  \ensuremath{\beta} decay rate}},\ }\href
  {https://doi.org/10.1103/PhysRevC.107.035503} {\bibfield  {journal} {\bibinfo
   {journal} {Phys. Rev. C}\ }\textbf {\bibinfo {volume} {107}},\ \bibinfo
  {pages} {035503} (\bibinfo {year} {2023}{\natexlab{a}})},\ \Eprint
  {https://arxiv.org/abs/2211.10214} {arXiv:2211.10214 [nucl-th]} \BibitemShut
  {NoStop}%
\bibitem [{\citenamefont {Seng}\ and\ \citenamefont
  {Gorchtein}(2024)}]{Seng:2023cvt}%
  \BibitemOpen
  \bibfield  {author} {\bibinfo {author} {\bibfnamefont {C.-Y.}\ \bibnamefont
  {Seng}}\ and\ \bibinfo {author} {\bibfnamefont {M.}~\bibnamefont
  {Gorchtein}},\ }\bibfield  {title} {\bibinfo {title} {{Toward ab-initio
  nuclear theory calculations of \ensuremath{\delta}C}},\ }\href
  {https://doi.org/10.1103/PhysRevC.109.044302} {\bibfield  {journal} {\bibinfo
   {journal} {Phys. Rev. C}\ }\textbf {\bibinfo {volume} {109}},\ \bibinfo
  {pages} {044302} (\bibinfo {year} {2024})},\ \Eprint
  {https://arxiv.org/abs/2304.03800} {arXiv:2304.03800 [nucl-th]} \BibitemShut
  {NoStop}%
\bibitem [{\citenamefont {Seng}\ and\ \citenamefont
  {Gorchtein}(2023{\natexlab{b}})}]{Seng:2022epj}%
  \BibitemOpen
  \bibfield  {author} {\bibinfo {author} {\bibfnamefont {C.-Y.}\ \bibnamefont
  {Seng}}\ and\ \bibinfo {author} {\bibfnamefont {M.}~\bibnamefont
  {Gorchtein}},\ }\bibfield  {title} {\bibinfo {title} {{Electroweak nuclear
  radii constrain the isospin breaking correction to Vud}},\ }\href
  {https://doi.org/10.1016/j.physletb.2022.137654} {\bibfield  {journal}
  {\bibinfo  {journal} {Phys. Lett. B}\ }\textbf {\bibinfo {volume} {838}},\
  \bibinfo {pages} {137654} (\bibinfo {year} {2023}{\natexlab{b}})},\ \Eprint
  {https://arxiv.org/abs/2208.03037} {arXiv:2208.03037 [nucl-th]} \BibitemShut
  {NoStop}%
\bibitem [{\citenamefont {Hardy}\ and\ \citenamefont
  {Towner}(2009)}]{Hardy:2008gy}%
  \BibitemOpen
  \bibfield  {author} {\bibinfo {author} {\bibfnamefont {J.~C.}\ \bibnamefont
  {Hardy}}\ and\ \bibinfo {author} {\bibfnamefont {I.~S.}\ \bibnamefont
  {Towner}},\ }\bibfield  {title} {\bibinfo {title} {{Superallowed 0+
  ---\ensuremath{>} 0+ nuclear beta decays: A New survey with precision tests
  of the conserved vector current hypothesis and the standard model}},\ }\href
  {https://doi.org/10.1103/PhysRevC.79.055502} {\bibfield  {journal} {\bibinfo
  {journal} {Phys. Rev. C}\ }\textbf {\bibinfo {volume} {79}},\ \bibinfo
  {pages} {055502} (\bibinfo {year} {2009})},\ \Eprint
  {https://arxiv.org/abs/0812.1202} {arXiv:0812.1202 [nucl-ex]} \BibitemShut
  {NoStop}%
\bibitem [{\citenamefont {Hardy}\ and\ \citenamefont
  {Towner}(2015)}]{Hardy:2014qxa}%
  \BibitemOpen
  \bibfield  {author} {\bibinfo {author} {\bibfnamefont {J.~C.}\ \bibnamefont
  {Hardy}}\ and\ \bibinfo {author} {\bibfnamefont {I.~S.}\ \bibnamefont
  {Towner}},\ }\bibfield  {title} {\bibinfo {title} {{Superallowed $0^+\to 0^+$
  nuclear \ensuremath{\beta} decays: 2014 critical survey, with precise results
  for $V_{ud}$ and CKM unitarity}},\ }\href
  {https://doi.org/10.1103/PhysRevC.91.025501} {\bibfield  {journal} {\bibinfo
  {journal} {Phys. Rev. C}\ }\textbf {\bibinfo {volume} {91}},\ \bibinfo
  {pages} {025501} (\bibinfo {year} {2015})},\ \Eprint
  {https://arxiv.org/abs/1411.5987} {arXiv:1411.5987 [nucl-ex]} \BibitemShut
  {NoStop}%
\bibitem [{\citenamefont {Seng}(2023)}]{Seng:2022inj}%
  \BibitemOpen
  \bibfield  {author} {\bibinfo {author} {\bibfnamefont {C.-Y.}\ \bibnamefont
  {Seng}},\ }\bibfield  {title} {\bibinfo {title} {{Model-Independent
  Determination of Nuclear Weak Form Factors and Implications for Standard
  Model Precision Tests}},\ }\href
  {https://doi.org/10.1103/PhysRevLett.130.152501} {\bibfield  {journal}
  {\bibinfo  {journal} {Phys. Rev. Lett.}\ }\textbf {\bibinfo {volume} {130}},\
  \bibinfo {pages} {152501} (\bibinfo {year} {2023})},\ \Eprint
  {https://arxiv.org/abs/2212.02681} {arXiv:2212.02681 [nucl-th]} \BibitemShut
  {NoStop}%
\bibitem [{\citenamefont {Hayen}\ \emph {et~al.}(2018)\citenamefont {Hayen},
  \citenamefont {Severijns}, \citenamefont {Bodek}, \citenamefont {Rozpedzik},\
  and\ \citenamefont {Mougeot}}]{Hayen:2017pwg}%
  \BibitemOpen
  \bibfield  {author} {\bibinfo {author} {\bibfnamefont {L.}~\bibnamefont
  {Hayen}}, \bibinfo {author} {\bibfnamefont {N.}~\bibnamefont {Severijns}},
  \bibinfo {author} {\bibfnamefont {K.}~\bibnamefont {Bodek}}, \bibinfo
  {author} {\bibfnamefont {D.}~\bibnamefont {Rozpedzik}},\ and\ \bibinfo
  {author} {\bibfnamefont {X.}~\bibnamefont {Mougeot}},\ }\bibfield  {title}
  {\bibinfo {title} {{High precision analytical description of the allowed
  $\beta$ spectrum shape}},\ }\href
  {https://doi.org/10.1103/RevModPhys.90.015008} {\bibfield  {journal}
  {\bibinfo  {journal} {Rev. Mod. Phys.}\ }\textbf {\bibinfo {volume} {90}},\
  \bibinfo {pages} {015008} (\bibinfo {year} {2018})},\ \Eprint
  {https://arxiv.org/abs/1709.07530} {arXiv:1709.07530 [nucl-th]} \BibitemShut
  {NoStop}%
\bibitem [{\citenamefont {Fermi}(1934)}]{Fermi:1934hr}%
  \BibitemOpen
  \bibfield  {author} {\bibinfo {author} {\bibfnamefont {E.}~\bibnamefont
  {Fermi}},\ }\bibfield  {title} {\bibinfo {title} {{An attempt of a theory of
  beta radiation. 1.}},\ }\href {https://doi.org/10.1007/BF01351864} {\bibfield
   {journal} {\bibinfo  {journal} {Z. Phys.}\ }\textbf {\bibinfo {volume}
  {88}},\ \bibinfo {pages} {161} (\bibinfo {year} {1934})}\BibitemShut
  {NoStop}%
\bibitem [{\citenamefont {Konopinski}\ and\ \citenamefont
  {Uhlenbeck}(1941)}]{konopinski1941fermi}%
  \BibitemOpen
  \bibfield  {author} {\bibinfo {author} {\bibfnamefont {E.}~\bibnamefont
  {Konopinski}}\ and\ \bibinfo {author} {\bibfnamefont {G.}~\bibnamefont
  {Uhlenbeck}},\ }\bibfield  {title} {\bibinfo {title} {On the fermi theory of
  $\beta$-radioactivity. ii. the" forbidden" spectra},\ }\href@noop {}
  {\bibfield  {journal} {\bibinfo  {journal} {Physical Review}\ }\textbf
  {\bibinfo {volume} {60}},\ \bibinfo {pages} {308} (\bibinfo {year}
  {1941})}\BibitemShut {NoStop}%
\bibitem [{\citenamefont {Schopper}\ \emph {et~al.}(1969)\citenamefont
  {Schopper}, \citenamefont {Behrens},\ and\ \citenamefont
  {J{\"a}necke}}]{behrens1969landolt}%
  \BibitemOpen
  \bibfield  {author} {\bibinfo {author} {\bibfnamefont {H.}~\bibnamefont
  {Schopper}}, \bibinfo {author} {\bibfnamefont {H.}~\bibnamefont {Behrens}},\
  and\ \bibinfo {author} {\bibfnamefont {J.}~\bibnamefont {J{\"a}necke}},\
  }\href@noop {} {\bibinfo {title} {Numerical tables for beta-decay and
  electron capture/numerische tabellen f{\"u}r beta-zerfall und
  elektronen-einfang}} (\bibinfo {year} {1969})\BibitemShut {NoStop}%
\bibitem [{\citenamefont {Calaprice}\ and\ \citenamefont
  {Holstein}(1976)}]{Calaprice:1976jbi}%
  \BibitemOpen
  \bibfield  {author} {\bibinfo {author} {\bibfnamefont {F.~P.}\ \bibnamefont
  {Calaprice}}\ and\ \bibinfo {author} {\bibfnamefont {B.~R.}\ \bibnamefont
  {Holstein}},\ }\bibfield  {title} {\bibinfo {title} {{Weak magnetism and the
  beta spectra of 12 B and 12 N}},\ }\href
  {https://doi.org/10.1016/0375-9474(76)90593-5} {\bibfield  {journal}
  {\bibinfo  {journal} {Nucl. Phys. A}\ }\textbf {\bibinfo {volume} {273}},\
  \bibinfo {pages} {301} (\bibinfo {year} {1976})}\BibitemShut {NoStop}%
\bibitem [{\citenamefont {Behrens}\ and\ \citenamefont
  {B{\"u}hring}(1982)}]{behrens1982electron}%
  \BibitemOpen
  \bibfield  {author} {\bibinfo {author} {\bibfnamefont {H.}~\bibnamefont
  {Behrens}}\ and\ \bibinfo {author} {\bibfnamefont {W.}~\bibnamefont
  {B{\"u}hring}},\ }\href@noop {} {\bibinfo {title} {Electron radial wave
  functions and nuclear betadecay}} (\bibinfo {year} {1982})\BibitemShut
  {NoStop}%
\bibitem [{\citenamefont {Wilkinson}(1993)}]{Wilkinson:1993hx}%
  \BibitemOpen
  \bibfield  {author} {\bibinfo {author} {\bibfnamefont {D.~H.}\ \bibnamefont
  {Wilkinson}},\ }\bibfield  {title} {\bibinfo {title} {{Methodology for
  superallowed Fermi beta decay. 2. Reduction of data}},\ }\href
  {https://doi.org/10.1016/0168-9002(93)90271-I} {\bibfield  {journal}
  {\bibinfo  {journal} {Nucl. Instrum. Meth. A}\ }\textbf {\bibinfo {volume}
  {335}},\ \bibinfo {pages} {182} (\bibinfo {year} {1993})}\BibitemShut
  {NoStop}%
\bibitem [{\citenamefont {Gershtein}\ and\ \citenamefont
  {Zeldovich}(1955)}]{Gershtein:1955fb}%
  \BibitemOpen
  \bibfield  {author} {\bibinfo {author} {\bibfnamefont {S.~S.}\ \bibnamefont
  {Gershtein}}\ and\ \bibinfo {author} {\bibfnamefont {Y.~B.}\ \bibnamefont
  {Zeldovich}},\ }\bibfield  {title} {\bibinfo {title} {{Meson corrections in
  the theory of beta decay}},\ }\href@noop {} {\bibfield  {journal} {\bibinfo
  {journal} {Zh. Eksp. Teor. Fiz.}\ }\textbf {\bibinfo {volume} {29}},\
  \bibinfo {pages} {698} (\bibinfo {year} {1955})}\BibitemShut {NoStop}%
\bibitem [{\citenamefont {Feynman}\ and\ \citenamefont
  {Gell-Mann}(1958)}]{Feynman:1958ty}%
  \BibitemOpen
  \bibfield  {author} {\bibinfo {author} {\bibfnamefont {R.~P.}\ \bibnamefont
  {Feynman}}\ and\ \bibinfo {author} {\bibfnamefont {M.}~\bibnamefont
  {Gell-Mann}},\ }\bibfield  {title} {\bibinfo {title} {{Theory of Fermi
  interaction}},\ }\href {https://doi.org/10.1103/PhysRev.109.193} {\bibfield
  {journal} {\bibinfo  {journal} {Phys. Rev.}\ }\textbf {\bibinfo {volume}
  {109}},\ \bibinfo {pages} {193} (\bibinfo {year} {1958})}\BibitemShut
  {NoStop}%
\bibitem [{\citenamefont {Holstein}(1974)}]{Holstein:1974zf}%
  \BibitemOpen
  \bibfield  {author} {\bibinfo {author} {\bibfnamefont {B.~R.}\ \bibnamefont
  {Holstein}},\ }\bibfield  {title} {\bibinfo {title} {{Recoil Effects in
  Allowed beta Decay: The Elementary Particle Approach}},\ }\href
  {https://doi.org/10.1103/RevModPhys.46.789} {\bibfield  {journal} {\bibinfo
  {journal} {Rev. Mod. Phys.}\ }\textbf {\bibinfo {volume} {46}},\ \bibinfo
  {pages} {789} (\bibinfo {year} {1974})},\ \bibinfo {note} {[Erratum:
  Rev.Mod.Phys. 48, 673--673 (1976)]}\BibitemShut {NoStop}%
\bibitem [{\citenamefont {Angeli}\ and\ \citenamefont
  {Marinova}(2013)}]{Angeli:2013epw}%
  \BibitemOpen
  \bibfield  {author} {\bibinfo {author} {\bibfnamefont {I.}~\bibnamefont
  {Angeli}}\ and\ \bibinfo {author} {\bibfnamefont {K.~P.}\ \bibnamefont
  {Marinova}},\ }\bibfield  {title} {\bibinfo {title} {{Table of experimental
  nuclear ground state charge radii: An update}},\ }\href
  {https://doi.org/10.1016/j.adt.2011.12.006} {\bibfield  {journal} {\bibinfo
  {journal} {Atom. Data Nucl. Data Tabl.}\ }\textbf {\bibinfo {volume} {99}},\
  \bibinfo {pages} {69} (\bibinfo {year} {2013})}\BibitemShut {NoStop}%
\bibitem [{\citenamefont {Li}\ \emph {et~al.}(2021)\citenamefont {Li},
  \citenamefont {Luo},\ and\ \citenamefont {Wang}}]{Li:2021fmk}%
  \BibitemOpen
  \bibfield  {author} {\bibinfo {author} {\bibfnamefont {T.}~\bibnamefont
  {Li}}, \bibinfo {author} {\bibfnamefont {Y.}~\bibnamefont {Luo}},\ and\
  \bibinfo {author} {\bibfnamefont {N.}~\bibnamefont {Wang}},\ }\bibfield
  {title} {\bibinfo {title} {{Compilation of recent nuclear ground state charge
  radius measurements and tests for models}},\ }\href
  {https://doi.org/10.1016/j.adt.2021.101440} {\bibfield  {journal} {\bibinfo
  {journal} {Atom. Data Nucl. Data Tabl.}\ }\textbf {\bibinfo {volume} {140}},\
  \bibinfo {pages} {101440} (\bibinfo {year} {2021})}\BibitemShut {NoStop}%
\bibitem [{\citenamefont {Miller}\ \emph {et~al.}(2019)\citenamefont {Miller},
  \citenamefont {Minamisono}, \citenamefont {Klose}, \citenamefont {Garand},
  \citenamefont {Kujawa}, \citenamefont {Lantis}, \citenamefont {Liu},
  \citenamefont {Maa{\ss}}, \citenamefont {Mantica}, \citenamefont {Nazarewicz}
  \emph {et~al.}}]{miller2019proton}%
  \BibitemOpen
  \bibfield  {author} {\bibinfo {author} {\bibfnamefont {A.~J.}\ \bibnamefont
  {Miller}}, \bibinfo {author} {\bibfnamefont {K.}~\bibnamefont {Minamisono}},
  \bibinfo {author} {\bibfnamefont {A.}~\bibnamefont {Klose}}, \bibinfo
  {author} {\bibfnamefont {D.}~\bibnamefont {Garand}}, \bibinfo {author}
  {\bibfnamefont {C.}~\bibnamefont {Kujawa}}, \bibinfo {author} {\bibfnamefont
  {J.}~\bibnamefont {Lantis}}, \bibinfo {author} {\bibfnamefont
  {Y.}~\bibnamefont {Liu}}, \bibinfo {author} {\bibfnamefont {B.}~\bibnamefont
  {Maa{\ss}}}, \bibinfo {author} {\bibfnamefont {P.}~\bibnamefont {Mantica}},
  \bibinfo {author} {\bibfnamefont {W.}~\bibnamefont {Nazarewicz}}, \emph
  {et~al.},\ }\bibfield  {title} {\bibinfo {title} {Proton superfluidity and
  charge radii in proton-rich calcium isotopes},\ }\href@noop {} {\bibfield
  {journal} {\bibinfo  {journal} {Nature physics}\ }\textbf {\bibinfo {volume}
  {15}},\ \bibinfo {pages} {432} (\bibinfo {year} {2019})}\BibitemShut
  {NoStop}%
\bibitem [{\citenamefont {Bissell}\ \emph {et~al.}(2014)\citenamefont {Bissell}
  \emph {et~al.}}]{Bissell:2014vva}%
  \BibitemOpen
  \bibfield  {author} {\bibinfo {author} {\bibfnamefont {M.~L.}\ \bibnamefont
  {Bissell}} \emph {et~al.},\ }\bibfield  {title} {\bibinfo {title}
  {{Proton-Neutron Pairing Correlations in the Self-Conjugate Nucleus K38
  Probed via a Direct Measurement of the Isomer Shift}},\ }\href
  {https://doi.org/10.1103/PhysRevLett.113.052502} {\bibfield  {journal}
  {\bibinfo  {journal} {Phys. Rev. Lett.}\ }\textbf {\bibinfo {volume} {113}},\
  \bibinfo {pages} {052502} (\bibinfo {year} {2014})}\BibitemShut {NoStop}%
\bibitem [{\citenamefont {Pineda}\ \emph {et~al.}(2021)\citenamefont {Pineda}
  \emph {et~al.}}]{Pineda:2021shy}%
  \BibitemOpen
  \bibfield  {author} {\bibinfo {author} {\bibfnamefont {S.~V.}\ \bibnamefont
  {Pineda}} \emph {et~al.},\ }\bibfield  {title} {\bibinfo {title} {{Charge
  Radius of Neutron-Deficient Ni54 and Symmetry Energy Constraints Using the
  Difference in Mirror Pair Charge Radii}},\ }\href
  {https://doi.org/10.1103/PhysRevLett.127.182503} {\bibfield  {journal}
  {\bibinfo  {journal} {Phys. Rev. Lett.}\ }\textbf {\bibinfo {volume} {127}},\
  \bibinfo {pages} {182503} (\bibinfo {year} {2021})},\ \Eprint
  {https://arxiv.org/abs/2106.10378} {arXiv:2106.10378 [nucl-ex]} \BibitemShut
  {NoStop}%
\bibitem [{\citenamefont {Plattner}\ \emph {et~al.}(2023)\citenamefont
  {Plattner} \emph {et~al.}}]{Plattner:2023fmu}%
  \BibitemOpen
  \bibfield  {author} {\bibinfo {author} {\bibfnamefont {P.}~\bibnamefont
  {Plattner}} \emph {et~al.},\ }\bibfield  {title} {\bibinfo {title} {{Nuclear
  Charge Radius of Al26m and Its Implication for Vud in the Quark Mixing
  Matrix}},\ }\href {https://doi.org/10.1103/PhysRevLett.131.222502} {\bibfield
   {journal} {\bibinfo  {journal} {Phys. Rev. Lett.}\ }\textbf {\bibinfo
  {volume} {131}},\ \bibinfo {pages} {222502} (\bibinfo {year} {2023})},\
  \Eprint {https://arxiv.org/abs/2310.15291} {arXiv:2310.15291 [nucl-ex]}
  \BibitemShut {NoStop}%
\bibitem [{\citenamefont {Fricke}\ \emph {et~al.}(2004)\citenamefont {Fricke},
  \citenamefont {Heilig},\ and\ \citenamefont {Schopper}}]{fricke2004nuclear}%
  \BibitemOpen
  \bibfield  {author} {\bibinfo {author} {\bibfnamefont {G.}~\bibnamefont
  {Fricke}}, \bibinfo {author} {\bibfnamefont {K.}~\bibnamefont {Heilig}},\
  and\ \bibinfo {author} {\bibfnamefont {H.~F.}\ \bibnamefont {Schopper}},\
  }\href@noop {} {\emph {\bibinfo {title} {Nuclear charge radii}}},\ Vol.\
  \bibinfo {volume} {454}\ (\bibinfo  {publisher} {Springer Berlin},\ \bibinfo
  {year} {2004})\BibitemShut {NoStop}%
\bibitem [{\citenamefont {De~Vries}\ \emph {et~al.}(1987)\citenamefont
  {De~Vries}, \citenamefont {De~Jager},\ and\ \citenamefont
  {De~Vries}}]{de1987nuclear}%
  \BibitemOpen
  \bibfield  {author} {\bibinfo {author} {\bibfnamefont {H.}~\bibnamefont
  {De~Vries}}, \bibinfo {author} {\bibfnamefont {C.}~\bibnamefont {De~Jager}},\
  and\ \bibinfo {author} {\bibfnamefont {C.}~\bibnamefont {De~Vries}},\
  }\bibfield  {title} {\bibinfo {title} {Nuclear charge-density-distribution
  parameters from elastic electron scattering},\ }\href@noop {} {\bibfield
  {journal} {\bibinfo  {journal} {Atomic data and nuclear data tables}\
  }\textbf {\bibinfo {volume} {36}},\ \bibinfo {pages} {495} (\bibinfo {year}
  {1987})}\BibitemShut {NoStop}%
\bibitem [{\citenamefont {Moreira}\ \emph {et~al.}(1971)\citenamefont
  {Moreira}, \citenamefont {Singhal},\ and\ \citenamefont
  {Caplan}}]{moreira1971charge}%
  \BibitemOpen
  \bibfield  {author} {\bibinfo {author} {\bibfnamefont {J.}~\bibnamefont
  {Moreira}}, \bibinfo {author} {\bibfnamefont {R.}~\bibnamefont {Singhal}},\
  and\ \bibinfo {author} {\bibfnamefont {H.}~\bibnamefont {Caplan}},\
  }\bibfield  {title} {\bibinfo {title} {Charge radii of 20, 22ne determined
  from elastic electron scattering},\ }\href@noop {} {\bibfield  {journal}
  {\bibinfo  {journal} {Canadian Journal of Physics}\ }\textbf {\bibinfo
  {volume} {49}},\ \bibinfo {pages} {1434} (\bibinfo {year}
  {1971})}\BibitemShut {NoStop}%
\bibitem [{\citenamefont {Knight}\ \emph {et~al.}(1981)\citenamefont {Knight},
  \citenamefont {Singhal}, \citenamefont {Arthur},\ and\ \citenamefont
  {Macauley}}]{knight1981elastic}%
  \BibitemOpen
  \bibfield  {author} {\bibinfo {author} {\bibfnamefont {E.}~\bibnamefont
  {Knight}}, \bibinfo {author} {\bibfnamefont {R.}~\bibnamefont {Singhal}},
  \bibinfo {author} {\bibfnamefont {R.}~\bibnamefont {Arthur}},\ and\ \bibinfo
  {author} {\bibfnamefont {M.}~\bibnamefont {Macauley}},\ }\bibfield  {title}
  {\bibinfo {title} {Elastic scattering of electrons from 20, 22ne},\
  }\href@noop {} {\bibfield  {journal} {\bibinfo  {journal} {Journal of Physics
  G: Nuclear Physics}\ }\textbf {\bibinfo {volume} {7}},\ \bibinfo {pages}
  {1115} (\bibinfo {year} {1981})}\BibitemShut {NoStop}%
\bibitem [{Be8()}]{Be85}%
  \BibitemOpen
  \href@noop {} {}\bibinfo {note} {J.C. Bergstrom, R. Neuhausen and G. Lahm,
  1985 (unpublished)}\BibitemShut {NoStop}%
\bibitem [{\citenamefont {Singhal}\ \emph {et~al.}(1970)\citenamefont
  {Singhal}, \citenamefont {Moreira},\ and\ \citenamefont
  {Caplan}}]{singhal1970rms}%
  \BibitemOpen
  \bibfield  {author} {\bibinfo {author} {\bibfnamefont {R.}~\bibnamefont
  {Singhal}}, \bibinfo {author} {\bibfnamefont {J.}~\bibnamefont {Moreira}},\
  and\ \bibinfo {author} {\bibfnamefont {H.}~\bibnamefont {Caplan}},\
  }\bibfield  {title} {\bibinfo {title} {Rms charge radii of o 1 6, 1 7, 1 8 by
  elastic electron scattering},\ }\href@noop {} {\bibfield  {journal} {\bibinfo
   {journal} {Physical Review Letters}\ }\textbf {\bibinfo {volume} {24}},\
  \bibinfo {pages} {73} (\bibinfo {year} {1970})}\BibitemShut {NoStop}%
\bibitem [{Av7()}]{Av74}%
  \BibitemOpen
  \href@noop {} {}\bibinfo {note} {H. Averdung, Internal Report KPH 3/74,
  Mainz, 1974 (unpublished)}\BibitemShut {NoStop}%
\bibitem [{\citenamefont {Li}\ \emph {et~al.}(1974)\citenamefont {Li},
  \citenamefont {Yearian},\ and\ \citenamefont {Sick}}]{li1974high}%
  \BibitemOpen
  \bibfield  {author} {\bibinfo {author} {\bibfnamefont {G.}~\bibnamefont
  {Li}}, \bibinfo {author} {\bibfnamefont {M.}~\bibnamefont {Yearian}},\ and\
  \bibinfo {author} {\bibfnamefont {I.}~\bibnamefont {Sick}},\ }\bibfield
  {title} {\bibinfo {title} {High-momentum-transfer electron scattering from mg
  24, al 27, si 28, and s 32},\ }\href@noop {} {\bibfield  {journal} {\bibinfo
  {journal} {Physical Review C}\ }\textbf {\bibinfo {volume} {9}},\ \bibinfo
  {pages} {1861} (\bibinfo {year} {1974})}\BibitemShut {NoStop}%
\bibitem [{\citenamefont {Lees}\ \emph {et~al.}(1976)\citenamefont {Lees},
  \citenamefont {Curran}, \citenamefont {Drake}, \citenamefont {Gillespie},
  \citenamefont {Johnston},\ and\ \citenamefont {Singhal}}]{lees1976elastic}%
  \BibitemOpen
  \bibfield  {author} {\bibinfo {author} {\bibfnamefont {E.}~\bibnamefont
  {Lees}}, \bibinfo {author} {\bibfnamefont {C.}~\bibnamefont {Curran}},
  \bibinfo {author} {\bibfnamefont {T.}~\bibnamefont {Drake}}, \bibinfo
  {author} {\bibfnamefont {W.}~\bibnamefont {Gillespie}}, \bibinfo {author}
  {\bibfnamefont {A.}~\bibnamefont {Johnston}},\ and\ \bibinfo {author}
  {\bibfnamefont {R.}~\bibnamefont {Singhal}},\ }\bibfield  {title} {\bibinfo
  {title} {Elastic electron scattering from the stable isotopes of magnesium},\
  }\href@noop {} {\bibfield  {journal} {\bibinfo  {journal} {Journal of Physics
  G: Nuclear Physics}\ }\textbf {\bibinfo {volume} {2}},\ \bibinfo {pages}
  {105} (\bibinfo {year} {1976})}\BibitemShut {NoStop}%
\bibitem [{\citenamefont {Lombard}\ and\ \citenamefont
  {Bishop}(1967)}]{Lombard:1967skb}%
  \BibitemOpen
  \bibfield  {author} {\bibinfo {author} {\bibfnamefont {R.~M.}\ \bibnamefont
  {Lombard}}\ and\ \bibinfo {author} {\bibfnamefont {G.~R.}\ \bibnamefont
  {Bishop}},\ }\bibfield  {title} {\bibinfo {title} {{The scattering of
  high-energy electrons by 27 Al}},\ }\href
  {https://doi.org/10.1016/0375-9474(67)90655-0} {\bibfield  {journal}
  {\bibinfo  {journal} {Nucl. Phys. A}\ }\textbf {\bibinfo {volume} {101}},\
  \bibinfo {pages} {601} (\bibinfo {year} {1967})}\BibitemShut {NoStop}%
\bibitem [{\citenamefont {Fey}\ \emph {et~al.}(1973)\citenamefont {Fey},
  \citenamefont {Frank}, \citenamefont {Sch{\"u}tz},\ and\ \citenamefont
  {Theissen}}]{fey1973nuclear}%
  \BibitemOpen
  \bibfield  {author} {\bibinfo {author} {\bibfnamefont {G.}~\bibnamefont
  {Fey}}, \bibinfo {author} {\bibfnamefont {H.}~\bibnamefont {Frank}}, \bibinfo
  {author} {\bibfnamefont {W.}~\bibnamefont {Sch{\"u}tz}},\ and\ \bibinfo
  {author} {\bibfnamefont {H.}~\bibnamefont {Theissen}},\ }\bibfield  {title}
  {\bibinfo {title} {Nuclear rms charge radii from relative electron scattering
  measurements at low energies},\ }\href@noop {} {\bibfield  {journal}
  {\bibinfo  {journal} {Zeitschrift f{\"u}r Physik}\ }\textbf {\bibinfo
  {volume} {265}},\ \bibinfo {pages} {401} (\bibinfo {year}
  {1973})}\BibitemShut {NoStop}%
\bibitem [{\citenamefont {Finn}\ \emph {et~al.}(1976)\citenamefont {Finn},
  \citenamefont {Crannell}, \citenamefont {Hallowell}, \citenamefont
  {O'brien},\ and\ \citenamefont {Penner}}]{finn1976elastic}%
  \BibitemOpen
  \bibfield  {author} {\bibinfo {author} {\bibfnamefont {J.}~\bibnamefont
  {Finn}}, \bibinfo {author} {\bibfnamefont {H.}~\bibnamefont {Crannell}},
  \bibinfo {author} {\bibfnamefont {P.}~\bibnamefont {Hallowell}}, \bibinfo
  {author} {\bibfnamefont {J.}~\bibnamefont {O'brien}},\ and\ \bibinfo {author}
  {\bibfnamefont {S.}~\bibnamefont {Penner}},\ }\bibfield  {title} {\bibinfo
  {title} {Elastic electron scattering from the isotopes 36ar and 40ar},\
  }\href@noop {} {\bibfield  {journal} {\bibinfo  {journal} {Nuclear Physics
  A}\ }\textbf {\bibinfo {volume} {274}},\ \bibinfo {pages} {28} (\bibinfo
  {year} {1976})}\BibitemShut {NoStop}%
\bibitem [{\citenamefont {Fivozinsky}\ \emph {et~al.}(1974)\citenamefont
  {Fivozinsky}, \citenamefont {Penner}, \citenamefont {Lightbody~Jr},\ and\
  \citenamefont {Blum}}]{fivozinsky1974electron}%
  \BibitemOpen
  \bibfield  {author} {\bibinfo {author} {\bibfnamefont {S.}~\bibnamefont
  {Fivozinsky}}, \bibinfo {author} {\bibfnamefont {S.}~\bibnamefont {Penner}},
  \bibinfo {author} {\bibfnamefont {J.}~\bibnamefont {Lightbody~Jr}},\ and\
  \bibinfo {author} {\bibfnamefont {D.}~\bibnamefont {Blum}},\ }\bibfield
  {title} {\bibinfo {title} {Electron scattering from sr 88 and y 89},\
  }\href@noop {} {\bibfield  {journal} {\bibinfo  {journal} {Physical Review
  C}\ }\textbf {\bibinfo {volume} {9}},\ \bibinfo {pages} {1533} (\bibinfo
  {year} {1974})}\BibitemShut {NoStop}%
\bibitem [{\citenamefont {Sinha}\ \emph {et~al.}(1973)\citenamefont {Sinha},
  \citenamefont {Peterson}, \citenamefont {Whitney}, \citenamefont {Sick},\
  and\ \citenamefont {McCarthy}}]{sinha1973nuclear}%
  \BibitemOpen
  \bibfield  {author} {\bibinfo {author} {\bibfnamefont {B.}~\bibnamefont
  {Sinha}}, \bibinfo {author} {\bibfnamefont {G.}~\bibnamefont {Peterson}},
  \bibinfo {author} {\bibfnamefont {R.}~\bibnamefont {Whitney}}, \bibinfo
  {author} {\bibfnamefont {I.}~\bibnamefont {Sick}},\ and\ \bibinfo {author}
  {\bibfnamefont {J.}~\bibnamefont {McCarthy}},\ }\bibfield  {title} {\bibinfo
  {title} {Nuclear charge distributions of isotone pairs. ii. k 39 and ca 40},\
  }\href@noop {} {\bibfield  {journal} {\bibinfo  {journal} {Physical Review
  C}\ }\textbf {\bibinfo {volume} {7}},\ \bibinfo {pages} {1930} (\bibinfo
  {year} {1973})}\BibitemShut {NoStop}%
\bibitem [{\citenamefont {Theissen}\ \emph {et~al.}(1969)\citenamefont
  {Theissen}, \citenamefont {Peterson}, \citenamefont {Alston~III},\ and\
  \citenamefont {Stewart}}]{theissen1969elastic}%
  \BibitemOpen
  \bibfield  {author} {\bibinfo {author} {\bibfnamefont {H.}~\bibnamefont
  {Theissen}}, \bibinfo {author} {\bibfnamefont {R.}~\bibnamefont {Peterson}},
  \bibinfo {author} {\bibfnamefont {W.}~\bibnamefont {Alston~III}},\ and\
  \bibinfo {author} {\bibfnamefont {J.}~\bibnamefont {Stewart}},\ }\bibfield
  {title} {\bibinfo {title} {Elastic and inelastic electron scattering from mn
  55},\ }\href@noop {} {\bibfield  {journal} {\bibinfo  {journal} {Physical
  Review}\ }\textbf {\bibinfo {volume} {186}},\ \bibinfo {pages} {1119}
  (\bibinfo {year} {1969})}\BibitemShut {NoStop}%
\bibitem [{La7()}]{La76}%
  \BibitemOpen
  \href@noop {} {}\bibinfo {note} {J.J. Lapikas, Master's thesis, University of
  Amsterdam, 1976 (unpublished)}\BibitemShut {NoStop}%
\bibitem [{\citenamefont {Shevchenko}\ \emph {et~al.}(1978)\citenamefont
  {Shevchenko}, \citenamefont {Polishchuk}, \citenamefont {Kasatkin},
  \citenamefont {Khomich}, \citenamefont {Buki}, \citenamefont {Mazanko},\ and\
  \citenamefont {Shula}}]{shevchenko1978charge}%
  \BibitemOpen
  \bibfield  {author} {\bibinfo {author} {\bibfnamefont {N.}~\bibnamefont
  {Shevchenko}}, \bibinfo {author} {\bibfnamefont {V.}~\bibnamefont
  {Polishchuk}}, \bibinfo {author} {\bibfnamefont {Y.}~\bibnamefont
  {Kasatkin}}, \bibinfo {author} {\bibfnamefont {A.}~\bibnamefont {Khomich}},
  \bibinfo {author} {\bibfnamefont {A.}~\bibnamefont {Buki}}, \bibinfo {author}
  {\bibfnamefont {B.}~\bibnamefont {Mazanko}},\ and\ \bibinfo {author}
  {\bibfnamefont {G.}~\bibnamefont {Shula}},\ }\bibfield  {title} {\bibinfo
  {title} {Charge-density distribution in the nuclei cr-50, cr-52, cr-53, cr-54
  and fe-54, fe-56},\ }\href@noop {} {\bibfield  {journal} {\bibinfo  {journal}
  {SOVIET JOURNAL OF NUCLEAR PHYSICS-USSR}\ }\textbf {\bibinfo {volume} {28}},\
  \bibinfo {pages} {139} (\bibinfo {year} {1978})}\BibitemShut {NoStop}%
\bibitem [{\citenamefont {Ficenec}\ \emph {et~al.}(1970)\citenamefont
  {Ficenec}, \citenamefont {Trower}, \citenamefont {Heisenberg},\ and\
  \citenamefont {Sick}}]{ficenec1970elastic}%
  \BibitemOpen
  \bibfield  {author} {\bibinfo {author} {\bibfnamefont {J.}~\bibnamefont
  {Ficenec}}, \bibinfo {author} {\bibfnamefont {W.}~\bibnamefont {Trower}},
  \bibinfo {author} {\bibfnamefont {J.}~\bibnamefont {Heisenberg}},\ and\
  \bibinfo {author} {\bibfnamefont {I.}~\bibnamefont {Sick}},\ }\bibfield
  {title} {\bibinfo {title} {Elastic electron-nickel scattering},\ }\href@noop
  {} {\bibfield  {journal} {\bibinfo  {journal} {Physics Letters B}\ }\textbf
  {\bibinfo {volume} {32}},\ \bibinfo {pages} {460} (\bibinfo {year}
  {1970})}\BibitemShut {NoStop}%
\bibitem [{Wo7()}]{Wo76}%
  \BibitemOpen
  \href@noop {} {}\bibinfo {note} {H.D. Wohlfahrt, Habilitationsschrift,
  University of Mainz, 1976 (unpublished)}\BibitemShut {NoStop}%
\bibitem [{\citenamefont {Kline}\ \emph {et~al.}(1975)\citenamefont {Kline},
  \citenamefont {Auer}, \citenamefont {Bergstrom},\ and\ \citenamefont
  {Caplan}}]{kline1975electron}%
  \BibitemOpen
  \bibfield  {author} {\bibinfo {author} {\bibfnamefont {F.}~\bibnamefont
  {Kline}}, \bibinfo {author} {\bibfnamefont {I.}~\bibnamefont {Auer}},
  \bibinfo {author} {\bibfnamefont {J.}~\bibnamefont {Bergstrom}},\ and\
  \bibinfo {author} {\bibfnamefont {H.}~\bibnamefont {Caplan}},\ }\bibfield
  {title} {\bibinfo {title} {Electron scattering from 70ge and 72ge},\
  }\href@noop {} {\bibfield  {journal} {\bibinfo  {journal} {Nuclear Physics
  A}\ }\textbf {\bibinfo {volume} {255}},\ \bibinfo {pages} {435} (\bibinfo
  {year} {1975})}\BibitemShut {NoStop}%
\bibitem [{\citenamefont {Garrett}\ and\ \citenamefont
  {Bhalla}(1967)}]{GarrettBhalla}%
  \BibitemOpen
  \bibfield  {author} {\bibinfo {author} {\bibfnamefont {W.~R.}\ \bibnamefont
  {Garrett}}\ and\ \bibinfo {author} {\bibfnamefont {C.~P.}\ \bibnamefont
  {Bhalla}},\ }\bibfield  {title} {\bibinfo {title} {Potential energy shift for
  a screnned coulomb potential},\ }\href {https://doi.org/10.1007/BF01325974}
  {\bibfield  {journal} {\bibinfo  {journal} {Z. Phys.}\ }\textbf {\bibinfo
  {volume} {198}},\ \bibinfo {pages} {453} (\bibinfo {year}
  {1967})}\BibitemShut {NoStop}%
\bibitem [{\citenamefont {Rose}(1936)}]{Rose:1936zz}%
  \BibitemOpen
  \bibfield  {author} {\bibinfo {author} {\bibfnamefont {M.~E.}\ \bibnamefont
  {Rose}},\ }\bibfield  {title} {\bibinfo {title} {{A Note on the Possible
  Effect of Screening in the Theory of Beta-Disintegration}},\ }\href
  {https://doi.org/10.1103/PhysRev.49.727} {\bibfield  {journal} {\bibinfo
  {journal} {Phys. Rev.}\ }\textbf {\bibinfo {volume} {49}},\ \bibinfo {pages}
  {727} (\bibinfo {year} {1936})}\BibitemShut {NoStop}%
\bibitem [{\citenamefont {Salvat}\ \emph {et~al.}(1987)\citenamefont {Salvat},
  \citenamefont {Martnez}, \citenamefont {Mayol},\ and\ \citenamefont
  {Parellada}}]{Salvat:1987zz}%
  \BibitemOpen
  \bibfield  {author} {\bibinfo {author} {\bibfnamefont {F.}~\bibnamefont
  {Salvat}}, \bibinfo {author} {\bibfnamefont {J.~D.}\ \bibnamefont {Martnez}},
  \bibinfo {author} {\bibfnamefont {R.}~\bibnamefont {Mayol}},\ and\ \bibinfo
  {author} {\bibfnamefont {J.}~\bibnamefont {Parellada}},\ }\bibfield  {title}
  {\bibinfo {title} {{Analytical Dirac-Hartree-Fock-Slater screening function
  for atoms (Z=1-92)}},\ }\href {https://doi.org/10.1103/PhysRevA.36.467}
  {\bibfield  {journal} {\bibinfo  {journal} {Phys. Rev. A}\ }\textbf {\bibinfo
  {volume} {36}},\ \bibinfo {pages} {467} (\bibinfo {year} {1987})}\BibitemShut
  {NoStop}%
\bibitem [{\citenamefont {Rinker}\ and\ \citenamefont
  {Speth}(1978)}]{Rinker:1978kh}%
  \BibitemOpen
  \bibfield  {author} {\bibinfo {author} {\bibfnamefont {G.~A.}\ \bibnamefont
  {Rinker}}\ and\ \bibinfo {author} {\bibfnamefont {J.}~\bibnamefont {Speth}},\
  }\bibfield  {title} {\bibinfo {title} {{Nuclear Polarization in Muonic
  Atoms}},\ }\href {https://doi.org/10.1016/0375-9474(78)90471-2} {\bibfield
  {journal} {\bibinfo  {journal} {Nucl. Phys. A}\ }\textbf {\bibinfo {volume}
  {306}},\ \bibinfo {pages} {397} (\bibinfo {year} {1978})}\BibitemShut
  {NoStop}%
\bibitem [{\citenamefont {Bhalla}\ and\ \citenamefont
  {Rose}(1962)}]{BhallaRose}%
  \BibitemOpen
  \bibfield  {author} {\bibinfo {author} {\bibfnamefont {C.~P.}\ \bibnamefont
  {Bhalla}}\ and\ \bibinfo {author} {\bibfnamefont {M.~E.}\ \bibnamefont
  {Rose}},\ }\bibfield  {title} {\bibinfo {title} {Finite nuclear size effects
  in $\ensuremath{\beta}$ decay},\ }\href
  {https://doi.org/10.1103/PhysRev.128.774} {\bibfield  {journal} {\bibinfo
  {journal} {Phys. Rev.}\ }\textbf {\bibinfo {volume} {128}},\ \bibinfo {pages}
  {774} (\bibinfo {year} {1962})}\BibitemShut {NoStop}%
\end{thebibliography}%

\end{document}